\documentclass[twoside,11pt,preprint]{article}

\usepackage{blindtext}

\usepackage{jmlr2e}

\usepackage[utf8]{inputenc} 
\usepackage[T1]{fontenc}    
\usepackage{hyperref}       
\usepackage{url}            
\usepackage{booktabs}       
\usepackage{amsfonts}       
\usepackage{nicefrac}       
\usepackage{microtype}      
\usepackage{xcolor}         
\usepackage{setspace}
\usepackage{algpseudocode,algorithm,algorithmicx}
\usepackage{amsmath}
\usepackage{amssymb}
\usepackage[export]{adjustbox}
\usepackage{caption}
\usepackage{subcaption}
\usepackage{wrapfig}
\usepackage{float,graphicx}
\usepackage{dsfont}
\usepackage{easyReview}
\usepackage{multirow}

\def \b{\beta}
\def \<{\langle}
\def \>{\rangle}

\newcommand{\cb}[1]{{\color{blue}{#1}}}

\usepackage{lastpage}
\jmlrheading{26}{2025}{1-\pageref{LastPage}}{9/24; Revised
5/25}{7/25}{24-1575}{Wei Zhang and Jr-Shin Li}
\ShortHeadings{Reinforcement Learning for Infinite-Dimensional Systems}{Zhang and Li}

\firstpageno{1}

\begin{document}

\title{
Reinforcement Learning for Infinite-Dimensional Systems
}

\author{\name Wei Zhang \email wei.zhang@wustl.edu \\
       \addr Department of Electrical \& Systems Engineering \\
       Washington University in St. Louis\\
       St. Louis, MO 63130, USA
       \AND
       \name Jr-Shin Li \email jsli@wustl.edu \\
       \addr Department of Electrical \& Systems Engineering\\
       Division of Computational \& Data Sciences\\
       Division of Biology \& Biomedical Sciences\\
       Washington University in St. Louis\\
       St. Louis, MO 63130, USA}

\editor{Nan Jiang}

\maketitle


\begin{abstract}
Interest in reinforcement learning (RL) for large-scale systems, comprising extensive populations of intelligent agents interacting with heterogeneous environments, has surged significantly across diverse scientific domains in recent years. However, the large-scale nature of these systems often leads to high computational costs or reduced performance for most state-of-the-art RL techniques. To address these challenges, we propose a novel RL architecture and derive effective algorithms to learn optimal policies for arbitrarily large systems of agents. In our formulation, we model such systems as parameterized control systems defined on an infinite-dimensional function space. We then develop a moment kernel transform that maps the parameterized system and the value function into a reproducing kernel Hilbert space. This transformation generates a sequence of finite-dimensional moment representations for the RL problem, organized into a filtrated structure. Leveraging this RL filtration, we develop a hierarchical algorithm for learning optimal policies for the infinite-dimensional parameterized system. To enhance the algorithm's efficiency, we exploit early stopping at each hierarchy, demonstrating the fast convergence property of the algorithm through the construction of a convergent spectral sequence. The performance and efficiency of the proposed algorithm are validated using practical examples in engineering and quantum systems.
\end{abstract}


\section{Introduction}
Reinforcement learning (RL), a prominent machine learning paradigm, has gained recognition as a powerful tool for intelligent agents to learn optimal policies. These policies guide the agents' decision-making processes toward achieving desired goals by maximizing the expected cumulative rewards. RL exhibits a wide range of applications across various domains,  encompassing robotics, game-playing, recommendation systems, autonomous vehicles, and control systems. In recent years, learning optimal policies to manipulate the behavior of large-scale systems of intelligent agents interacting with different environments has attracted increasing attention and is gradually becoming a recurrent research theme in the RL community. Existing and potential applications of ``large-scale RL'' include targeted coordination of robot swarms for motion planning in robotics \citep{Becker12,Shahrokhi2018},
desynchronization of neuronal ensembles with abnormal oscillatory activity for the treatment of movement disorders such as Parkinson's disease, epilepsy, and essential tremor in neuroscience and brain medicine \citep{Marks05,Wilson05,Shoeb09,Zlotnik12,Li13,Vu2024},
and the robust excitation of spin ensembles for nuclear magnetic resonance (NMR) spectroscopy, imaging (MRI), and quantum computation in quantum science and technology \citep{Glaser98,Li_PNAS11,Dong08,Chen14}.

The fundamental challenge in these RL tasks undoubtedly lies in the massive scale of agent systems and the dynamic environments in which the agents take actions. For instance, a neuronal ensemble in the human brain may comprise up to $10^{11}$ neurons \citep{Suzana2012,Ching13}, and a quantum ensemble in NMR experiments consists of spins on the order of Avogadro's number $10^{23}$ \citep{Li_thesis,Cavanagh10,Li_TAC11}. Although, mathematically speaking, these systems are composed of finitely many agents, it is more appropriate to treat them as infinite populations. This is because their massive scale makes it infeasible 
to specifically identify and make decisions for each individual agent in these populations. These restrictions particularly imply that such policy learning tasks exceed the capabilities of classical multi-agent RL methods, which focus on 
learning joint policies that provide each agent with a customized action based on the states of all the other agents \citep{Cohen1994,Lee2016,Gupta2017,Zhang2021,Albrecht2024}. Consequently, all the agents are required to take the same action, necessitating the learning and implementation of policies at the population level. In addition, natural models describing the dynamic behavior of these intelligent agents typically take the form of continuous-time deterministic control systems. Learning optimal policies for this category of systems also lies outside the scope of most state-of-the-art and benchmark RL algorithms, notably Trust Region Policy Optimization (TRPO) and Proximal Policy Optimization (PPO) \citep{Schulman2017,Schulman15}, which are based on Markov decision processes (MDPs) and operate as discrete-time stochastic control processes. This, in turn, underscores the urgent demand for a more inclusive RL framework that accommodates policy learning tasks for infinite populations of agents in a continuous-time deterministic setting.

\paragraph{Our contributions.} 
This work is devoted to developing a novel RL architecture that enables the derivation of effective algorithms to learn optimal policies for population systems consisting of infinitely many intelligent agents interacting with heterogeneous dynamic environments. We formulate such an agent system as a parameterized control system, where each individual system indexed by a specific parameter value represents the environment of an agent in the population. We then evolve this parameterized system on an infinite-dimensional function space and carry out a functional setting for RL of this infinite-dimensional dynamical system. The primary tool we develop to tackle this RL problem is the moment kernel transform. This transformation maps the parameterized system and the value function to a control system and a value function defined on a reproducing kernel Hilbert space (RKHS) consisting of moment sequences, thus yielding a kernel parameterization of the RL problem. The use of moment sequences for this kernelization directly enables finite-dimensional truncation representations of the infinite-dimensional RL problem. We then organize these representations into a filtrated structure with respect to the truncation order. Leveraging this RL filtration, we develop a hierarchical policy learning algorithm, wherein each hierarchy consists of an RL problem for a finite-dimensional truncated moment-kernelized system. To enhance the computational efficiency of the proposed algorithm, we also develop early stopping criteria for the finite-dimensional RL problem in each hierarchy and prove convergence of the hierarchical algorithm with early-stopped hierarchies in terms of spectral sequences. The performance and efficiency of the proposed hierarchical policy learning algorithm are demonstrated using examples arising from practical applications. The contributions of our work are summarized as follows. 

\begin{itemize}
    \item Formulation of an infinite population of intelligent agents as a parameterized control system defined on an infinite-dimensional function space. 
    \item Development of the moment kernel transform that gives rise to a kernel parameterization of RL problems for parameterized systems in terms of moment sequences in an RKHS. 
    \item Design of a filtrated RL algorithm for learning optimal policies of parameterized systems with convergence guarantees. 
    \item Exploration of early stopping criteria for each hierarchy in the proposed hierarchical algorithm and proof of the spectral sequence convergence of the hierarchical algorithm with early-stopped hierarchies.
    \end{itemize}

\paragraph{Related works.} Learning control policies to coordinate large populations of intelligent agents using RL approaches has attracted increasing attention in recent years. Significant achievements have been recognized in the field of control theory and engineering. Prominent examples include control of multi-agent systems \citep{Lucian2010,Mou2019,Jing21,Zhang2021,Albrecht2024} and many-body quantum systems \citep{Dong08,Lamata2017,Bukov2018,Haug2021}, as well as planning and regulation for multi-robot systems \citep{Mataric1997,Long2018,Yang2020,Yang2024,Lu2024}. Furthermore, large-scale RL has also gained great success in various emerging applications that are generally considered to be outside the traditional control-theoretic scope, notably, strategic gaming 
\citep{Silver2017,Silver2018,Vinyals2019,OpenAI2019,Schrittwieser2020,Kaiser2020} and human-level decision making \citep{Mnih2015,Liu2022,Baker2020}.


One of the most active research focuses regarding large-scale and high-dimensional policy learning problems is deep RL, which incorporates deep learning techniques, particularly the use of deep neural networks, into RL algorithms \citep{Bertsekas1996,Vincent2018,Bellemare2020,Tenavi2021,Le2022,Sarang23}. Despite the advantage of exceptional generalizability, training deep neural networks to tackle large-scale and/or high-dimensional learning problems is widely known to suffer from the notorious curse of dimensionality, leading to expensive computational costs and scalability issues. Various approaches have been proposed to mitigate the impact of these phenomena, primarily including the development of distributed and multi-agent RL algorithms \citep{Mou2019,Heredia20,Yazdanbakhsh20,Heredia2022,Xie2024} and the search for compact representations of high-dimensional measurement data. Notable among these are the successor representation \citep{Momennejad2017}, latent space representation \citep{Chaudhuri2019}, contrastive unsupervised representation \citep{Daume2020}, and invariant representation \citep{Zhang2021_RL}. These works have achieved considerable success in learning optimal policies for agent populations on a scale ranging from tens to thousands. 

To further push the boundaries of RL towards addressing larger agent populations, tools in mean-field theory - particularly mean-field games and control - have drawn increasing attention and have been adopted in the RL setting in recent years. Instead of learning a policy for each individual agent in a population, mean-field RL focuses on the mean-field approximation of the population, where each collection of interacting agents is replaced by a single agent representing their averaged behavior \citep{Yang2018,Subramanian2019,Laurire2022,Pasztor2023,Bensoussan2013,Carmona2019,Fu2020,Carmona2020}. However, the prerequisite of the mean-field approximation, arising from the
fundamental principles of statistical physics, places the primary focus of mean-field RL on populations of identical agents \citep{Pathria2021}. In this context, policy learning for arbitrarily large, and in the limit, infinite populations of heterogeneous intelligent agents, as considered in this work, remains under-explored with sparse literature in the RL community. On the other hand, the formulation of RL problems over infinite-dimensional spaces has only been proposed in the stochastic setting, aiming to learn feedback control policies for stochastic partial differential equation systems using variational optimization methods \citep{Evans2020}.

The major technical tool developed in this work to overcome the curse of dimensionality is the moment kernel transform, which is inspired by the method of moments. This method was developed by the Russian mathematician P. L. Chebyshev in 1887 to prove the law of large numbers and the central limit theorem \citep{Mackey80}. Since then, the method has been extensively studied under different settings, notably the Hausdorff, Hamburger, and Stieltjes moment problems \citep{Hamburger20,Hamburger20_2,Hamburger21,Hausdorff23,Stieltjes93}. The most general formulation in modern terms was proposed by the Japanese mathematician Kōsaku Yosida \citep{Yosida80}. Recently, the method of moments was introduced to control theory for establishing dual representations of ensemble systems \citep{Narayanan2024,Li_SICON25} and in machine learning-aided medical decision making as a feature engineering technique \citep{Yu2023}. These prior works 
lay the foundation for the development of the moment kernel transform in this work.

In addition to establishing the filtrated RL architecture, the moment kernel transform also gives rise to a reduced kernel representation of (infinite) agent populations over an RKHS. It is widely known that RKHS theory forms the building blocks for kernel methods in machine learning, notably support vector machines (SVMs) and kernel principal component analysis (kernel PCA) \citep{Hastie2009,Paulsen2016}. In the context of RL, elements of RKHSs are commonly used as function approximators for MDPs, through which the learning targets, including policies, value functions, and/or transition maps, are
estimated in terms of linear combinations of reproducing kernels \citep{Guy2015,Yang2019,Yang2020_NIPS,Alec2021}. This kernel approximation is typically enabled by imposing the condition that these learning targets belong to an RKHS consisting of functions defined on the state space of the MDPs. In our work, we explore the use of RKHS-theoretic techniques within a functional setup rather than the traditional MDP setup, thus relaxing the condition on the learning targets. More importantly, along with enabling kernel approximation, the developed moment kernel transform also serves as a model reduction machine for arbitrarily large agent populations.

\section{Policy Learning for Massive-Scale Dynamic Populations}
\label{sec:ensemble_RL}

This section is primarily dedicated to establishing a general formulation for RL of arbitrarily large populations of intelligent agents, which interact with heterogeneous environments in continuous time and are regulated by policies taking values in continuous spaces. We begin by demonstrating the challenges associated with such RL tasks, and then introduce our formulation that models these populations as continuous-time parametrized control systems defined on infinite-dimensional function spaces. We subsequently delve into imposing conditions that guarantee the existence of optimal policies for RL problems involving these infinite-dimensional control systems over function spaces.

\subsection{Challenges to reinforcement learning of large-scale population agents}
\label{sec:challenges}
Learning control policies for a large-scale population of intelligent agents, effectively a continuum in the limit, described by continuous-time deterministic dynamical systems, presents a significant challenge to RL. The primary obstacle is the enormous population size, which forces RL algorithms to operate in high-dimensional spaces. This inevitably leads to the curse of dimensionality, resulting in high computational costs and reduced learning accuracy \citep{Bellman1957,Bellman1961,Sutton18}. Additionally, the continuous-time deterministic system formulation is inconsistent with the setting of most state-of-the-art RL algorithms, which are typically based on MDPs modeled by  
discrete-time stochastic control processes.

To illustrate these challenges with a concrete example, we consider a population system consisting of $N$ dynamic agents regulated by a common control policy, given by 
\begin{align}
\label{eq:motivating_example}
    \frac{d}{dt}x_i(t)=a_ix_i(t)+u(t), \quad i=1,\dots,N,
\end{align}
where $x_i(t)\in\mathbb{R}$ is the state of agent $i$, $u(t)\in\mathbb{R}$ is the control policy, and $a_i\in\mathbb{R}$ for all $i$. The entire size-$N$ population can be represented as an $N$-dimensional system, $\frac{d}{dt}x(t)=A_Nx(t)+B_Nu(t)$, where $x(t)=[\,x_1(t), \dots, x_N(t) \,]'\in\mathbb{R}^N$ is the population state with '$\prime$' denoting the transpose of vectors (and matrices), $A_N\in\mathbb{R}^{N\times N}$ is the diagonal matrix with the $(i,i)$-entry given by $a_i$ for all $i=1,\dots,N$, and $B_N\in\mathbb{R}^{N}$ is a vector of ones. To put the analysis in a simplified setting, we choose $-1=a_0<a_1<\dots<a_N=1$ to be the uniform partition of the interval $[-1,1]$, i.e., $a_i=-1+2(i-1)/(N-1)$ for each $i=1,\dots,N$. We aim to learn the infinite-time horizon linear quadratic regulator (LQR) with the state-value function (cumulative reward or cost-to-go), 
$V(x(t))=\int_t^\infty e^{-2.5 t} \Big[\frac{1}{N}\sum_{i=1}^Nx_i^2(t)+u^2(t)\Big]dt=\int_t^\infty e^{-2.5 t}\Big[\frac{1}{N}x'(t)x(t)+u^2(t)\Big]dt$. It is well-known that the LQR value function $V^*(x(t))=\inf_uV(x(t))$ is in the quadratic form $V^*(x(t))=x'(t)Qx(t)$,  parameterized by a positive definite matrix $Q\in\mathbb{R}^{N\times N}$ \citep{Brockett15}. Therefore, the value iteration stands out as the prime algorithm to tune the $N(N+1)/2$ training parameters in $Q$ to learn the LQR policy and value function \citep{Sutton18,Bertsekas19}. 

\begin{figure}[H]
     \centering
     \begin{subfigure}[b]{0.49\textwidth}
     \centering
     \includegraphics[width=\textwidth]{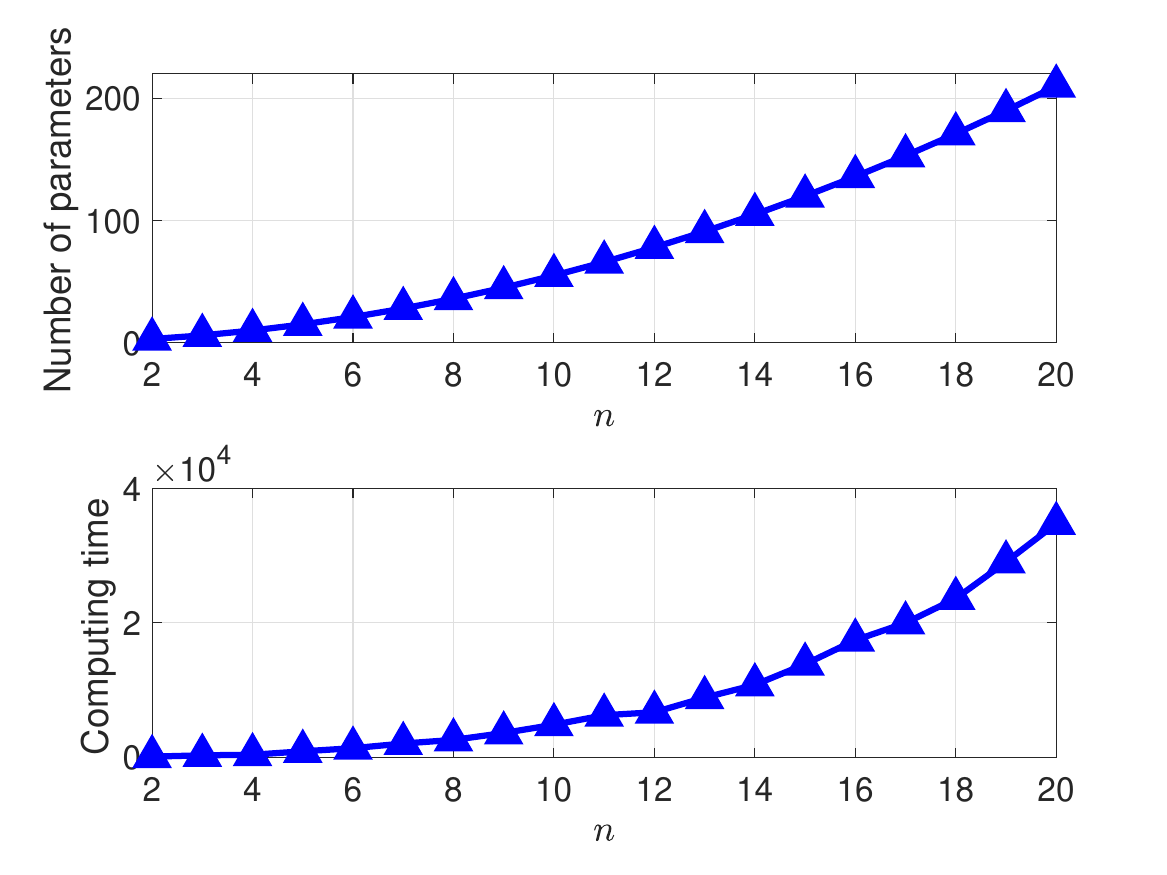}
     \caption{}
     \label{fig:curse_of_dim}
     \end{subfigure}
     \begin{subfigure}[b]{0.48\textwidth}
     \centering
     \includegraphics[width=\textwidth]{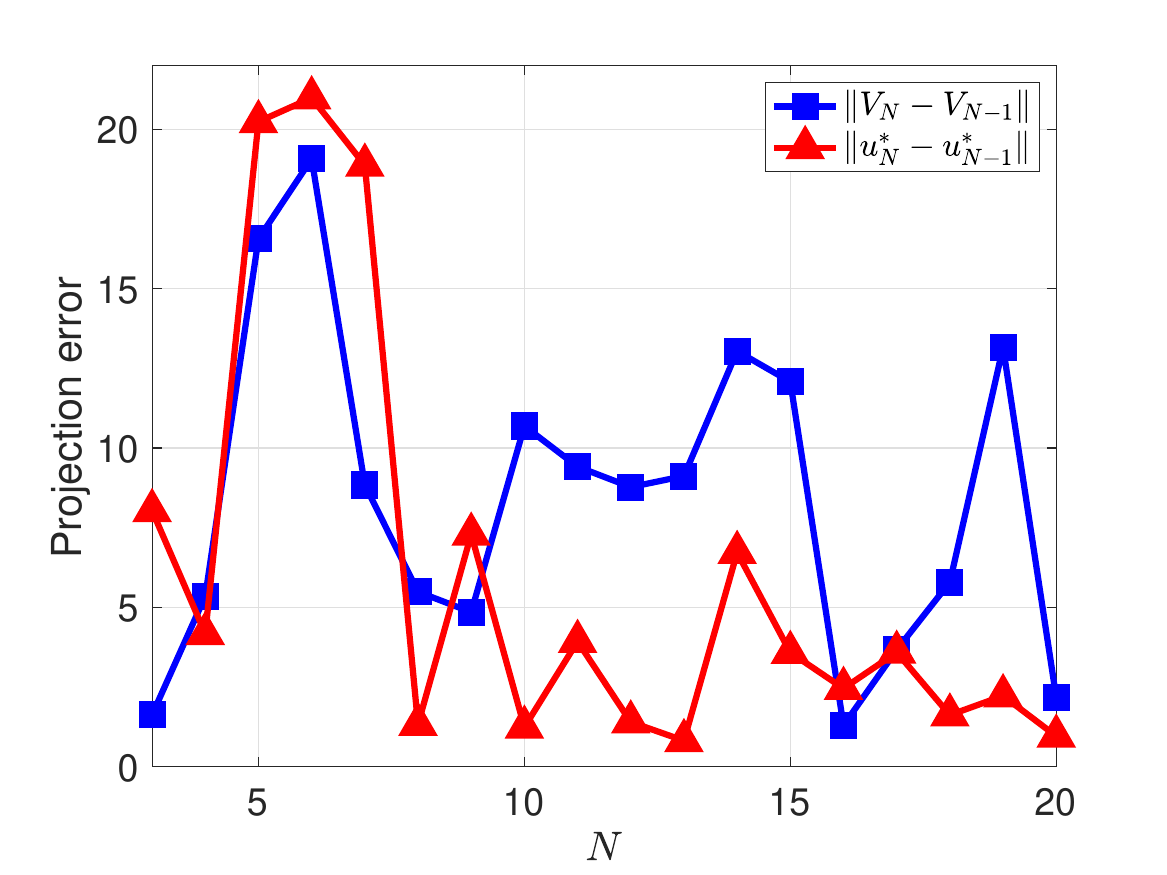}
     \caption{}
      \label{fig:non_convergence}
     \end{subfigure}
     \caption{\footnotesize Illustration of the challenges to RL for high-dimensional systems. The standard value iteration is applied to learn the LQR policy and value function for an $N$-dimensional time-invariant deterministic linear system, with $N$ ranging from 2 to 20. (a) shows the number of training parameters (top) and average computing time (bottom) with respect to the system dimension \cb{$N$}. (b) plots the average distance between successive terms in the sequences of the learned value functions (blue) and LQR policies (red) with respect to $N$.}
     \label{fig:challenge}
 \end{figure}

\paragraph{Curse of dimensionality.} In the simulation, we varied the system dimension $N$ from 2 to 20. For each $N$, we independently ran the value iteration 5 times with random initial conditions. The simulation results are shown in Figure \ref{fig:challenge}. In particular, Figure \ref{fig:curse_of_dim} shows the number of training parameters (top panel) and the average computational time over the 5 runs of the value iteration versus $N$ (bottom panel). From these plots, we observe dramatic increases in both quantities. In addition to these common phenomena associated with the curse of dimensionality, we also observed convergence issues in this high-dimensional system policy learning problem.

\paragraph{Convergence issue.} Note that by the definition of Riemann integrals \citep{Rudin76}, for each $t$, the term $\frac{1}{N}\sum_{i=1}^Nx_i^2(t)$ in the state-value function $V$ is essentially a Riemann sum of the real-value function $\frac{1}{2}x(t,\cdot)$ over the interval $[-1,1]$. Therefore, by the Dominated Convergence theorem \citep{Folland13}, $V$ possesses the convergence property,
\begin{align}
\label{eq:challenge_convergence}
   V(x(t))=\int_t^\infty e^{-2.5 t}\Big[\frac{1}{N}\sum_{i=1}^Nx_i^2(t)+u^2(t)\Big]dt\rightarrow\int_t^\infty\frac{e^{-2.5 t}}{2}\Big[\int_{-1}^1x^2(t,\beta)d\beta+u^2(t)\Big]dt 
\end{align}
as $N\rightarrow\infty$, where $x(t,\beta)$ satisfies the linear parameterized system on $\mathbb{R}$, given by
\begin{align}
\label{eq:challenge_system}
    \frac{d}{dt}x(t,\beta)=\beta x(t,\beta)+u(t),\quad \beta\in\Omega= [-1,1]. 
\end{align}
To see that the value function also inherits this convergence property, we notice that the state-value function is continuous, and hence upper semi-continuous, for any control policy $u$. This implies that the values function $V(x(t))=\inf_u V(x(t))$ is upper semi-continuous \citep{Folland13}. Together with the non-negativity of $V$, we obtain the convergence of the state-value function, $\lim_{N\rightarrow\infty}\inf_u V(x(t))=\inf_u\lim_{N\rightarrow\infty}V(x(t))=\inf_u\int_t^\infty\frac{e^{-2.5 t}}{2}\Big[\int_{-1}^1x^2(t,\beta)d\beta+u^2(t)\Big]dt$ (see Theorem \ref{thm:existence} below for the justification that the right-hand side is well-defined) \citep{Folland13}. Let $V_N^*$ and $u_N^*$ denote the value function and LQR policy learned from the $N$-dimensional system. As shown in Figure \ref{fig:non_convergence}, neither $\|V_N^*-V_{N-1}^*\|=\sup_{t\geq0}|V_N^*(x^*(t))-V_{N-1}^*(x^*(t))|$ nor $\|u_N^*-u_{N-1}^*\|=\sup_{t\geq0}|u_N^*(t)-u_{N-1}^*(t)|$ shows a trend of converging to 0, where $x^*(t)$ denotes the optimal trajectory. This particularly fails to verify the convergence of the value function as illustrated in \eqref{eq:challenge_convergence}.

To further elaborate on this convergence issue, we treat the parameterized system in \eqref{eq:challenge_system} as a continuum ensemble of linear systems indexed by $\beta\in[-1,1]$. The finite ensemble in \eqref{eq:motivating_example} can be seen as essentially a size-$N$ sample of this infinite ensemble. From this perspective, the convergence issue arises due to the ill-posed problem stemming from the  
sampling (discretizing) of the continuum ensemble. Specifically, the optimal policy for a size-$n$ sample may be highly suboptimal for a size-$m$ ensemble for $m>n$, so that $u_m$ is significantly different from $u_n$. To carry out a quantitative analysis in the simplest setting, we consider the case of learning a policy that steers the entire infinite ensemble to the origin $0$ with minimal energy. Specifically, the cumulative reward function is given by $V(t,x(t))=\int_t^1u^2(t)dt$. The minimum energy policy for the size-$n$ sample is given by $u_n(t)=B_n'e^{-A_nt}W_n^{-1}(0,1)x_{0n}$ so that $V_n(t,x(t))=x_{0n}'W_n^{-1}(t,1)x_{0n}$, where $W_n(t_1,t_2)=\int_{t_1}^{t_2}e^{-A_ns}B_nB_n'e^{-A_n's}ds$ for $0\leq t_1\leq t_2\leq 1$ is referred to as the \emph{controllability Gramian}, and $x_{0n}$ is the initial condition of the size-$n$ ensemble system (Brockett, 2015; Liberzon, 2012). It is not hard to see that the analytic properties of $V_n$ and $u_n$ primarily depend on those of $W_n$. Specifically, the $(i,j)$-entry of $W_n(t_1,t_2)$ is given by $\frac{e^{-(a_i+a_j)t_1}-e^{-(a_i+a_j)t_2}}{a_i+a_j}$. Its trace, ${\rm tr}\big(W_n(t_1,t_2)\big)=\sum_{i=1}^n\frac{e^{-2a_it_1}-e^{-2a_it_2}}{2a_i}$, satisfies $\frac{1-e^{-2}}{2}\leq\frac{1}{n}{\rm tr}\big(W_n(t_1,t_2)\big)\leq\frac{e^2-1}{2}$. Hence, all the eigenvalues of $W(t_1,t_2)$ are bounded below and above by $\lambda_{\rm min}=\frac{1-e^{-2}}{2}$ and $\lambda_{\rm max}=\frac{e^2-1}{2}$, respectively. As a result, for any different sample sizes $m$ and $n$, there are initial conditions $x_{0n}$ and $x_{0m}$ such that $\sup_t|V_n(t,x(t))-V_m(t,x(t))|\geq\lambda_{\rm min}^{-1}-\lambda_{\rm max}^{-1}=2$ and $\sup_t|u_n(t)-u_m(t)|\geq2$, highlighting the convergence issue. 

To overcome the presented challenges, we propose a new kernel parameterization technique that will serve as a foundational element 
for developing a novel RL architecture. To facilitate this, 
we will first introduce a principled RL formulation that accounts for large-scale population systems, regardless of their size.

\subsection{Reinforcement learning for parameterized systems on function spaces}

The parameterized representation in \eqref{eq:challenge_system} of the ``linear agents'' population in the limiting case inspires the modeling of agent populations of any size as parameterized differential equation systems, also referred to as \emph{ensemble systems}, of the form 
\begin{align}
    \label{eq:ensemble}
    \frac{d}{dt}x(t,\beta)=F(t,\beta,x(t,\beta),u(t)),
\end{align}
where $\beta$ is the system parameter taking values in  $\Omega\subseteq\mathbb{R}^d$, $x(t,\beta)\in M$ is the state of the system indexed by the particular value of $\beta$ (the environmental state of agent $\b$)
in the population with $M\subseteq\mathbb{R}^n$ being a differentiable manifold, $u(t)\in\mathbb{R}^r$ is the control policy, and $F(t,\beta,\cdot,u(t))$ is a (time-varying) vector field on $M$ for each $\beta\in\Omega$ and $u(t)\in\mathbb{R}^r$, characterizing the environment of the agent $\beta$. Similarly, as motivated by the Riemann sum convergence illustrated in \eqref{eq:challenge_convergence}, we associate the parametrized ensemble system in \eqref{eq:ensemble} with a state-value function in the integral form, given by, 
\begin{align}
\label{eq:ensemble_rewards} V(t,x_t)=\int_\Omega\Big[\int_t^Tr(s,x(s,\beta),u(s))ds+K(T,x(T,\beta))\Big]d\beta,
\end{align}
where $x_t(\cdot)\doteq x(t,\cdot)$, and $r(s,x(s,\beta),u(s))$ and $K(T,x(T,\beta))$ are the running and terminal costs of the agent $\beta$, respectively. One of the major advantages of the parameterized system formulation is the ability to model agent populations of arbitrarily large size, in the limit a continuum of agents when the parameter space $\Omega$ has an uncountable cardinality, as in the case of \eqref{eq:challenge_system}. However, due to practical limitations on sensing capability and computing power, it is impossible to collect comprehensive measurement data documenting the state and reward information for all the agents in such a massive-scale agent population. This particularly disables the design and implementation of feedback control policies, which lays this policy learning problem beyond the scope of many existing RL and optimal control methods.

To develop a new RL paradigm inclusive for this type of policy learning tasks, we view the parameterized system in \eqref{eq:ensemble} as well as its state-value function in \eqref{eq:ensemble_rewards} from a different perspective. Indeed, the ensemble state of the parameterized system in is a function $x(t,\cdot):\Omega\rightarrow M$ so that the system is evolving on a space $\mathcal{F}(\Omega,M)$ of $M$-valued functions defined on $\Omega$; the state-value function is a functional on $\mathcal{F}(\Omega,M)$, essentially characterizing the ``average'' of the cumulative rewards of all the agents over the entire population. This functional viewpoint in turn places the policy learning tasks over the function space $\mathcal{F}(\Omega,M)$. When $\Omega$ is an infinite space, $\mathcal{F}(\Omega,M)$ is an infinite-dimensional manifold. Therefore, some regularity conditions on the system dynamics $F$, immediate reward $r$, and terminal cost $K$, more stringent than the canonical setup RL, are expected to guarantee solvability of this policy learning tasks.

\paragraph{Assumption S1.} (Boundedness of control policies) The ensmeble control policy $u:[0,T]\rightarrow\mathbb{R}^r$ is a measurable function and takes values on a compact subset $U$ of $\mathbb{R}^r$. \\


\paragraph{Assumption S2.} (Lipschitz continuity of system dynamics) The vector field $F:\mathbb{R}\times\Omega\times M\times U\rightarrow TM$ is continuous in all of the variables and Lipschitz continuous in $z\in M$ uniformly for $(t,\b,a)\in [0,T]\times\Omega\times U$, that is, there exists a constant $C$ (independent of $t\in[0,T]$, $\beta\in\Omega$, and $a\in U$) such that $|(\varphi_*F|_V)(t,\beta,\varphi(p),a)-(\varphi_*F|_V)(t,\beta,\varphi(q),a)|\leq C|\varphi(p)-\varphi(q)|$ for any $p,q\in V$ and coordinate chart $(V,\varphi)$ on $M$, where $TM$ denotes the tangent bundle of $M$, $\varphi_*F|_V$ denotes the pushforword of $F|_V$, the restriction of the vector field $F$ on $V\subseteq M$, and $|\cdot|$ denotes a norm on $\mathbb{R}^n$. \\

According to the theory of ordinary differential equations, Assumption S2 guarantees that, driven by any admissible control policies, each individual system, say the one indexed by $\beta$, in the ensemble in \eqref{eq:ensemble} has a unique and Lipschitz continuous solution $t\mapsto x(t,\beta)$ \citep{Arnold78,Lang99}. Correspondingly, on the population level with $\b$ varying on $\Omega$, the ensemble system has a unique solution on $\mathcal{F}(\Omega,M)$, given by $t\mapsto x(t,\beta)$. 

\paragraph{Assumption C1.} (Integrability of the state-value function) There exists an ensemble control policy $u\in\mathcal{U}$ such that $\int_\Omega\int_0^T|r(t, x(t,\beta),u(t))|dtd\b+\int_\Omega|K(T,x(T,\b))|d\b<\infty$, where $x(t,\cdot)\in\mathcal{F}(\Omega,M)$ is the solution of the ensemble system in \eqref{eq:ensemble} driven by $u$.

\paragraph{Assumption C2.} (Lipschitz continuity of priviate costs) Both the running cost $r:\mathbb{R}\times M\times U\rightarrow\mathbb{R}$ and terminal cost $K:\mathbb{R}\times M\rightarrow\mathbb{R}$ are continuous functions in all the variables and Lipschitz continuous in $z\in M$ for any $(t,a)\in[0,T]\times U$.\\

In the sequel, these regularity assumptions will be exploited to prove the existence of a solution to the functional policy learning problem formulated in \eqref{eq:ensemble} and \eqref{eq:ensemble_rewards}. In other words, the value function $V^*(t,x_t)=\inf_uV(t,x_t)$ is a well-defined real-valued function on $[0,T]\times\mathcal{F}(\Omega,M)$. Notationally, to emphasize the dependence of the state-value function $V$ on the policy $u$, we denote the total cost by $V(0,x_0)=J(u)$, which then defines a function $J:\mathcal{U}\rightarrow\mathbb{R}$, referred to as the cost functional, with $\mathcal{U}=\{u:[0,T]\rightarrow\mathbb{R}^r\mid u(t)\in U\}$ the space of admissible policies. The solvability of the learning problem then boils down to proving the compactness of $\mathcal{U}$ and continuity of $J$.

\begin{lemma}
\label{lem:compact_control}
The space of admissible policies $\mathcal{U}$ is compact.
\end{lemma}
\begin{proof}
Topologically, $\mathcal{U}$, equipped with the topology of pointwise convergence, is the product space $\prod_{t\in[0,T]}U$ under the product topology. Because $U\subset\mathbb{R}^m$ is a compact by Assumption S1, the compactness of $\mathcal{U}$ directly follows from Tychonoff's theorem \citep{Munkres00}.
\end{proof}

\begin{lemma}
\label{lem:sequential_continuity_cost}
The cost functional $J:\mathcal{U}\rightarrow\mathbb{R}$ is sequentially continuous, that is, $J(u_k)\rightarrow J(u)$ for any admissible policy sequence $(u_k)_{k\in\mathbb{N}}$ such that $u_k\rightarrow u$ in $\mathcal{U}$.
\end{lemma}
\begin{proof}
See Appendix \ref{appd:sequential_continuity_cost}.
\end{proof}


\begin{theorem}[Existence of optimal policies]
\label{thm:existence}
Given a parameterized ensemble system defined on the function space $\mathcal{F}(\Omega,M)$ as in \eqref{eq:ensemble} satisfying Assumptions S1 and S2 and the state-value function defined in \eqref{eq:ensemble_rewards} satisfying Assumptions C1 and C2. Then, the value function $V^*:[0,T]\times\mathcal{F}(\Omega,M)\rightarrow\mathbb{R}$, given by $V^*(t,x_t)=\inf_uV(t,x_t)$, is well-defined, equivalently, an optimal policy exists. 
\end{theorem}
\begin{proof}
By the uniqueness of the solution to the parameterized ordinary differential equation in (4), guaranteed by Assumptions S1 and S2, if there is a policy $u^*\in\mathcal{U}$ minimizing $J$, then it necessarily minimizes $V(t,x_t)$ for all $t\in[0,T]$ as well. In addition, we have $J(u^*)=\min_{u\in \mathcal{U}}J(u)=V^*(0,x_0)$ by the definition of $J$, which is unique as the infimum (greatest lower bound) of the set $J(\mathcal{U})\subset\mathbb{R}$ (Rudin, 1976). As a result, $V^*(0,x_0)$, and hence $V^*(t,x_t)$ for each $t\in[0,T]$, has a definite real value, indicating that $V^*$ is a well-defined (i.e., single-valued) real-valued function on $[0,T]\times\mathcal{F}(\Omega,M)$. Therefore, it suffices to show the existence of $u^*\in\mathcal{U}$.

Without loss of generality, we assume that the integrability condition in Assumption C1 is satisfied for all $u\in\mathcal{U}$. We then show that the range $J(\mathcal{U})$ of the cost functional $J$ is a compact subspace of $\mathbb{R}$, equivalently, any sequence in $J(\mathcal{U})$ has a convergent subsequence, because of $J(\mathcal{U})\subseteq\mathbb{R}$ \citep{Munkres00}. To show this, we pick an arbitrary sequence $(J_k)_{k\in\mathbb{N}}$ in $J(\mathcal{U})$, then we claim that any policy sequence $(u_k)_{k\in\mathbb{N}}$ satisfying $J_k=J(u_k)$ for all $k\in\mathbb{N}$ has an accumulation point $u\in\mathcal{U}$. To see this, we define $T_k=\{u_n:n\geq k\}$, then $\bigcap_{k\in\mathbb{N}}\overline T_k\neq\varnothing$ holds, where $\overline T_n$ denotes the closure of $T_n$. Otherwise, $\{\mathcal{U}\backslash\overline T_k\}_{k\in\mathbb{N}}$ forms an open cover of $\mathcal{U}$, which has a finite subcover, say $\mathcal{U}=\bigcap_{k=1}^N(\mathcal{U}\backslash\overline T_k)$, due to the compactness of $\mathcal{U}$ shown in Lemma \ref{lem:compact_control}, leading to the contradiction $\varnothing=\mathcal{U}\backslash\big(\bigcap_{k=1}^N(\mathcal{U}\backslash\overline T_k)\big)=\bigcap_{k=1}^N\overline T_k=\overline T_N$. Now, let $u\in\bigcap_{k\in\mathbb{N}}\overline T_k$ and $W$ be a neighborhood of $u$, then $W\cap T_k\neq\varnothing$ for every $k$, i.e., $W$ contains some $u_k$ for any arbitrarily large $k$, and hence $u$ is necessarily an accumulation point of the sequence $(u_k)_{k\in\mathbb{N}}$. The sequential continuity of $J$ proved in Lemma \ref{lem:sequential_continuity_cost} implies that $J$ necessarily maps accumulation points of policy sequences to accumulation points in $J(u)$. As a result, there is a subsequence of $(J_k)_{k\in\mathbb{N}}$ converges to $J(u)$, showing the sequential compactness, and hence also compactness, of $J(\mathcal{U})$.

The compactness of $J(\mathcal{U})$ particularly implies $J^*=\inf J(\mathcal{U})=\inf_{u\in\mathcal{U}}J(u)\in J(\mathcal{U})$. Therefore, there exists $u^*\in\mathcal{U}$ such that $J(u^*)=J^*$, meaning, $u^*$ is an optimal policy. 
\end{proof}

\begin{remark}[Topological actor-critic algorithm]
    Although Theorem \ref{thm:existence} is to verify the solvability of the policy learning problem over the infinite-dimensional function space from the theoretical perspective, the main idea of the employed topological argument exactly coincides with the actor-critic algorithm in RL. Specifically, the ``critic'' $J$ constantly evaluates the ``actor'' $u$ to iterative improve its performance, generating a sequence of cost $J_k$ converging to the minimal cost $J^*$, and the corresponding policy $u^*$ with $J^*=J(u^*)$ is then a desired optimal policy. Moreo importantly, the way of approaching $u^*$ through a cost sequence instead of a policy sequence deliberately avoids a technical issue. Although $\mathcal{U}$ is compact, a policy sequence $u_k$ in $\mathcal{U}$ may not contain any convergent subsequence since $\mathcal{U}$ is not a first-countable space \citep{Munkres00}.
\end{remark}

\section{Reinforcement Learning for Parameterized Systems via Moment Parameterization}
\label{sec:moment}

In this section, we will focus on developing an RL framework for learning optimal policies for parameterized ensemble systems defined on an infinite-dimensional function space. Our initial step, which is also essential to most learning problems, is to explore an appropriate parameterization for the learning targets. To this end, we will introduce a moment kernel transform, which generates kernel representations of parameterized systems and state-value functions over a reproducing kernel Hilbert space (RKHS).



\subsection{Moment kernelization of parameterized systems and value functions} 
\label{sec:moment_function}

Our theoretical development is based on leveraging and extending the method of moments in functional analysis and probability theory \citep{Mackey80}. The central idea is to represent the time-varying state functions of a parameterized system as time-dependent sequences of real numbers. To put this into a formal setting, we impose the following assumption: 


\paragraph{Assumption K1.} The state space $\mathcal{F}(\Omega,M)$ of the parameterized ensemble system 
in \eqref{eq:ensemble}, given by $\frac{d}{dt}x(t,\b)=F(t,\b,x(t,\b),u(t))$ with $\b\in\Omega\subset\mathbb{R}^d$, is a Hilbert space $\mathcal{H}$ contained in $L^2(\Omega,\mathbb{R}^n)$, the space of $\mathbb{R}^n$-valued square-integrable functions defined on $\Omega$.

\paragraph{Ensemble moments and moment kernel transform.}
To motivate our idea of dynamic moment kernelization, we consider the scalar-valued parameterized system defined on the Hilbert space $\mathcal{H}_0\subset L^2(\Omega,\mathbb{R})$; namely, the ensemble state $x(t,\cdot)\doteq x_t(\cdot)\in\mathcal{H}_0$. Because $L^2(\Omega,\mathbb{R})$ is separable, $\mathcal{H}_0$ is also separable as a linear subspace of $\mathcal{H}$ \citep{Yosida80}. Hence, $\mathcal{H}_0$ possesses a countable orthonormal basis, denoted $\{\Phi_k\}_{k\in\mathbb{N}}$. We define the \emph{$k^{\rm th}$ moment} of the parameterized system for each $k\in\mathbb{N}$ with respect to $\Phi_k$ as    
\begin{align}
\label{eq:moment}
m_k(t)=\<\Phi_k,x_t\>, 
\end{align}
where $\<\cdot,\cdot\>:\mathcal{H}_0\times\mathcal{H}_0\rightarrow\mathbb{R}$ is the inner product on $\mathcal{H}_0$. Then, the moment sequence $m(t)\in\mathcal{M}_0$ associated with $x_t\in\mathcal{H}_0$ is denoted by $m(t)=(m_k(t))_{k\in\mathbb{N}}$, with $\mathcal{M}_0$ being the space of all moment sequences, referred to as the \emph{moment space}. In addition, the inner product on $\mathcal{H}_0$, as a subspace of $L^2(\Omega,\mathbb{R})$ illustrated in Assumption K1, is specifically given by $m_k(t)=\<\Phi_k,x_t\>=\int_\Omega\Phi_k(\beta)x_t(\beta)d\beta$. 
The choice of $\{\Phi_k\}_{k\in\mathbb{N}}$ as an orthonormal basis yields  $\sum_{k\in\mathbb{N}}|m_k(t)|^2=\|m(t)\|_{\mathcal{M}_0}=\|x_t\|_{\mathcal{H}_0}=\int_\Omega|x_t(\beta)|^2d\beta<\infty$ by Parseval's identity \citep{Folland13}, where $\|\cdot\|_{\mathcal{M}_0}$ and $\|\cdot\|_{\mathcal{H}_0}$ denote the norms on $\mathcal{M}_0$ and $\mathcal{H}_0$, respectively. This implies that $\mathcal{M}_0$ is contained in the $\ell^2$-space, 
consisting of square-summable sequences. Essentially, $m_k(t)$ is the $k^{\rm th}$ Fourier coefficient of the function $x_t\in\mathcal{H}$, and thus the moment sequence $m(t)$ provides a coordinate representation of $x_t$ with respect to the basis $\{\Phi_k\}_{k\in\mathbb{N}}$.

In the general case where $n>1$, the state space $\mathcal{H}$ of the parameterized system admits a decomposition as a direct sum of $n$ copies of $\mathcal{H}_0$, i.e., $\mathcal{H}=\mathcal{H}_0\oplus\cdots\oplus\mathcal{H}_0$. Equivalently, each component $x^i_t$ of the ensemble state $x_t=\big(x^1_t,\dots,x^n_t\big)$ is an element of $\mathcal{H}_0$. As a result, the definition of moments in \eqref{eq:moment} can be extended to $x_t\in\mathcal{H}$ in a component-wise manner as $m_k(t)=\<\Phi_k,x_t\>=\big(\<\Phi_k,x_t^1\>,\dots,\<\Phi_k,x_t^n\>\big)$. Then, $m_k(t)$ becomes an $\mathbb{R}^n$-valued sequence so that the moment space $\mathcal{M}=\mathcal{M}_0\oplus\cdots\oplus\mathcal{M}_0$ is a Hilbert subspace of $\ell^2\oplus\cdots\oplus\ell^2$, the $n$-fold direct sum of $\ell^2$-spaces. More importantly, this indicates that $\mathcal{M}$ is a reproducing kernel Hilbert space as shown below.

\begin{proposition}
    The moment space $\mathcal{M}$ is a reproducing kernel Hilbert space (RKHS) on $\mathbb{N}$, and the \emph{moment kernel transform} $\mathcal{K}:\mathcal{H}\rightarrow\mathcal{M}$, given by $x_t\mapsto m(t)$, is an isometric isomorphism between Hilbert spaces. 
\end{proposition}
\begin{proof}
     Following the analysis above, we know that the moment space $\mathcal{M}_0$ of functions in $\mathcal{H}_0$ is contained in $\ell^2$, it remains to show that $\ell^2\subseteq\mathcal{M}_0$. To this end, we pick any $m(t)\in\ell^2$. According to the Pythagorean theorem, we have $\big|\sum_{k=0}^\infty m_k(t)\Phi_k\big|=\sum_{k=0}^\infty |m_k(t)|^2<\infty$, implying that the partial sum $\sum_{k=0}^N m_k(t)\Phi_k$ forms a Cauchy sequence. Therefore, $x_t=\sum_{k=0}^\infty m_k(t)\Phi_k$ is a well-defined element (function) in $\mathcal{H}_0$ and satisfies $\<\Phi_k,x_t\>=m_k(t)$, yielding $\ell^2\subseteq\mathcal{M}_0$. This concludes $\mathcal{M}_0=\ell^2$. The restriction of the moment transform to each component of $x_t\in\mathcal{H}$ is thus an isometric isomorphism from $\mathcal{H}_0$ to $\mathcal{M}_0$. Together with the fact that $\mathcal{H}=\mathcal{H}_0\oplus\cdots\oplus\mathcal{H}_0$, we obtain $\mathcal{M}=\mathcal{M}_0\oplus\cdots\oplus\mathcal{M}_0=\ell^2\oplus\cdots\oplus\ell^2$, and the moment kernel transform from $\mathcal{H}$ to $\mathcal{M}$ is an isometric isomorphism as well. 

     Lastly, to show that $\mathcal{M}$ is an ($\mathbb{R}^n$-valued) RKHS on $\mathbb{N}$, it suffices to demonstrate that the point evaluation map $E_k:\mathcal{M}\rightarrow\mathbb{R}^n$, given by $m(t)\mapsto m_k(t)$, is bounded \citep{Paulsen2016}. This follows from the estimate,
     \begin{align*}
         |m_k(t)|^2=\sum_{i=1}^n|\<\Phi_k,x_t^i\>|^2\leq\sum_{i=1}^n\|\Phi_k\|_{\mathcal{H}_0}^2\|x_t^i\|_{\mathcal{H}_0}^2=\sum_{i=1}^n\|x_t^i\|_{\mathcal{H}_0}^2=\|x_t\|^2_{\mathcal{H}}=\|m(t)\|^2_{\mathcal{M}},
     \end{align*}
     where $|\cdot|$, $\|\cdot\|_{\mathcal{H}}$, and $\|\cdot\|_{\mathcal{M}}$ denote the norms on $\mathbb{R}^n$, $\mathcal{H}$, and $\mathcal{M}$, respectively.    
\end{proof}


\paragraph{Moment kernelization of ensemble systems.} 
Having kernelized the ensemble state of the parametrized system, the next step is to exploit the kernelized state and the moment transform to kernelize the system dynamics, i.e., the temporal evolution of the ensemble state over $\mathcal{H}$. Intuitively, this requires taking the time-derivative of the moments, which yields a differential equation system governing the evolution of the moment sequence, given by
\begin{align}
\label{eq:moment_system_kth}
\frac{d}{dt}m_k(t)=\frac{d}{dt}\<\Phi_k,x_t\>=\Big\<\Phi_k,\frac{d}{dt}x_t\Big\>=\<\Phi_k,F(t,\cdot,x_t,u(t))\>.
\end{align}
Here, the change of the order of the time-derivative and the inner product operation follows from the dominant convergence theorem \citep{Folland13}. Note that because the state-space $\mathcal{H}$ of the parameterized system is a vector space, the vector field $F(t,\beta,x_t(\beta),u(t))$ governing the system dynamics, considered as a function in $\beta$, is an element of $\mathcal{H}$ as well. 
Following the definition in \eqref{eq:moment}, we observe that $\<\Phi_k,F(t,\cdot,x_t,u(t))\>$ in \eqref{eq:moment_system_kth}  is essentially the $k^{\rm th}$ moment of $F(t,\cdot,x_t,u(t))$. 
Let $\bar F(t,m(t),u(t))$ denote the moment sequence of $F(t,\cdot,x_t,u(t))=F(t,\cdot,\mathcal{K}^{-1}m(t),u(t))$, we obtain a concrete representation of the moment kernelized ensemble system defined on $\mathcal{M}$ as
\begin{align}
\label{eq:moment_system}
\frac{d}{dt}m(t)=\bar F(t,m(t),u(t)).
\end{align}

\begin{remark}\label{rmk:countability}
    Note that the moment kernelized system in \eqref{eq:moment_system} always consists of countably many components, even though the parameterized ensemble system in \eqref{eq:ensemble} may be composed of a continuum of (uncountably many) intelligent agents. This implies that the moment kernelization not only defines a kernel parameterization but also provides model reduction for parameterized systems. This feature 
    will be fully exploited in the development of the filtrated RL architecture for learning optimal policies of parameterized systems. 
\end{remark}

\paragraph{Moment kernelization of value functions.} 
The last piece of the exploration of the moment kernel representation of the proposed policy learning problem is to kernelize the state-value function $V:[0,T]\times\mathcal{H}\rightarrow\mathbb{R}$. To this end, supported by the integrability condition in Assumption C1, we apply Fubini's theorem to change the order of the two integrals in the first summand of $V$, resulting in
$V(t,x_t)=\int_t^T\int_\Omega r(s,\beta,x(s,\b),u(s))d\b ds+\int_\Omega K(T,\beta,x(T,\b))d\b$ \citep{Folland13}.
Observe that $\int_\Omega r(t,\beta,x(t,\b),u(t))d\b$ and $\int_\Omega K(T,\beta,x(T,\b))d\b$ are nothing but the $0^{\rm th}$ moments of $r(t,\beta,x_t,u(t))$ and $K(T,\beta,x_T)$ as real-valued functions defined on $\Omega$, provided that $\Phi_0$ is a constant function. Denoting them by $\bar r(t,m(t),u(t))$ and $\bar K(T,m(T))$, respectively, we obtain the desired moment kernelized state-value function as
\begin{align}
\label{eq:moment_V}
V(t,m(t))=\int_t^T\bar r(s,m(s),u(s)) ds+\bar K(T,m(T)).
\end{align}

\begin{remark}
    The key step in kernelizing the state-value function above is the change of the order of the integrals with respect to $\beta\in\Omega$ and $t\in[0,T]$ using the integrability condition in Assumption C1. This derivation remains valid under a relatively weaker condition, that is, the immediate reward function $r$ is nonnegative for any admissible control policy, as a consequence of Tonelli's theorem \citep{Folland13}. 
    This nonnegative condition also occurs commonly in practice, e.g., the LQR problem presented in Section \ref{sec:challenges}, demonstrating the general applicability of the proposed moment kernelization approach.
\end{remark}

As the pointwise infimum of the state-value function over the space of admissible policies, the value function naturally admits the moment kernel representation $V^*:[0,T]\times\mathcal{M}\rightarrow\mathbb{R}$, given by,
\begin{align}
\label{eq:DP}
V^*(t,m(t))=\inf_{u\in\mathcal{U}}V(t,m(t))=\inf_{u\in\mathcal{U}}\Big\{\int_t^{t+h}\bar r(s,m(s),u(s)) ds+V^*(t+h,m(t+h))\Big\}
\end{align}
for any $h>0$ such that $t+h\leq T$, where the second equality follows from the dynamic programming principle \citep{Evans10}. The integral representation of the value function in \eqref{eq:DP} leads to the following regularity property, which plays a crucial role in establishing RL approaches to policy learning of parameterized systems over the moment domain. 


\begin{proposition}
\label{prop:V_continuity}
The value function $V^*:[0,T]\times\mathcal{M}\rightarrow\mathbb{R}$ is Lipschitz continous. 
\end{proposition}
\begin{proof}
See Appendix \ref{appd:V_continuity}.
\end{proof}


\subsection{Moment convergence for reinforcement learning of kernelized ensemble systems}
\label{sec:moment_RL}
As the moment kernelized system in \eqref{eq:moment_system} consists of countable state variables,  
this enables the use of truncated moment systems to facilitate RL of parameterized systems. This section is dedicated to laying the theoretical foundation for this approach. In particular, the main focus is to show that value functions of truncated moment systems converge to those of moment kernelized parameterized systems in an appropriate sense.

To rigorously formulate the corresponding policy learning problem for a truncated moment system, we use the hat notation ` $\hat\cdot$ ' to denote the truncation operation and identify the order-$N$ truncated moment sequence with the projection of the infinite moment sequence onto the first $N$ components, e.g., $\widehat m_N(t)=(m_0(t),\dots,m_N(t),0,0,\dots)'$ and $\widehat F_N=(\bar F_0,\dots,\bar F_N,0,0,\dots,)'$. This then constructs the order-$N$ truncated moment system as
\begin{align}
\label{eq:moment_system_truncated}
\frac{d}{dt} \widehat m_N(t)=\widehat F_N(t,\widehat m_N(t),u(t)),
\end{align}
which is a control system defined on  $\widehat{\mathcal{M}}_N=\{\hat z_N\in\mathcal{M}\mid z\in\mathcal{M}\}$, the space of all order-$N$ truncated moment sequences. With a slight abuse of the hat notion, we denote the state-value and value functions of the system in \eqref{eq:moment_system_truncated} by $\widehat V_N[0,T]\times\widehat{\mathcal{M}}_N\rightarrow\mathbb{R}$ and $\widehat V_N^*:[0,T]\times\widehat{\mathcal{M}}_N\rightarrow\mathbb{R}$, respectively. Clearly, the dynamic programming principle is also satisfied by this ``truncated'' value function as
\begin{align}
\label{eq:value_function_truncated}
    \widehat V_N^*(t,\widehat m_N(t))=\inf_{u\in\mathcal{U}}\widehat V_N(t,\widehat m_N(t))=\inf_{u\in\mathcal{U}}\int_t^T \bar r(s,\widehat m_N(s),u(s))ds+\bar K(T,\widehat m_N(T)).
\end{align}

The RKHS structure on the moment space $\mathcal{M}$ particularly implies the converges of the truncated moment sequence $\widehat m_N(t)$ to the entire moment sequence $m(t)$ as $N\rightarrow\infty$. This can be observed by the vanishing of the truncation error as
\begin{align}
    \|m(t)-\widehat m_N(t)\|^2=\sum_{k=N}^\infty m_k^2(t)\rightarrow0 \label{eq:moment_convergence}
\end{align} 
as $N\rightarrow\infty$, because $m_k\rightarrow0$ as $k\rightarrow\infty$ following from $m(t)$ being square summable. We then have
\begin{align*}
    \|\bar F(t,m(t),u(t))-\widehat F_N(t,\widehat m_N(t),u(t))\|\leq \|&\bar F(t,m(t),u(t)) -\widehat F_N(t,m(t),u(t))\|\\
    &+\|\widehat F_N(t,m(t),u(t))-\widehat F_N(t,\widehat m_N(t),u(t))\|\rightarrow0 
\end{align*}
Here, because $\bar F$ is also a moment sequence as discussed previously, the convergence of the first term to 0 is essentially the same as \eqref{eq:moment_convergence}, together with which the continuity of $\widehat F_N$ gives the convergence of the second term to 0.

 To reinforce the convergence property from truncated moment sequences to ``truncated'' value functions, it is crucial to be alerted to the fact that $\widehat V^*_N$ is barely the restriction of $\widehat V^*$ on the subspace $\widehat{\mathcal{M}}_N\subset\mathcal{M}$, which we denote by $V^*|_{\widehat{\mathcal{M}}_N}$. This is because the state trajectory of the order-$N$ truncated moment system in \eqref{eq:moment_system_truncated} has to stay in $\widehat{\mathcal{M}}_N$, while the optimal trajectory of the entire moment system \eqref{eq:moment_system} starting from an initial condition in $\widehat{\mathcal{M}}_N$ may leave $\widehat{\mathcal{M}}_N$. Therefore, it is necessary that $\widehat V^*_N\big(t,\widehat m_N(t)\big)\geq V^*|_{\widehat{\mathcal{M}}_N}\big(t,\widehat m_N(t)\big)$, but this minor annoyance will not destroy the desired convergence of $\widehat V^*_N$ to $V^*$ as $N\rightarrow\infty$. To make this convergence argument in a mathematically rigorous manner, we extend the domain of $\widehat V_N^*$ from $[0,T]\times\widehat{\mathcal{M}}_N$ to $[0,T]\times\mathcal{M}$ by defining $\widehat {V}_N^*(t,m(t))\doteq\widehat{V}_N(t,\widehat{m}_N(t))$.

 \begin{theorem}[Moment convergence of value functions]
 \label{thm:V_uniform_convergence}
 The value function $\big(\widehat V_N^*\big)_{N\in\mathbb{N}}$ of the truncated moment system in \eqref{eq:moment_system_truncated} converges locally uniformly to $V^*$, the value function of the moment system in \eqref{eq:moment_system}, on $[0,T]\times\mathcal{M}$ as $N\rightarrow\infty$. 
 \end{theorem}
 \begin{proof}
 Because the spaces of truncated moment sequences form an ascending chain of subspaces of the moment space, meaning $\widehat{\mathcal{M}}_{0}\subset\widehat{\mathcal{M}}_{1}\subset\dots\subset\mathcal{M}$, the sequence of value functions $\big(\widehat V_N^*\big)_{N\in\mathbb{N}}$ forms a decreasing chain $\widehat V_{0}^*\big(t,z\big)\geq\widehat V_{1}^*\big(t,z\big)\geq\dots\geq V^*\big(t,z\big)$ for any $z\in\mathcal{M}$. This particularly implies that $\big(\widehat V_N^*\big)_{N\in\mathbb{N}}$ is locally uniformly bounded, i.e., there exists a neighborhood $\mathcal{N}$ of $(t,z)$ in $[0,T]\times\mathcal{M}$ and a real number $M$ so that $\big|\widehat V_N^*(s,w)\big|\leq M$ for all $(s,w)\in\mathcal{N}$ and $N\in\mathbb{N}$ with $M$ independent of $(s,w)$ and $N$. On the other hand, by the aforementioned fact that $\widehat V_N^*\geq V^*$, together with the Lipschitz continuity of $V^*$ shown in Lemma \ref{prop:V_continuity}, all the functions in the sequence $\big(\widehat V_N^*\big)_{N\in\mathbb{N}}$ are Lipschitz continuous and have the same Lipschitz constant equal to that of $V^*$. Consequence, this sequence of value functions is equicontinuous \cite{Rudin76}. A direct application of the Arzelà–Ascoli theorem then shows that $\widehat V_N^*\rightarrow V^*$ on $\mathcal{N}$ uniformly as $N\rightarrow\infty$ \citep{Folland13}. Since $(t,z)\in[0,T]\times\mathcal{M}$ is arbitrary, we obtain $\widehat V_N^*\rightarrow V^*$ locally uniformly on $[0,T]\times\mathcal{M}$ as $N\rightarrow\infty$ as desired.
 \end{proof}

 According to the dynamic programming principle illustrated in \eqref{eq:DP}, value functions are obtained by evaluating the corresponding state-value functions along optimal trajectories. It then becomes intuitive that, accompanied by the convergence of the ``truncated'' value function, the optimal trajectory of the truncated moment system also converges to that of the entire moment system.

\begin{theorem}[Moment convergence of optimal trajectories]
\label{thm:moment_convergence}
Let $\widehat m_N^*(t)$ and $m^*(t)$ be the optimal trajectories of the order-$N$ truncated and entire moment systems in \eqref{eq:moment_system_truncated} and \eqref{eq:moment_system}, respectively, then $\widehat m^*_N(t)\rightarrow m^*(t)$ on $\mathcal{M}$ as $N\rightarrow\infty$ for all $t\in[0,T]$, and equivalently, $\widehat x_N^*(t,\cdot)\rightarrow x^*(t,\cdot)$ on $\mathcal{H}$ as $N\rightarrow\infty$ for all $t\in[0,T]$, where $\widehat x_N^*(t,\cdot)$ and $x^*(t,\cdot)$ are the trajectories of the parameterized system in \eqref{eq:ensemble} driven by the optimal control policies for the truncated and entire moment systems, respectively. 
\end{theorem} 
\begin{proof}
By the dynamic programming principle in \eqref{eq:DP}, the optimal trajectory $m^*(t)$ on $[0,T]$ remains optimal when restricted to any subinterval $[s,T]$ for $0<s<T$, and the same result holds for $\widehat m_N^*(t)$ for each truncation order $N\in\mathbb{N}$. Because the sequence of real numbers $\big(\widehat V_N(t,\widehat m^*_N(t))\big)_{N\in\mathbb{N}}$ is monotonically decreasing and bounded from below by $V(t,m^*(t))$ as shown in the proof of Theorem \ref{thm:V_uniform_convergence}, it is necessary that $\widehat V_N(t,\widehat m^*_N(t))\rightarrow V(t,m^*(t))$ as $N\rightarrow\infty$ for each $t\in[0,N]$ \citep{Rudin76}. Together with the continuity of $V$ and the locally uniform convergence of the value function sequence $\big(\widehat V_N\big)_{N\in\mathbb{N}}$ by Theorem \ref{thm:V_uniform_convergence}, we obtain $V(t,m^*(t))=\lim_{N\rightarrow\infty}\widehat V_N\big(t,\widehat m^*_N(t)\big)=V\big(t,\lim_{N\rightarrow\infty}\widehat{m}^*_N(t)\big)$ so that $\widehat m_N^*(t)\rightarrow m^*(t)$ on $\mathcal{M}$ as $N\rightarrow\infty$. The isometrically isomorphic property of the moment kernel transform $\mathcal{K}:\mathcal{H}\rightarrow\mathcal{M}$ then implies that $\mathcal{K}^{-1}\widehat m_N^*(t)\rightarrow\mathcal{K}^{-1} m^*(t)$, i.e., $\widehat x_N^*(t,\cdot)\rightarrow x^*(t,\cdot)$, on $\mathcal{H}$ as $N\rightarrow\infty$.
\end{proof}


A fundamental property of value functions crucial to RL is that they are viscosity solutions of Hamiltonian-Jacobi-Bellman (HJB) equations \citep{Evans10}, which enables the design of RL algorithms to learn optimal control policies. The moment convergence shown in Theorems \ref{thm:V_uniform_convergence} and \ref{thm:moment_convergence} then leads to an extension of HJB equations to value functions defined on infinite-dimensional spaces.

\begin{corollary}[Moment convergence of Hamilton-Jacobi-Bellman equations]\hfill
\label{cor:HJB}
The value function $V^*:[0,T]\times\mathcal{M}\rightarrow\mathbb{R}$ is the unique viscosity solution of the Hamilton-Jacobi-Bellman equation, given by,
\begin{align}
\label{eq:HJB}
\frac{\partial}{\partial t}V^*(t,z)+\min_{a\in U}\Big\{\<DV^*(t,z),\bar F(t,z,a)\>+\bar r(t,z,a)\Big\}=0,
\end{align}
on $(0,T)\times\mathcal{M}$ with the boundary condition $V=K$ on $\{t=T\}\times\mathcal{M}$, where 
$DV^*(t,z)$ is the (Gateaux) differential of $V^*$ respect to $z\in\mathcal{M}$.
\end{corollary}
\begin{proof}
The uniqueness directly follows from the Lipschtiz continuity of the \emph{Hamiltonian} $H(t,z,p)=\min_{a\in U}\big\{\<p,\bar F(t,z,a)\>+\bar r(t,z,a)\big\}$ in $(z,p)\in\mathcal{M}\times\mathcal{M}^*$ uniformly in $t\in[0,T]$, as a consequence of Assumption C2, where $\mathcal{M}^*$ is the dual space of $\mathcal{M}$ \citep{Evans10}.
Then, it remains to examine that $V^*$ is a viscosity solution of the first order partial differential equation in \eqref{eq:HJB}.

To show this, for any $(t,z)\in[0,T]\times\mathcal{M}$, we pick a continuously differentiable function $v:[0,T]\times\mathcal{M}\rightarrow\mathbb{R}$ such that $V-v$ has a local maximum at $(t,z)$. Without loss of generality, we can assume $(t,z)$ to be a strict local maximum, i.e., there is a neighborhood $\mathcal{N}$ whose closure $\overline{\mathcal{N}}$ contains $(t,z)$ such that $V^*(t,z)-v(t,z)>V^*(s,w)-v(t,z)$ for any $(s,w)\in\overline{\mathcal{N}}\backslash\{(t,z)\}$. Let $(t_N,z_N)$ be a maximum of $\widehat V_N^*-v$ on $\overline{\mathcal{N}}$, then we have 
\begin{align}
\label{eq:HJB_truncated}
\frac{\partial}{\partial t}v(t_N,z_N)+\min_{a\in U}\Big\{\<Dv(t_N,z_N),\widehat F_N(t_N, z_N,a)\>+\bar r(t_N,z_N,a)\Big\}\geq0
\end{align}
following from the fact that $\widehat V_N^*$ is the value function for he truncated moment system in \eqref{eq:moment_system_truncated} defined on the finite-dimensional space $\widehat{\mathcal{M}}_N$ \citep{Evans10}. 

By passing to a subsequence and shrinking the neighborhood $\mathcal{N}$ if necessary, we have $(t_N, z_N)\rightarrow(\widetilde t,\widetilde z)$ as $N\rightarrow\infty$ for some $\tilde t\in[0,T]$ and $\widetilde z\in\overline{\mathcal{N}}$, which leads to the convergences $\widehat{V}_N^*(t_N, z_N)-v(t_N, z_N)\rightarrow V^*(\widetilde t,\widetilde z)-v(\widetilde t,\widetilde z)$ by the locally uniform convergence of $\widehat V_N^*$ to $V^*$ shown in Theorem \ref{thm:V_uniform_convergence}. Because $\widehat V_N^*(t_N, z_N)-v(t_N, z_N)\geq\widehat V_N^*(s,w)-v(s,w)$ for any $(s,w)\in\overline{\mathcal{N}}$ and $N\in\mathbb{N}$ by the choice of $(t_N, z_N)$, we obtain $V^*(\widetilde t,\widetilde z)-v(\widetilde t, z)\geq V^*(s,w)-v(s,w)$ for any $(s,w)\in\overline{\mathcal{N}}$ by letting $N\rightarrow\infty$, particularly, $V^*(\widetilde t,\widetilde z)-v(\widetilde t,\widetilde z)\geq V^*(t,z)-v(t,z)$. This shows that $(\widetilde t,\widetilde z)=(t,z)$ and hence $(t_N, z_N)\rightarrow(t,z)$ as $N\rightarrow\infty$. 

Passing to the limit as $N\rightarrow\infty$ in \eqref{eq:HJB_truncated}  yields
\begin{align}
\label{eq:HJB_max}
\frac{\partial}{\partial t}v(t,z)+\min_{a\in U}\big\{\<Dv(t,z),\bar F(t,z,a)\>+\bar r(t,z,a)\big\}\geq0,
\end{align}
where we use the uniform convergence of $\widehat F_N$ to $\bar F$, following from a similar proof as Theorem \ref{thm:V_uniform_convergence} by replacing $\widehat V_N^*$ and $V^*$ with $\widehat F_N$ and $\bar F$, respectively, and the continuity of the Hamitonian $H(t,z,p)=\min_{a\in U}\big\{\<p,\bar F(t,z,a)\>+\bar r(t,z,a)\big\}$. A similar argument shows 
\begin{align}
\label{eq:HJB_min}
\frac{\partial}{\partial t}w(t,z)+\min_{a\in U}\big\{\<Dw(t,z),\bar F(t,z,a)\>+\bar r(t,z,a)\big\}\leq0
\end{align}
if $V^*-w$ attains a local minimum at $(t,z)\in[0,T]\times\mathcal{M}$. The equations in \eqref{eq:HJB_max} and \eqref{eq:HJB_min} imply that $V^*$ is a viscosity solution of the Hamilton-Jacobi-Bellman equation in \eqref{eq:HJB}.
\end{proof}

\subsection{Moment Kernelization of Parameterized Systems in Stochastic Environments}
The proposed parametrized model formulation and moment kernelization framework naturally extend to stochastic settings. We explore the adoption of our approach in three types of stochastic environments.

\paragraph{Model uncertainty.} 
The parametrized ensemble system in \eqref{eq:ensemble} also gives a natural formulation describing systems with model uncertainties. In this interpretation, the system parameter $\beta$ characterizes uncertainty and can be regarded as a random variable drawn from a probability distribution $\mathbb{P}$ on the parameter space $\Omega$. Consequently, the trajectory of the parameterized system $x_t$ is a stochastic process on the probability space $(\Omega,\mathbb{P})$, and the value function becomes $$V(t,x_t)=\mathbb{E}\Big(\int_t^Tr(s,x_s,u(s))ds+K(T,x_T)\Big)=\int_\Omega\Big(\int_t^Tr(s,x_s,u(s))ds+K(T,x_T)\Big)d\mathbb{P},$$
where $\mathbb{E}$ denotes expectation with respect to the probability measure $\mathbb{P}$. Suppose that $x_t$ possesses finite variance for all $t$, then the $k^{\rm th}$ moment of the parameterized system can be defined as
\begin{align}
    \label{eq:moment_uncertainty}
    m_k(t)=\<\Phi_k,x_t\>=\mathbb{E}\big(\Phi_kx_t\big)=\int_\Omega\Phi_kx_td\mathbb{P},
\end{align}
where $\{\Phi_k\}_{k\in\mathbb{N}}$ is an orthonormal basis of $L^2(\Omega,\mathbb{P})$. Leveraging this definition of moments, the kernelized system and value function follow the same forms as in \eqref{eq:moment_system} and \eqref{eq:moment_V}, respectively, implying that all the developments and conclusions in Sections \ref{sec:moment_function} and \ref{sec:moment_RL} remain hold. 

\paragraph{Background noise.} Background noise is frequently inherent in the agents' environments 
and typically appears as additive noise present in measurements of the states of agents.
Here, we consider two types of background noise commonly encountered in practice: independent noise and common noise.

\begin{itemize}
\item Independent noise: In this case, each agent in the parameterized family in \eqref{eq:ensemble} experiences parameter-dependent independent additive noise of the form
$x_t(\b)+\varepsilon(\b)$, where $\{\varepsilon(\b)\}_{\b\in\Omega}$ is a family of pairwise independent random variables. Suppose that $\varepsilon(\b)$ has zero mean and finite variance for all $\b\in\Omega$, then Fubini's theorem \citep{Folland13} gives 
    \begin{align*}
        &\mathbb{E}\Big(\int_\Omega \varepsilon(\beta)d\beta\Big)^2=\mathbb{E}\Big(\int_\Omega\varepsilon(\beta)d\beta\int_\Omega\varepsilon(\gamma)d\gamma\Big)=\mathbb{E}\Big(\int_{\Omega^2}\varepsilon(\beta)\varepsilon(\gamma)d\beta d\gamma\Big)\\
        &=\int_{\Omega^2}\mathbb{E}\big(\varepsilon(\beta)\varepsilon(\gamma)\big)d\beta d\gamma=\int_{D}\mathbb{E}\big(\varepsilon^2(\beta)\big)d\beta d\gamma+\int_{\Omega^2\backslash D}\mathbb{E}\big(\varepsilon(\beta)\big)\mathbb{E}\big(\varepsilon(\gamma)\big)d\beta d\gamma.
    \end{align*} 
    Here, $D=\{(\beta,\gamma)\in\Omega^2:\beta=\gamma\}$ is the diagonal subset of $\Omega^2=\Omega\times\Omega$, and the last equality follows from the pairwise independence of $\{\varepsilon(\beta)\}_{\beta\in\Omega}$. We further observe that $D$ has Lebesgue measure 0, and together with $\mathbb{E}(\varepsilon^2(\beta))<\infty$, this leads to $\int_{D}\mathbb{E}\big(\varepsilon^2(\beta)\big)d\beta d\gamma=0$. On the other hand, the zero mean property gives $\int_{\Omega^2\backslash D}\mathbb{E}\big(\varepsilon(\beta)\big)\mathbb{E}\big(\varepsilon(\gamma)\big)d\beta d\gamma=0$. As a result, $\mathbb{E}\Big(\int_\Omega \varepsilon(\beta)d\beta\Big)^2=0$ holds, indicating that $\int_\Omega\varepsilon(\beta)d\beta=0$ almost surely (\emph{a.s.}). 
    
    Following the same definition as in \eqref{eq:moment}, we have the $k^{\rm th}$ moment satsifies
    \begin{align*}
        m_k(t)=\<\Phi_k,x_t+\varepsilon\>=\int_\Omega\Phi_k(\beta)x_t(\beta)d\beta+\int_\Omega\Phi_k(\beta)\varepsilon(\beta)d\beta=\<\Phi_k,x_t\> \quad a.s.
    \end{align*}
    for all $k\in\mathbb{N}$. Here, we use the claim that $\int_\Omega\Phi_k(\beta)\varepsilon(\beta)d\beta=0$ \emph{a.s.} To see this, because of the square-integrability of $\Phi_k$, the set $L=\{\beta\in\Omega:\Phi_k(\beta)=\infty\}$ has Lebesgue measure 0. This gives $\int_\Omega\Phi_k(\beta)\varepsilon(\beta)d\beta=\int_L\Phi_k(\beta)\varepsilon(\beta)d\beta+\int_{\Omega\backslash L}\Phi_k(\beta)\cdot0d\beta=0\ a.s.$ as claimed.

    \item Common noise: In this case, each agent in the parameterized ensemble system in \eqref{eq:ensemble} experiences a common noise $\varepsilon$ so that its environmental state follows the form $x_t(\beta)+\varepsilon$. Suppose that $\varepsilon$ is a random variable with zero mean and finite variance, 
    then the $k^{\rm th}$ moment,
    $$m_k(t)=\<\Phi_k,x_t+\varepsilon\>=\int_\Omega\Phi_k(\beta)x_t(\beta)d\beta+\varepsilon \int_\Omega\Phi_k(\beta)d\beta,$$
    is a random variable with the mean given by $$\mathbb{E}\big(m_k(t)\big)=\int_\Omega\Phi_k(\beta)x_t(\beta)d\beta+\int_\Omega\Phi_k(\beta)d\beta\cdot\mathbb{E}(\varepsilon)=\int_\Omega\Phi_k(\beta)x_t(\beta)d\beta=\<\Phi_k,x_t\>.$$ 
    Statistically, this implies that the  moments of the parameterized system in the presence of common background noise is an unbiased estimator of the moment without noise. More importantly, the moment kernelization reduces the variance: following from the Hölder's inequality, we have
    \begin{align*}
        {\rm Var}(m_k(t))=\Big(\int_\Omega\Phi_k(\beta)d\beta\Big)^2{\rm Var}(\varepsilon)\leq|\Omega|\|\Phi_k\|^2{\rm Var}(\varepsilon)<{\rm Var}(\varepsilon),
    \end{align*}
    provided, without loss of generality, the Lebesgue measure of $\Omega$ satisfies $|\Omega|<1$. 
\end{itemize}

\paragraph{Stochastic dynamics.} When the parameterized system in \eqref{eq:ensemble} is driven by a noise process, e.g., a Brownian motion $W_t$ on a probability space $(\mathbb{R}^n,\mathbb{P})$, the evolution of its state obeys the stochastic differential equation
$$dx_t(\b)=F(t,\b,x_t(\b),u(t))dt+G(t,\b,x_t(\b))dW_t,$$
with the state-value function given by
$$V(t,x_t)=\mathbb{E}\Big(\int_\Omega\big[\int_t^Tr(s,x_s,u(s))ds+K(T,x_T)\big]d\beta\Big),$$ 
where $F$ and $G$ satisfy the Lipchitz and linear growth conditions in the system state variable, $\|F(t,\beta,x,u)-F(t,\beta,y,u)\|+\|G(t,\beta,x,u)-G(t,\beta,y)\|\leq K_1\|x-y\|$ and $\|F(t,\beta,x,u)\|^2+\|G(t,\beta,x,u\|^2\leq K_2(1+\|x\|^2)$, respectively, uniformly in $(t,\beta,u)\in[0,T]\times\Omega\times\mathcal{U}$. These conditions guarantee that, for each $\beta\in\Omega$, the system trajectory is a finite variance stochastic process on the probability space $(\mathbb{R}^n,\mathbb{P})$ \citep{Oksendal2003}. Following the definition in \eqref{eq:moment_uncertainty}, 
the moments of this parameterized stochastic system can be defined as 
$$m_{kl}(t)=\mathbb{E}\big(\Phi_k\Psi_l\circ x_t\big)=\mathbb{E}\Big(\int_\Omega\Phi_k(\b)\Psi_l(x_t(\b))d\b\Big)$$
for all $(k,l)\in\mathbb{N}^2$, where $(\Psi_l)_{l\in\mathbb{N}}$ is an orthonormal basis of $L^2(\mathbb{R}^n,\mathbb{P})$. Note that, in this case, the moment sequence $(m_{kl}(t))_{k,l\in\mathbb{N}}$ is a deterministic double sequence, demonstrating again the ability of the moment kernel transform to mitigate the noise effect. \\

It should be commented that, within a stochastic environment, the open-loop requirement for the control policy may cause the loss of the Markov property to the state evolution of the parameterized system. This severely limits the application of MDP-based RL algorithms to the parameterized system, which in turn stresses the necessity for the proposed moment kernelization technique to reduce the randomness involved in the system. In the next section, we will develop a new RL architecture by fully exploiting the algebraic structure of moment kernelized systems to facilitate policy learning for parameterized systems.

\section{Filtrated Reinforcement Learning Architecture for Policy Learning of Parameterized Ensemble Systems}

Built upon the theoretical foundations established in the previous sections, we now turn our attention to algorithmic approaches to policy learning of parameterized ensemble systems evolving on infinite-dimensional function spaces. In this section, we will develop a novel RL architecture with effective algorithms through organizing truncated moment kernelized systems with increasing truncation orders into a filtrated structure. We will then adopt spectral sequence techniques to conduct convergence analysis of the proposed filtrated RL (FRL) algorithms. Meanwhile, computational and sample efficiency of FRL will also be investigated quantitatively. We will demonstrate the performance and efficiency of the FRL algorithms by using examples arising from practical applications and comparing with baseline deep RL models.

\subsection{Filtrated policy search for moment kernelized systems}

\begin{figure}[h]
      \centering
      \includegraphics[width=1\textwidth,keepaspectratio]{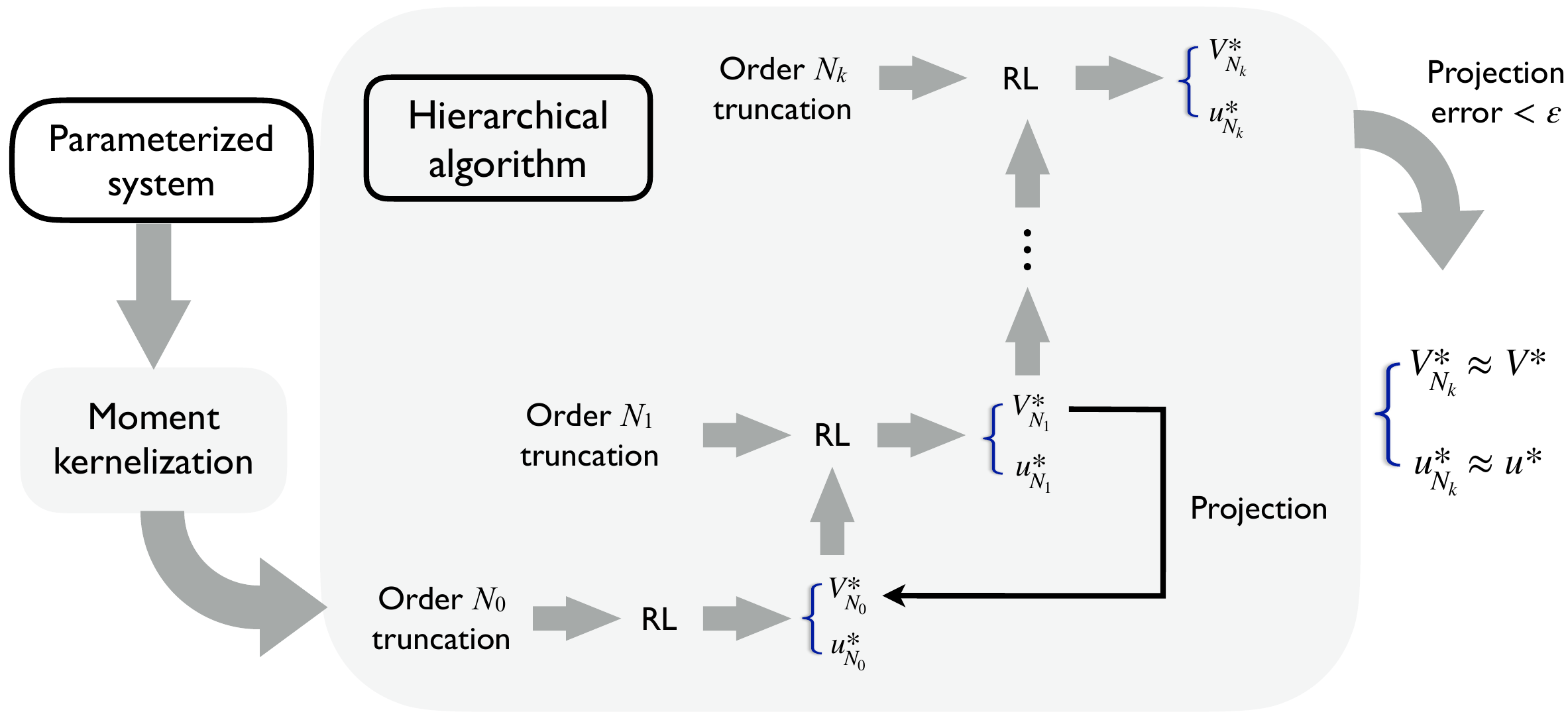}
      \caption{\footnotesize \small \noindent Workflow of FRL for parameterized ensemble systems defined on infinite-dimensional function spaces.} 
       \vspace{-15pt}
      \label{fig:FRL}
  \end{figure}

 After transforming the parameterized ensemble system to the moment domain, it is not hard to observe that the order-$N$ truncated moment system contains any truncated moment system with a lower truncation order $N'<N$ as a subsystem. Interpreting this from the learning perspective, the corresponding RL problem for the order-$N'$ truncated moment system is a subproblem of that for the order-$N$ truncated moment system. In the general situation, along any increasing sequence of truncation orders $N_0<N_1<\cdots$, a filtration consisting the ascending chain of RL problems is revealed. To further elaborate on this FRL architecture, starting from an initial moment truncation order $N_0$, an optimal policy $u^*_{N_0}(t)$ and the value function $\widehat V_{N_0}^*(t,\widehat m_{N_0}(t))$ can be learned for the order-$N_0$ truncated moment system. Next, the truncation order is increased to $N_1$ to learn $u^*_{N_1}(t)$ and  $\widehat{V}_{N_1}^*\big(t,\widehat{m}_{N_1}(t)\big)$ of the order-$N_1$ truncated moment system. Continuing this procedure, we generate a sequence of control policy $\big(u_{N_k}^*(t)\big)_{k\in\mathbb{N}}$ and value function $\big(\widehat V^*(t,\widehat m_{N_k}(t))\big)_{k\in\mathbb{N}}$ pairs. Guaranteed by the moment convergence proved in Theorem \ref{thm:moment_convergence}, the value function sequence $\big(\widehat V^*(t,\widehat m_{N_k}(t))\big)_{k\in\mathbb{N}}$ necessarily converges to the value function $V^*(t,m(t))$ of the entire moment system. Algorithmically, when the \emph{projection error} $\sup_{t\in[0,T]}\big|\widehat{V}^*_{N_k}\big(t,\widehat{m}^*_{N_k}(t)\big)-\widehat{V}^*_{N_{k-1}}\big(t,\widehat{m}^*_{N_{k-1}}(t)\big)\big|<\varepsilon$ is satisfied for the prescribed tolerance $\varepsilon>0$, the optimal policy $u_{N_k}^*(t)$ of the order-$N_k$ truncated moment system gives a sufficiently good approximation to that $u^*(t)$ of the entire moment system, equivalently the parameterized ensemble system. The workflow of this FRL approach is shown in Figure \ref{fig:FRL}.

\begin{remark}[Optimality preserving filtration]
\label{rmk:filtration}
The observation made above Theorem \ref{thm:V_uniform_convergence}, that is, $\widehat V_N^*\geq V^*|_{\widehat{M}_{N}}$ for any moment truncation order $N\in\mathbb{N}$, directly extends to the established RL filtration as $\widehat V_{N_i}^*\geq\widehat V_{N_j}^*|_{\widehat{M}_{N_i}}\geq \widehat V_{N_j}^*$ for any $i<j$. This means that the optimal policy learned from one hierarchy of the filtrated RL problem is also optimal for, and may even decrease the values of the cumulative reward function learned from, all the lower-level hierarchies. This demonstrates that moving up the hierarchy in the RL filtration preserves the optimality of every hierarchy. 
\end{remark}

A direct observation on the truncated and entire moment systems in \eqref{eq:moment_system_truncated} and \eqref{eq:moment_system} together reveals that they are regulated by the same control policy, which announces policy search algorithms as the prime candidate for learning the optimal policy at each hierarchy in the RL filtration. Moreover, the optimality preserving property revealed in Remark \ref{rmk:filtration} strongly suggests a specific algorithmic approach: the policy learned from the current hierarchy is always promised to be a good initial condition for the successive hierarchy. Consequently, as the truncation order increases, the initial condition becomes closer to the optimal solution, effectively reducing the computational cost for leaning optimal policies of high-dimensional systems. This filtrated policy search algorithm is summarized in Algorithm \ref{alg:FRL}.

 \begin{algorithm}
 \caption{Filtrated policy search for learning optimal policies for parameterized systems}
 \label{alg:FRL}
 \begin{algorithmic}[2]
 \renewcommand{\algorithmicrequire}{\textbf{Input:}}
 \renewcommand{\algorithmicensure}{\textbf{Output:}}
\Require Initial state $x_0$, final time $T$, projection error tolerance $\varepsilon$       
 \Ensure  Optimal policy $u^*$ 
 \State \emph{Initialization}: truncation order $N_0$, initial policy $u^{(0)}$, projection error $P>\varepsilon$, hierarchy level $i=0$
\While {$P>\varepsilon$}
\If {$i>0$}
\State $i\gets i+1$
\State Pick $N_i>N_{i-1}$ and $u^{(0)}\gets u_{N_{i-1}}^*$
\EndIf
\State Generate data by solving the parameterized system with the input $u^{(0)}$ and collect the rewards
\State Compute order $N_i$ truncated moment kernelization: $\widehat m_{N_i}(t)$, $\widehat F_{N_i}(t,\widehat m_{N_i}(t),u^{(0)})$, $\bar r(t,\widehat m_{N_i}(t),u^{(0)})$, $\bar K(T,\widehat m_{N_i}(T))$
\State Solve $u_{N_i}^*(t)={\rm argmin}_u \widehat V_{N_i}(t,\widehat m_{N_i}(t))$ and compute $\widehat V^*_{N_i}(t,\widehat m^*_{N_i}(t))=\min_u\widehat V^*_{N_i}(t,\widehat m_{N_i}(t))$ 
\State Compute $P=\sup_{t\in[0,T]}\big|\widehat{V}^*_{N_i}\big(t,\widehat{m}^*_{N_i}(t)\big)-\widehat{V}^*_{N_{i-1}}\big(t,\widehat{m}^*_{N_{i-1}}(t)\big)\big|$ with $\widehat V^*_{N_{-1}}\equiv0$
\EndWhile
\State $u^*\gets u^*_{N_i}$
\Return $u^*$
 \end{algorithmic} 
 \end{algorithm}

\subsubsection{Spectral sequence convergence of filtrated policy search}

Let $\widehat m_{N_i}^{(j)}(t)$ denote the state trajectory of the order-$N_i$ truncated moment system driven by the policy $u_{N_i}^{(j)}$ resulting from the $j^{\rm th}$ iteration of the policy search (PS) algorithm, then the filtrated policy search shown in Algorithm \ref{alg:FRL} generate a \emph{spectral sequence}, given by,

\begin{align*}
\begin{array}{ccccccc}
                                                 & \text{PS iteration }0                               & \text{PS iteration }1                             & \cdots & \text{PS iteration }j                               & \cdots \\
 \text{Hierarchy 0}\ &  \widehat V_{N_0}(t,\widehat m^{(0)}_{N_0}(t)) & \widehat V_{N_0}(t,\widehat m^{(1)}_{N_0}(t)) & \cdots & \widehat V_{N_0}(t,\widehat m^{(j)}_{N_0}(t)) & \longrightarrow & \widehat V_{N_0}^*(t,\widehat m_{N_0}(t)) \\
 \text{Hierarchy 1}\ &  \widehat V_{N_1}(t,\widehat m^{(0)}_{N_1}(t)) & \widehat V_{N_1}(t,\widehat m^{(1)}_{N_1}(t)) & \cdots & \widehat V_{N_1}(t,\widehat m^{(j)}_{N_1}(t)) & \longrightarrow & \widehat V_{N_1}^*(t,\widehat m_{N_1}(t)) \\
 \vdots                                    \  &    \vdots                                            &     \vdots                                          &   &    \vdots                                         &          & \vdots     \\
 \text{Hierarchy $i$}\ &  \widehat V_{N_i}(t,\widehat m^{(0)}_{N_i}(t)) & \widehat V_{N_i}(t,\widehat m^{(1)}_{N_i}(t)) & \cdots & \widehat V_{N_i}(t,\widehat m^{(j)}_{N_i}(t)) & \longrightarrow & \widehat V_{N_i}^*(t,\widehat m_{N_i}(t)) \\
 \vdots                            \          &    \downarrow                                          &     \downarrow                                         &   &    \downarrow                                         &            &  \downarrow  \\
 &  V(t, m^{(0)}(t)) & V(t, m^{(1)}(t)) & \cdots & V(t, m^{(j)}(t)) & \longrightarrow & V^*(t,m(t))
\end{array}
\end{align*}
In this sequence, the $i^{\rm th}$ row converges to the value function $\widehat V^*_{N_i}(t,\widehat m_{N_i}(t))$ of order-$N_i$ truncated moment system, and the $j^{\rm th}$ column converges to the state-value function $V(t,m^{(j)}(t))$ of the entire moment system. Specifically, the row convergence naturally follows from the convergence of the PS algorithm \citep{Sutton18}, while the column convergence results from the continuity of the state-value function and the convergence of the truncated moment sequence to the infinite moment sequence. The row and column convergence together imply a stronger convergence property of Algorithm 1, that is, spectral sequence convergence, as defined and proved below.

\begin{theorem}[Spectral sequence convergence of FRL]
    \label{thm:spectral_sequence}
    Let $\big(\widehat V_{N_i}(t,\widehat m_{N_i}^{(j)})\big)_{i,j\in\mathbb{N}}$ be the spectral sequence generated by Algorithm \ref{alg:FRL}. Then, this sequence converges to the value function $V^*(t,m(t))$ of the moment kernelized system, i.e., $\widehat V_{i}(t,\widehat m_{N_i}^{(\phi(i))})\rightarrow V^*(t,m(t))$ as $i\rightarrow\infty$ for any monotonically increasing function $\phi:\mathbb{N}\rightarrow\mathbb{N}$.
\end{theorem}
\begin{proof}
    It suffices to show that both iterated limits, $\lim_{i\rightarrow\infty}\lim_{i\rightarrow\infty}\widehat V_{N_i}(t,\widehat m_{N_i}^{(j)})$ and 
    $\lim_{j\rightarrow\infty}$ $\lim_{i\rightarrow\infty}\widehat V_{N_i}(t,\widehat m_{N_i}^{(j)})$, exist and are equal to $V^*(t,m(t))$. 
    
    We first compute $\lim_{i\rightarrow\infty}\lim_{i\rightarrow\infty}\widehat V_{N_i}(t,\widehat m_{N_i}^{(j)})$. For each $j\in\mathbb{N}$, following a similar proof as for Theorem \ref{thm:moment_convergence} by replacing $\widehat m_{N}^*(t)$ and $m^*(t)$ with $\widehat m_{N_i}^{(j)}(t)$ and $m^{(j)}(t)$, respectively, we obtain $\widehat V_{N_i}(t,\widehat m_{N_i}^{(j)})\rightarrow V(t,m^{(j)}(t))$. Next, the convergence of $V(t,m^{(j)}(t))$ to $V^*(t,m(t))$ is essentially the convergence of the policy search algorithm \citep{Sutton18}, concluding that $\lim_{i\rightarrow\infty}\lim_{i\rightarrow\infty}\widehat V_{N_i}(t,\widehat m_{N_i}^{(j)})=V^*(t,m(t))$. 
    
    On the other hand, to compute $\lim_{j\rightarrow\infty}\lim_{i\rightarrow\infty}\widehat V_{N_i}(t,\widehat m_{N_i}^{(j)})$, we first note that for each $i$, the convergence of $\widehat V_{N_i}(t,\widehat m)_{N_i}(t)^{(j)})$ to $\widehat V^*_{N_i}(t,\widehat m_{N_i}(t))$ as $j\rightarrow\infty$ again follows from the convergence of the policy search algorithm. Since each $V^*_{N_i}(t,\widehat m_{N_i}(t))$, as a value function, is a viscosity solution of the Hamilton-Jacobi-Bellman equation associated with the order-$N_i$ truncated moment system, Corollary \ref{cor:HJB} implies that $V^*_{N_i}(t,\widehat m_{N_i}(t))\rightarrow V^*(t,m(t))$ as $i\rightarrow\infty$. This shows that $\lim_{j\rightarrow\infty}\lim_{i\rightarrow\infty}\widehat V_{N_i}(t,\widehat m_{N_i}^{(j)})=V^*(t,m(t))$, concluding the proof.     
\end{proof}

\paragraph{Spectral sequence-enabled early stopping for FRL.} Interpreted using the spectral sequence above, Algorithm \ref{alg:FRL} approaches the optimal policy $u^*$ of the parameterized ensemble system ``along the rightmost column'', meaning the learning sequence consists of the optimal policies of all the truncated moment systems. The spectral sequence convergence of FRL shown in Theorem \ref{thm:spectral_sequence} guarantees that learning sequences along any paths towards the bottom right element $V^*$ all converge to $u^*$. Algorithmically, such a learning sequence is obtained by employing early stopping to the PS algorithm at each hierarchy. For example, in the extreme case, the learning sequence along the diagonal of the spectral sequence is generated by executing only one iteration of the PS algorithm at each hierarchy. However, this ``naive'' diagonal learning sequence is generally not effective in terms of computational efficiency and learning accuracy, which motives the exploration of a sophisticated early stopping criterion for fully exploiting the advantage of the distinguished spectral sequence convergence of FRL. 

To collect some thoughts about the design of early stopping criteria, we devote our attention to policy gradient (PG) methods, the most popular policy search approach in RL literature. In the context of the spectral sequence, an PG algorithm applied at the $i^{\rm th}$ hierarchy generates a policy sequence in the form of $u_{N_i}^{(j)}=u_{N_i}^{(j-1)}+\delta u_{N_i}^{(j-1)}$. The update rule $\delta u_{N_i}^{(j-1)}$ is generally chosen to be proportional to $D\widehat J_{N_i}(u_{N_i}^{(j-1)})$, the gradient of the total reward function of the order-$N_i$ truncated moment system evaluated at $u_{N_i}^{(j-1)}$, to ensure $u_{N_i}^{(j)}\rightarrow u_{N_i}^*$. In practice, to keep the agents' behavior under control, e.g., staying in the safety region, all the time, the PG algorithm is generally implemented in the ``clipped'' manner to bound the amplitude of $\delta u_{N_i}^{(j)}$, such as Proximal Policy Optimization (PPO) and Trust Region Policy Optimization (TRPO) \citep{Schulman2017,Schulman15}. Inspired by this, we choose the the early stopping criterion, for starting the successive hierarchy of FRL, to be a threshold $\delta>0$ for the variation of the state-value function as $\sup_{t\in[0,T]}\big|\widehat V_{N_k}(t,\widehat m_k^{(i)}(t))-\widehat V_{N_k}(t,\widehat m_k^{(i)}(t))\big|>\delta$. It is worth mentioning that the use of the state-function threshold, instead of a policy threshold as in TRPO and PPO, takes into consideration the possible failure of the convergence of the generated policy sequence, as pointed out at the end of Section \ref{sec:ensemble_RL}. Before integrating the hierarchy-wise early stopping criterion into the FRL algorithm, we first carry out a detailed investigation into the computational and sample efficiency of FRL.

\paragraph{Convergence rate.}  It is intuitive that the rate of convergence of the spectral sequence depends on both the row and column convergence rates. As described in the proof of Theorem \ref{thm:spectral_sequence}, each row is generated by a standard policy search algorithm, and hence the rate of row convergence is entirely determined by the applied algorithm. Meanwhile, the column convergence is guaranteed by the moment convergence of the value function shown in Theorem \ref{thm:V_uniform_convergence}, which is a consequence of the convergence of the truncated moment sequence to the entire infinite moment sequence. Hence, we first evaluate the convergence rate of the truncate moment sequence. To this end, we note that $\|\widehat m_N(t)-m(t)\|^2=\sum_{k=N+1}^\infty|m_k|^2$ is essentially the tail of the moment sequence. Therefore, this convergence rate coincides with the rate of convergence of $m_k(t)$ to 0. Owing to the $\ell^2$-convergence of the moment sequence $\sum_{k=0}^\infty|m_k(t)|^2<\infty$, by the comparison test \citep{Rudin76}, it is necessary that $|m_k(t)|^2<k^{-1}$ for large enough $k$ since the harmonic series $\sum_{k=1}^\infty k^{-1}$ fails to converge. As a result, the convergence rate of $\widehat{m}_N(t)$ to $m(t)$ is bounded above by $O(N^{-{1/2}})$. 

To leverage this to compute the convergence rate of the column convergence, we have
\begin{align*}
    &\big|V(t, m^{(j)}(t))-\widehat V_{N_i}(t,\widehat m_j(t))\big| \\
    &=\Big|\int_0^T\bar r(s,m^{(j)}(s),u(s))ds+\bar K(T,m^{(j)}(s))
    -\int_0^T\bar r(s,\widehat m_{N_i}^{(j)}(s),u_{N_i}(s))ds+\bar K(T,m^{(j)}_{N_i}(s))\Big|\\
    &\leq \int_0^T\big|r(s,m^{(j)}(s),u(s))-\bar r(s,\widehat m_{N_i}^{(j)}(s),u_{N_i}(s))\big|ds+\big|\bar K(T,m^{(j)}(s))-\bar K(T,m^{(j)}_{N_i}(s))\big|\\
    &\leq\int_0^TL_r\|m^{(j)}(s)-m^{(j)}_{N_i}(s)\|ds+L_K\|m^{(j)}(T)-m^{(j)}_{N_i}(T)\|\\
    &\leq L_rTO(N_i^{-1/2})+L_KO(N_i^{-1/2})\sim O(N_i^{-1/2}),
\end{align*}
where we used the Lipchitz continuity of the running cost $\bar r$ and terminal cost $\bar K$, as presented in Assumption C2, with $L_r$ and $L_K$ denoting their Lipchitz constants, respectively. 

We can now integrate the rates of the column and row convergence. Let $O(\alpha(N))$ be the convergence rate of the policy search algorithm applied to the order-$N$ truncated moment system, then we have 
\begin{align*}
    \big|V^*(t,m(t))-V_{N}(t,& \widehat m_N^{(\lceil 1/\alpha(N)\rceil)}(t))\big|\leq\big|V^*(t,m(t))-V_{N}^*(t,\widehat m_N(t))\big| \\ 
    &+\big|V_{N}^*(t,\widehat m_N(t))-V_{N}(t,\widehat m_N^{(\lceil 1/\alpha(N)\rceil)}(t)\big|\sim O(N^{-1/2})+O(\alpha(N))
\end{align*}
by the triangle inequality, where $\lceil 1/\alpha(N)\rceil$ denotes the smallest integer greater than or equal to $1/\alpha(N)$. This concludes that, in the worst-case scenario, the convergence rate of the spectral sequence is bounded above by $O(N^{-1/2})+O(\alpha(N))$. To further elaborate on it, if the stopping criterion is set to be $\varepsilon$, then we should choose $N$ satisfying $N^{-1/2}+\alpha(N)<\varepsilon$ as the moment truncation order and run $\lceil 1/\alpha(N)\rceil$ iterations of the policy search algorithm on the truncated moment system.

\paragraph{Computational and sample efficiency.} As it is impractical to collect comprehensive measurement data from an parameterized ensemble system, practical implementations of the proposed FRL method require estimations of moments and moment systems using finite data points, even in cases where the mathematical model of the parameterized system is known. Suppose that only 
$q$ agents in the parameterized ensemble system in \eqref{eq:ensemble}, say $x(t,\beta_1)$, $\dots$, $x(t,\beta_q)$, can be measured. We then define the 
\emph{sample moments} by
\begin{align}
    \mathfrak{m}_k(t)=\frac{|\Omega|}{M}\sum_{i=1}^q\Phi_k(\beta_i)x(t,\beta_i)
    \label{eq:sample_moment}
\end{align}
for all $k\in\mathbb{N}$, where $|\Omega|$ is the Lebesgue measure (volume) of $\Omega$.

\begin{proposition}
    Suppose that at each time $t$, measurement data for $q$ randomly selected systems in the parameterized ensemble in \eqref{eq:ensemble} are available. Then, the sample moments satisfy $\mathfrak{m}_k(t)\rightarrow m_k(t)$ for all $k\in\mathbb{N}$ as $q\rightarrow\infty$. Moreover, the convergence rate is bounded above by $O(M^{-{1}})$, provided that each $\Phi_k$ is (essentially) bounded.

\end{proposition}
\begin{proof}
    Because the measured systems are randomly selected, $\b_i$, $i=1,\dots,q$ can be considered as a sequence of independent random variables, each of which has the uniform distribution on $\Omega$. This indicates that $\Phi_k(\beta_i)x(t,\beta_i)$, $i=1,\dots,q$, are independent and identically distributed random variables on $\mathbb{R}^n$. The expectation of each $\Phi_k(\beta_i)x(t,\beta_i)$, that is, $\mathbb{E}(\Phi_kx_t)=\frac{1}{|\Omega|}\int\Phi_k(\beta)x_t(\beta)d\beta$, coincides with $\frac{1}{|\Omega|}m_k(t)$ and satisfies $\mathbb{E}\big|\Phi_k x_t\big|=\frac{1}{|\Omega|}\int_\Omega|\Phi_k(\beta)x_t(\beta)|d\beta\leq\frac{1}{|\Omega|}\|\Phi_k\|_{\mathcal{H}}\|x_t\|_{\mathcal{H}}=\frac{1}{|\Omega|}\|x_t\|_{\mathcal{H}}<\infty$. The strong law of large numbers then implies $\mathfrak{m}_k(t)\rightarrow m_k(t)$ almost surely as $q\rightarrow\infty$  \citep{Billingsley95}.

    To derive the bound of the convergence rate, it is necessary to compute the tail probability $\mathbb{P}\big(|\mathfrak{m}_k(t)-m_k(t)|>\varepsilon\big)$. We claim that when $\Phi_k$ is essentially bounded by some constant $C$, then the random variable $\Phi_kx_t$ has a finite variance. This follows from $\mathbb{E}|\Phi_kx_t|^2=\frac{1}{|\Omega|}\int_\Omega\Phi_k^2(\beta)x_t^2(\beta)d\beta\leq \frac{C^2}{|\Omega|}\int_\Omega x_t^2(\beta)d\beta=\frac{C^2\|x_t\|_{\mathcal{H}}^2}{|\Omega|}<\infty$. By Chebyshev's inequality, 
    we have
    \begin{align*}
        &\mathbb{P}\big(|\mathfrak{m}_k(t)-m_k(t)|>\varepsilon\big)\leq\frac{\mathbb{E}|\mathfrak{m}_k(t)-m_k(t)|^2}{\varepsilon^2}=\frac{|\Omega|^2}{(M\varepsilon)^2}\mathbb{E}\Big|\sum_{i=1}^M\Big(\Phi_k(\beta_i)x_t(\beta_i)-\frac{m_k(t)}{|\Omega|}\Big)\Big|^2\\
        &=\frac{|\Omega|^2}{(M\varepsilon)^2}\mathbb{E}\Big|\sum_{i=1}^M\Big(\Phi_k(\beta_i)x_t(\beta_i)-\mathbb{E}(\Phi_kx_t)\Big)\Big|^2=\frac{|\Omega|^2}{(M\varepsilon)^2}\sum_{i=1}^M\mathbb{E}\Big|\Big(\Phi_k(\beta_i)x_t(\beta_i)-\mathbb{E}(\Phi_kx_t)\Big)\Big|^2\\
        &=\frac{|\Omega|^2}{(M\varepsilon)^2}\sum_{i=1}^M{\rm Var}(\Phi_kx_t)\leq\frac{C^2|\Omega|\|x_t\|_{\mathcal{H}}^2}{M\varepsilon^2}\sim O\Big(\frac{1}{M}\Big),
    \end{align*}
    where the second and third equalities follow from the independence of $\b_i$, giving the desired bound. 
\end{proof}

Because the bound on the convergence rate is independent of the order of moments, the entire sample moment sequence $\mathfrak{m}(t)=(\mathfrak{m}_k(t))_{k\in\mathbb{N}}$ converges to the moment sequence $m(t)$ with the rate $O(M^{-1})$ as well. In practice, we will compute the sample moments up to a finite order $N$, constituting an order-$N$ truncated sample moment sequence $\widehat{\mathfrak{m}}_N(t)$. This computation only requires basic linear algebra operations with low computational complexity. To illustrate this, we concatenate the states of the $q$ measured systems into a vector $\widehat X_M(t)=[x'(t,\beta_1),\dots,x'(t,\beta_M)]'\in\mathbb{R}^{Mn}$ and define the \emph{moment kernel matrix} $\widehat \Psi_N\in\mathbb{R}^{(N+1)\times M}$, whose $(i,j)$-entry is given by $\Phi_i(\beta_j)$. By the definition of sample moments in \eqref{eq:sample_moment}, the order-$N$ truncated sample moment sequence, represented as a vector $\widehat{\mathfrak{m}}_N(t)=[\mathfrak{m}_0'(t),\dots,\mathfrak{m}_{N}(t)]'\in\mathbb{R}^{(N+1)n}$, is given by
\begin{align}
    \label{eq:sample_moment_kernel}
    \widehat{\mathfrak{m}}_N(t)=\big(\widehat\Psi_N\otimes I_n\big)\cdot\widehat X_M(t),
\end{align}
where $I_n$ denotes the $n$-by-$n$ identity matrix and $\otimes$ is the Kronecker product of matrices. In terms of computations, $\widehat\Psi_N\otimes I_n$ essentially arranges the $(N+1)M$ pre-determined numbers $\Psi_i(\beta_j)$ into an $(N+1)n$-by-$Mn$ (block) matrix with $(N+1)M$ $n$-by-$n$ diagonal blocks, which is of $O(1)$ complexity. The matrix multiplication in \eqref{eq:sample_moment_kernel}, in the worst case, requires $(N+1)Mn$ scalar multiplications and $(N+1)(M-1)n$ additions, resulting in a complexity of $O(NMn)$. In total, the complexity of computing the order-$N$ truncated moment sequence using \eqref{eq:sample_moment_kernel} is $O(NMn)$ in the worst-case scenario. 

In addition, because the vector field $\bar F$ of the moment kernelized system in \eqref{eq:moment_system} is the moment sequence of that $F$ of the parameterized ensemble system in \eqref{eq:ensemble} as discussed in Section \ref{sec:moment_function}, the above computational and sample efficiency analysis is directly applicable to learning the moment kernelized system.


\subsubsection{Second-order policy search}
\label{sec:PIA}

To showcase the FRL algorithm with early stopped hierarchies, in this section, we propose a second-order PS algorithm and integrate it into the FRL structure with a early stopping criterion. The development of the PS algorithm is based on the theory of differential dynamic programming \citep{Jacobson70,Mayne66}. As a major advantage, the algorithm does not involve time discretization and is directly applicable to infinite-dimensional continuous-time systems, particularly, the moment kernelized ensemble system in \eqref{eq:moment_system}. The main idea is to expand the value function into a Taylor series up to the second-order term and then derive an update rule represented in the form of differential equations. In the following, we only highlight the key steps in the development of the algorithm (see Appendix \ref{appd:PIA_derivation} for the detailed derivation).

\paragraph{Quadratic approximation of Hamilton-Jacobi-Bellman equations.} 
Let $\<\cdot,\cdot\>:\mathcal{M}\times\mathcal{M}\rightarrow\mathbb{R}$ be the inner product on the RKHS $\mathcal{M}$, then, for any variation $\delta m\in\mathcal{M}$ at $m\in\mathcal{M}$, applying Taylor's theorem to the ($\mathcal{M}$-component of) value function $V^*:[0,T]\times\mathcal{M}\rightarrow\mathbb{R}$ yields $V^*(t,m+\delta m)=V^*(t,m)+\<DV^*(t,m),\delta m\>+\frac{1}{2}\<D^2V^*(t,m)\cdot \delta m,\delta m\>+o(\delta m^2)$, where the ``Hessian'' $D^2V^*(t,m)$ is identified with a bounded linear operator from $\mathcal{M}\rightarrow\mathcal{M}$ and $D^2V^*(t,m)\cdot \delta m$ denotes the evaluation of $D^2V^*(t,m)$ at $\delta m$ \citep{Lang99}. The quadratic approximation of the HJB equation in \eqref{eq:HJB} is subsequently obtained by replacing $V^*$ by its Taylor expansion with the high-order term $o(\delta m^2)$ neglected, that is,
\begin{align}
    \frac{\partial}{\partial t}&V^*(t,m)+\Big\<\frac{\partial}{\partial t}DV^*(t, m),\delta m\Big\>+\frac{1}{2}\Big\<\frac{\partial}{\partial t}D^2V^*(t, m)\cdot\delta m,\delta m\Big\>\nonumber\\
&+\min_{u}\Big\{H(t, m+\delta m, u,DV^*(t, m))+\Big\<D^2V^*(t, m)\cdot\delta m,\bar F(t, m+\delta m, u)\Big\>\Big\}=0, \label{eq:HJB_Taylor}
\end{align}
where $H(t,m+\delta m,u,DV^*(t,m))=\bar r(t,m,u)+\<DV^*(t,m),\bar F(t,m,u)\>$ is the \emph{Hamiltonian} of the moment kernelized system. 

\paragraph{Second-order policy search.} The next step is to design an iterative algorithm to solve \eqref{eq:HJB_Taylor}. The main idea is to generate a policy learning sequence $u^{(k)}$ using the update rule $u^{(k+1)}=u^{(k)}+\delta u^{(k)}$ so that $m^*=\lim_{k\rightarrow\infty}m^{(k)}$, where $m^{(k)}$ and $m^*$ are the trajectories of the moment system steered by $u^{(k)}$ and $u^*$, respectively. Then, the $(m^*,u^*)$ solves \eqref{eq:HJB_Taylor}. To find $\delta u^{(k)}$, we note that because the pair $(m^{(k)},u^{(k)})$ satisfies the moment system in \eqref{eq:moment_system}, $\delta m^{(k)}=0$ holds so that the equation in \eqref{eq:HJB_Taylor} reduces to $\frac{\partial}{\partial t} V^*(t,m^{(k)})+\min_{u} H(t,m^{(k)}, u,DV(t,m^{(k)}))=0$, in which the solution $\tilde u^{(k)}={\rm argmin}_uH(t,m^{(k)}, u,DV(t,m^{(k)}))$ is represented in terms of $m^{(k)}$ and $DV^*(t,m^{(k)})$. Steered by this policy $\tilde u^{(k)}$, the trajectory of the moment system is not $m^{(k)}$ any more, and hence a variation $\delta m^{(k)}$ on $m^{(k)}$ is produced so that $m^{(k)}+\delta m^{(k)}$ satisfies the moment system, which also updates the equation in \eqref{eq:HJB_Taylor} to
\begin{align}
\frac{\partial}{\partial t}& V^*(t,m^{(k)})+\Big\<\frac{\partial}{\partial t}DV^*(t,m^{(k)}),\delta m^{(k)}\Big\>+\frac{1}{2}\Big\<\frac{\partial}{\partial t}D^2V^*(t, m^{(k)})\cdot\delta m^{(k)},\delta m^{(k)}\Big\>\nonumber\\
&\qquad\qquad\qquad\quad+\min_{\delta u}\Big\{H(t,m^{(k)}+\delta m^{(k)},\tilde u^{(k)}+\delta u,DV^*(t,m^{(k)}))\nonumber\\
&\qquad\qquad\qquad\qquad+\Big\<D^2V^*(t,m^{(k)})\cdot\delta m^{(k)},\bar F(t,m^{(k)}+\delta m^{(k)},\tilde u^{(k)}+\delta u)\Big\>\Big\}=0. \label{eq:HJB_update_1}
\end{align}
To solve the minimization problem in \eqref{eq:HJB_update_1}, we expand the objective function into the Taylor series with respect to ($\delta m^{(k)},\delta u$) up to second-order, and then compute the critical point $\delta u^{(k)}$ in terms of $DV^*(t,m^{(k)})$ and $D^2V^*(t,m^{(k)})$. Following from this, it remains to find $DV^*(t,m^{(k)})$ and $D^2V^*(t,m^{(k)})$. To this end, we observe that, after the aforementioned Taylor expansion, the equation in \eqref{eq:HJB_update_1} becomes an algebraic second-order polynomial equation in $\delta m^{(k)}$, which is satisfied if all the coefficients are 0. This yields a systems of three ordinary differential equations, given by,
\begin{align}
\frac{d}{dt}\delta V(t,m^{(k)}(t))&=H(t,m^{(k)},{u^{(k)}},DV^*(t,m^{(k}))-H(t,m^{(k)},\tilde u^{(k)},DV^*(t,m^{(k)})), \label{eq:dynamics_deltaV_1}\\
\frac{d}{dt}DV^*(t,m^{(k)}(t))&=-DH-D^2V\cdot(\bar F-\bar F(t,m^{(k)},{u^{(k)}})),\label{eq:dynamics_DV_1}\\
\frac{d}{dt}D^2V^*(t,m^{(k)}(t))&=-D^2H-D\bar F'\cdot D^2V^*-D^2{V^*}'\cdot D\bar F\nonumber\\
&\,+\left[\frac{\partial DH}{\partial u}+\frac{\partial \bar F'}{\partial u}\cdot D^2V^*\right]'\cdot\Big(\frac{\partial^2 H}{\partial u^2}\Big)^{-1}\cdot\left[\frac{\partial DH}{\partial u}+\frac{\partial \bar F'}{\partial u}\cdot D^2V^*\right] \label{eq:dynamics_D^2V_1}
\end{align}
with the terminal conditions $DV^*(T,m^{(k)}(T))=D\bar K(T,m^{(k)}(T))$ and $D^2V^*(T,m^{(k)}(T))=D\bar K(T,m^{(k)}(T))$, where $\delta V(t,m^{(k)}(t))=V^*(t,m^{(k)}(t))-V(t,m^{(k)}(t))$ and $V(t,m^{(k)}(t))$ is the state-value function of the moment system driven by the policy $u^{(k)}$. Moreover, to simplify the notations, all the functions in the equations in \eqref{eq:dynamics_DV_1} and \eqref{eq:dynamics_D^2V_1} without arguments are evaluated at $(t,m^{(k)},\tilde u^{(k)})$, and `$\prime$' denotes the transpose of linear operators. 

\paragraph{Early stopping criterion.} Because the development of the second-order PS algorithm is based on the Taylor series approximation, it is required that the amplitudes of $\delta m^{(k)}$ and $\delta u^{(k)}$ are small enough for all $k$. A necessary condition to guarantee this is to bound $\delta V^{(k)}=\sup_{t}|\delta V(t,m^{(k)}(t))|$ by a threshold $\eta$. 
When putting the proposed second-order PS algorithm into each hierarchy of the FRL structure, say the $i^{\rm th}$ hierarchy, the algorithm will be terminated if $\sup_{t}|\widehat V_{N_i}(t,\widehat m_{N_i}^{(k)}(t))-\widehat V_{N_i}(t,\widehat m_{N_i}^{(k-1)}(t))|>\eta$. Then, the resulting policy $u^{(k)}$ will be input to the $(i+1)^{\rm th}$ hierarchy as the initial condition. The FRL with the early stopped second-order policy search is shown in Algorithm \ref{alg:FRL_PIA}.

 \begin{algorithm}
 \caption{Filtrated reinforcement learning for parameterized systems with early stopped second-order policy search hierarchies}
 \label{alg:FRL_PIA}
 \begin{algorithmic}[2]
 \renewcommand{\algorithmicrequire}{\textbf{Input:}}
 \renewcommand{\algorithmicensure}{\textbf{Output:}}
\Require Initial state $x_0$, final time $T$, projection error tolerance $\varepsilon$, value function variation \\ \qquad tolerance $\eta$, maximum number of policy search  iterations $K$
 \State \emph{Initialization}: truncation order $N_0$, control policy $u^{(0)}$, projection error $P>\varepsilon$, value function variation $\delta V=0$, hierarchy level $i=0$, policy search iteration number $j=0$
 \While {$P>\varepsilon$}
 \While {$\delta V\leq\eta$ \textbf{and} $j\leq K$}
 \State Generate data by solving the parameterized system with the input $u^{(j)}$ and collect the rewards
 \State Compute order $N_i$ truncated moment kernelization: $\widehat m_{N_i}(t)$, $\widehat F_{N_i}(t,\widehat m_{N_i}(t),u^{(0)})$, $\bar r(t,\widehat m_{N_i}(t),u^{(0)})$, $\bar K(T,\widehat m_{N_i}(T))$
 \State Compute $\delta\widehat V_{N_i}(t,\widehat m_{N_i}^{(j)}(t))$ $D\widehat{V}_{N_i}^*(t,\widehat m_{N_i}^{(j)}(t))$ and $D^2\widehat{V}_{N_i}^*(t,\widehat m_{N_i}^{(j)}(t))$ for $0\leq t\leq T$ by solving the system of differential equations in \eqref{eq:dynamics_deltaV_1}, \eqref{eq:dynamics_DV_1} and \eqref{eq:dynamics_D^2V_1}
\State ${u^{(j+1)}}(t)\gets {\rm argmin}_a H(t,\widehat m_{N_i}^{(j)},a,D\widehat{V}_{N_i}(t,\widehat m_{N_i}^{(j)}))$ for all $0\leq t\leq T$
\State $\delta V\gets\max_{t}|\delta\widehat V(t,\widehat m^{(j)}_{N_i})|$
\State $j\gets j+1$
 \EndWhile
 \State $\widehat m_{N_i}^*\gets \widehat m_{N_i}^{(j-1)}$
 \If {$i\geq 1$}
 \State $P=\max_{t\in[0,T]}\big|\widehat V_{N_i}(t,\widehat m_{N_i}^*(t))-\widehat V_{N_{i-1}}(t,\widehat m_{N_{i-1}}^*(t))\big|$
 \EndIf
 \State $N_{i+1}\gets N_{i}+1$, $u^{(0)}\gets u^{(j-1)}$
 \State $i\gets i+1$
 \EndWhile
 \State $u^*\gets u^{(j-1)}$ \Return $u^*$
 \end{algorithmic}
 \end{algorithm}

\subsection{Examples and simulations}
\label{sec:ex}

In this section, we will demonstrate the performance and efficiency of the proposed FRL algorithm using parameterized ensemble systems arising from practical applications. All the simulations were run on an Apple M1 chip with 16 GB memory.

\subsubsection{Infinite-dimensional linear–quadratic regulators}
\label{sec:LQR}

Linear-quadratic (LQ) problems, those are, linear systems with state-value functions given by quadratic forms in both of the system state and control variables, are the most fundamental control problems, which have been extensively studied for finite-dimensional linear systems \citep{Brockett15,Kwakernaak72,Sontag98}. However, LQ problems for parameterized ensemble systems defined on infinite-dimensional function spaces remain barely understood. In this example, we will fill in this literature gap to approach such type of infinite-dimensional LQ problems using the proposed FRL algorithm. 

To illuminate the main idea as well as to demonstrate how FRL addresses the curse of dimensionality, we revisit the scalar linear parameterized system in \eqref{eq:challenge_system}, i.e. $\frac{d}{dt}x(t,\b)=\b x(t,\b)+u(t)$, $\beta\in[-1,1]$, with a finite-time horizon state-vale function given by $V(t,x_t)=\int_{-1}^1\Big[\int_t^T\big(x^2(t,\b)+u^2(t)\big)dtd\b+x^2(T,\beta)\Big]d\b$. To apply the moment kernel transform, we evolve the system on the Hilbert space $L^2([0,1],\mathbb{R})$ and choose the basis $\{\Phi_k\}_{k\in\mathbb{N}}$ to be the set of Chebyshev polynomials. In this case, the moment kernelized system and state-value function can actually be analytically derived as
$\frac{d}{dt}m(t)=Am(t)+bu(t)$ and $V(t,m(t))=\int_0^T\big[\|m(t)\|^2+2u^2(t)\big]dt+\|m(T)\|^2$, where $A=\frac{1}{2}(L+R)$ with $L$ and $R$ the left- and right-shift operators, given by $(m_0(t),m_1(t),\dots)\mapsto (m_1(t),m_2(t),\dots)$ and $(m_0(t),m_1(t),\dots)\mapsto (0,m_0(t),\dots)$, respectively (see Appendix \ref{appd:derive_moment_system_linear} for the detailed derivation). However, the analytic form of the moment kernelized system and state-value function are not required in the implementation of FRL.

\begin{figure}[H]
     \centering
     \begin{subfigure}[b]{0.5\textwidth}
     \centering
     \includegraphics[width=\textwidth]{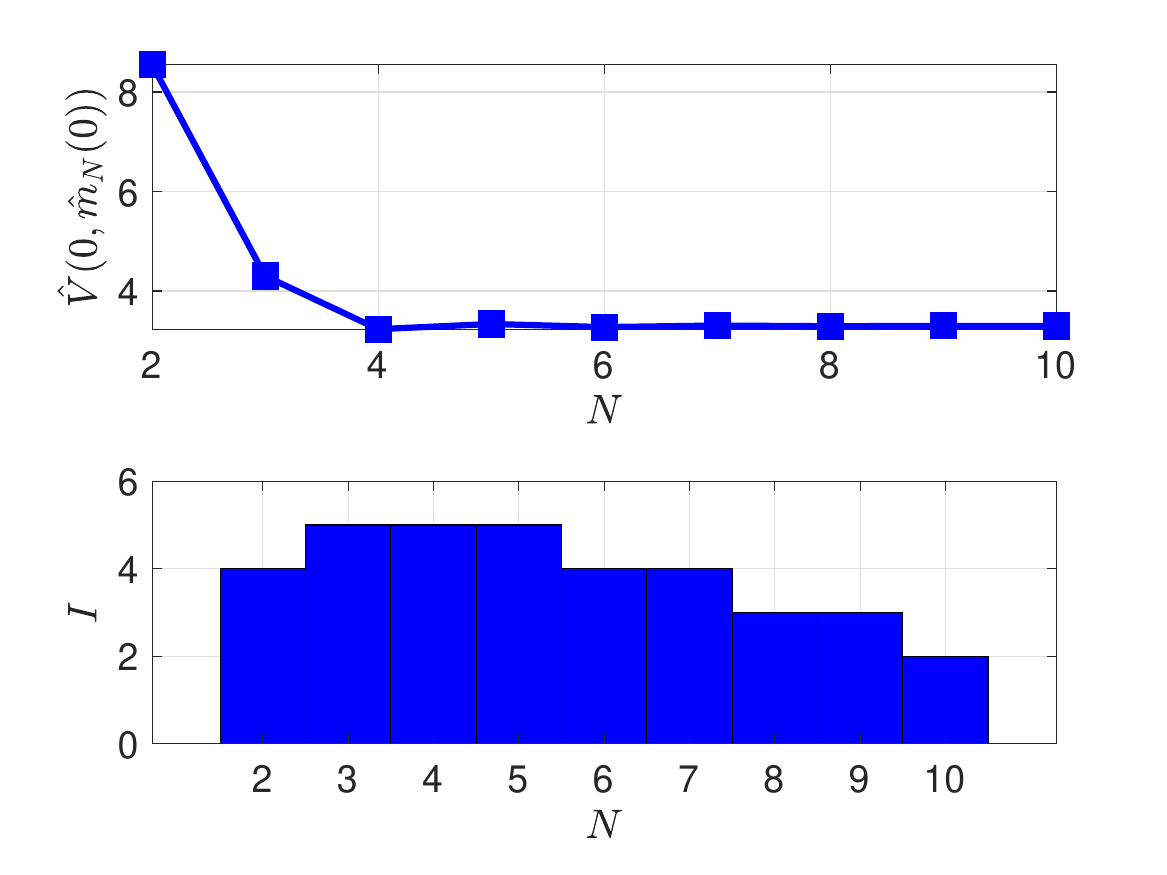}
     \caption{}
     \label{fig:cost_linear}
     \end{subfigure}
     \begin{subfigure}[b]{0.49\textwidth}
     \centering
     \includegraphics[width=\textwidth]{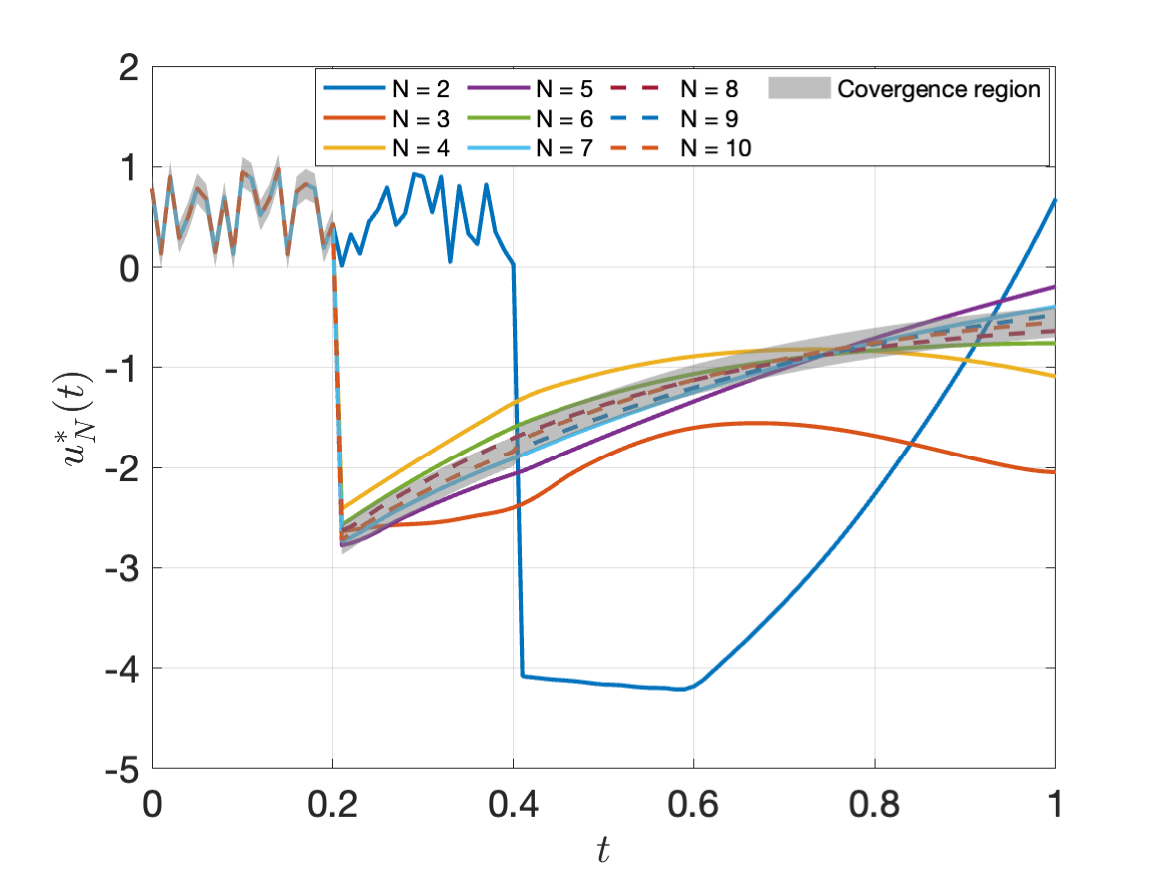}
     \caption{}
      \label{fig:control_linear}
     \end{subfigure}
     \caption{\footnotesize Learning the finite-time horizon LQR policy for the parameterized ensemble system in \eqref{eq:challenge_system} using FRL with early stopped hierarchies shown in Algorithm \ref{alg:FRL_PIA}. In particular, (a) shows the total reward (top panel) and the number of the second-order PS iterations with respect to the truncation order (hierarchy level) $N$ ranging from 2 to 20, and (b) plots the learned policy from each hierarchy of FRL.}
     \label{fig:linear}
 \end{figure}

 In the simulation, we choose the final time and initial condition for the parameterized system to be $T=1$ and $x_0=1$, the constant function on $[-1,1]$, respectively, and the tolerance for the value function variation at each hierarchy is set to be $\eta=1$. Moreover, the truncation order $N$ is varied from $2$ to $10$, and in each case the evolution of the truncated moment kernelized system is approximated using the sample moments computed from the measurement data for 500 systems with their system parameters $\beta$ uniformly sampled from $[-1,1]$. The simulation results generated by Algorithm \ref{alg:FRL_PIA} are shown in Figure \ref{fig:linear}. Specifically, Figure \ref{fig:cost_linear} shows the total reward $\widehat V(0,\widehat m_N(0))$ (top panel) and the number of the policy search iterations (bottom panel) with respect to the truncation order (hierarchy level) $N$. We observe that the total reward converges to the minimum value after only 4 hierarchies of the algorithm, which demonstrates the high efficiency of FRL. Correspondingly, Figure \ref{fig:control_linear} plots the policy $u_N$ learned from each hierarchy of FRL, which stabilizes to the shadowed region starting from $N=6$. In addition, it is worth mentioning that the computational time for running 10 hierarchies of the algorithm is only 3.97 seconds, indicating the low computational cost of FRL. As a result, the curse of dimensionality is effectively mitigated. 


\begin{figure}[H]
     \centering
     \begin{subfigure}[b]{0.493\textwidth}
     \centering
     \includegraphics[width=\textwidth]{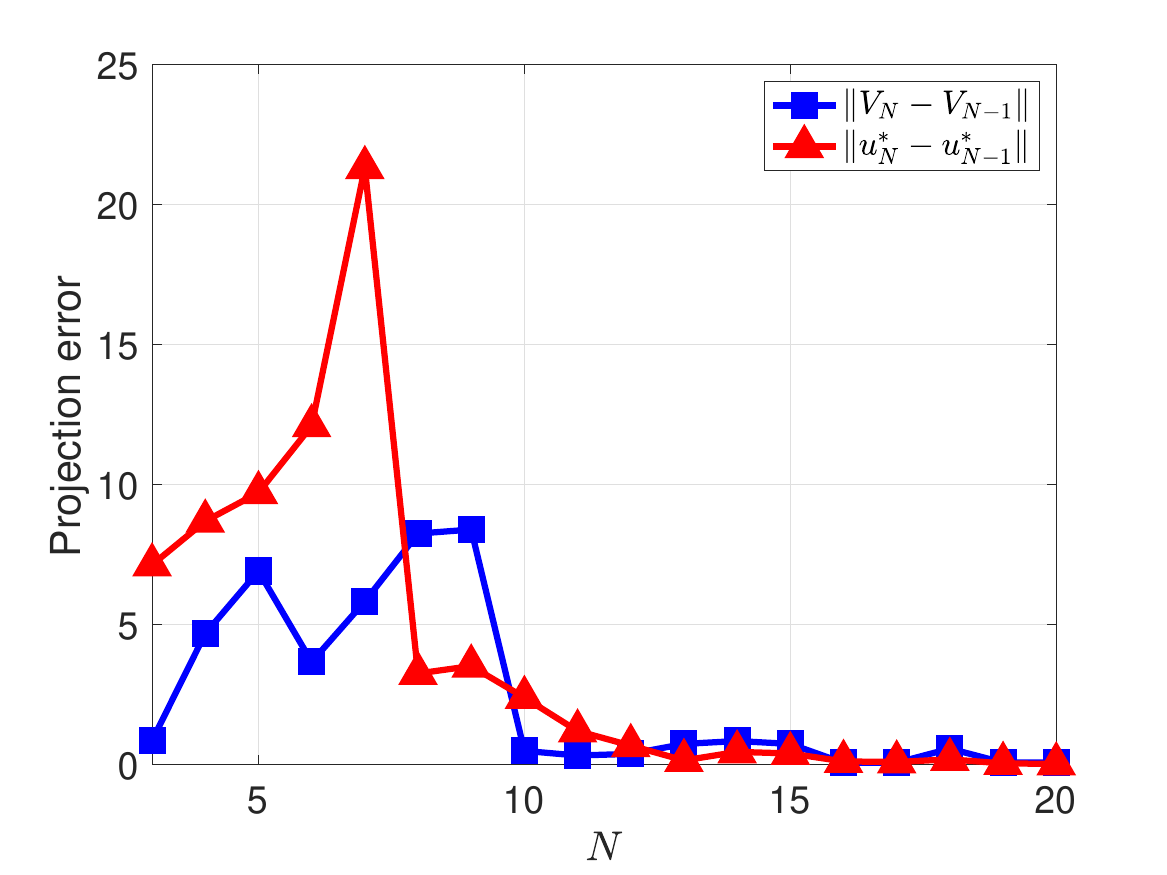}
     \caption{}
     \label{fig:HRL_error}
     \end{subfigure}
     \begin{subfigure}[b]{0.495\textwidth}
     \centering
     \includegraphics[width=\textwidth]{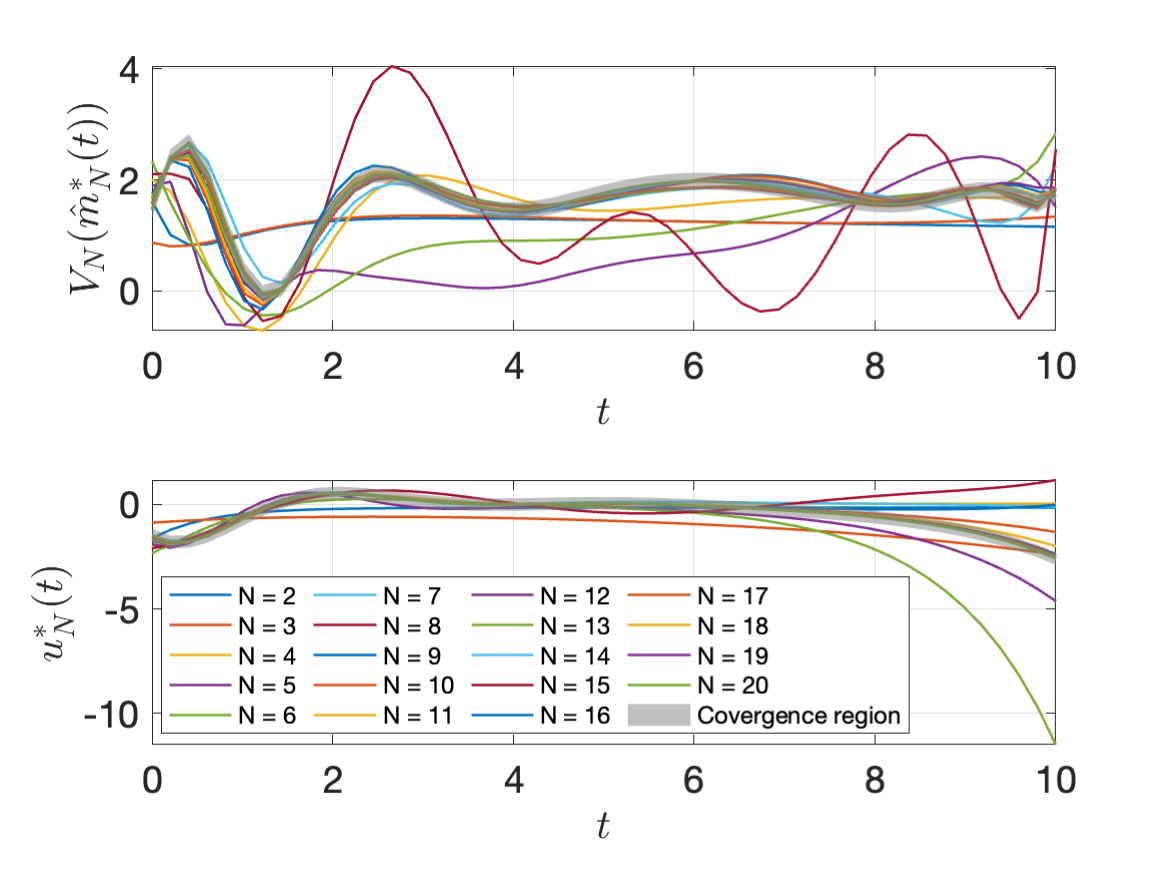}
     \caption{}
      \label{fig:HRL_V_and_u}
     \end{subfigure}
     \caption{\footnotesize FRL resolves the convergence issue caused by applying classical RL algorithms to sampled parameterized systems. Algorithm \ref{alg:FRL} with the standard value iteration applied to each hierarchy is used to learn the infinite-time horizon LQR policy and value function of the parameterized system in \eqref{eq:challenge_system}. Specifically, (a) plots the projection error of the learned value functions and optimal policies, and (b) shows the learned value function (top) optimal policy (bottom) at each hierarchy.}
     \label{fig:HRL_ex}
 \end{figure}

To further demonstrate the advantages of FRL, we will show that it also resolves the convergence issue caused by applying classical RL algorithms to sampled parameterized systems as pointed out in Section \ref{sec:challenges}. To this end, we revisit the infinite-time horizon LQR problem presented there, that is, the same system in \eqref{eq:challenge_system} with the state-value function given by $V(x_t)=\int_{-1}^1\int_0^\infty e^{-2.5t}\big[x^2(t,\beta)+u^2(t)\big]dtd\beta$. In this case, we use Algorithm \ref{alg:FRL} with the standard value iteration applied to each hierarchy, and the simulation results are shown in Figure \ref{fig:HRL_ex}. A comparison between Figures \ref{fig:HRL_error} and \ref{fig:non_convergence} reveals that now the projection error for both of the learned value functions and optimal policies converge to 0. Meaning, the sequences of value functions and optimal policies generated by FRL in Algorithm \ref{alg:FRL} are Cauchy sequences, and hence necessarily converge to those of the parameterized system in \eqref{eq:challenge_system} \citep{Rudin76}. As a further verification, Figure \ref{fig:HRL_V_and_u} shows that the learned value functions and optimal policies stabilize to the corresponding shadowed regions. 

\subsubsection{Filtrated reinforcement learning for parameterized nonlinear ensemble systems}
\label{eq:Bloch}

In this section, we apply FRL to learn the optimal policy for a parameterized nonlinear ensemble system arising from quantum mechanics and quantum control. The policy learning problem in our study is concerned with robust excitation of a nuclear spin sample, typically consisting of as many as spins in the order of the Avogadro’s 
number ($\sim 10^{23}$). It is also referred to as the pulse design problem, which is crucial in quantum science and technology. For example, it enables all the applications of nuclear magnetic resonance (NMR) spectroscopy, including magnetic resonance imaging (MRI), quantum computing, quantum optics, and quantum information processing \citep{Li09,Li22,Silver85,Molmer04,Li_SICON11,Li_PRA_Transport11,Li_TAC14}. The dynamics of nuclear spins immersed in a static magnetic field with respect to the rotating frame is governed by the Bloch equation
\begin{align}
\frac{d}{dt}x(t,\beta)=\left[\begin{array}{ccc} 0 & 0 & -\beta u(t) \\ 0 & 0 & \beta v(t) \\ \beta u(t) & -\beta v(t) & 0 \end{array}\right]x(t,\beta),\label{eq:Bloch}
\end{align}
which was derived by the Swiss-American physicist Felix Bloch in 1946 \citep{Cavanagh10}. In this system, the state variable $x(t,\beta)$ denotes the bulk magnetization of spins, $u(t)$ and $v(t)$ represent the external radio frequency (rf) fields, and the system parameter $\beta\in\Omega=[1-\delta,1+\delta]$ with $0<\delta<1$ is referred to as the \emph{rf inhomogeneity}, characterizing the phenomenon that spins in different positions of the sample receive different strength of the rf fields. In practice, the inhomogeneity can be up to $\delta=40\%$ of the strength of the applied rf fields \citep{Nishimura2001}. A typical policy learning task is to design the rf fields $u(t)$ and $v(t)$, with the minimum energy, to steer the parameterized Bloch system in \eqref{eq:Bloch} from the \emph{equilibrium state} $x_0(\beta)=(0,0,1)'$ to the \emph{excited state} $x_F(\beta)=(1,0,0)'$ for all $\beta$. We formulate this task as an RL problem over the function space $L^2(\Omega,\mathbb{R}^3)$, i.e., $x(t,\cdot)\in L^2(\Omega,\mathbb{R}^3)$, with the state-value function defined as $V(t,x_t)=\int_t^T\big(u^2(t)+v^2(t)\big)dt+\int_{1-\delta}^{1+\delta}|x(T,\beta)-x_F(\beta)|^2d\beta$, where $|\cdot|$ denotes a norm on $\mathbb{R}^3$. 

We still use the Chebyshev polynomial basis $\{\Phi_k\}_{k\in\mathbb{N}}$ to define the moment transform for kernelizing the Bloch system and the state-value function. To guarantee the orthonomality of $\{\Phi_k\}_{k\in\mathbb{N}}$, we first rescale the range of the rf inhomogenity from $[1-\delta,1+\delta]$ to $[-1,1]$ by the linear transformation $\beta\mapsto\eta=(\beta-1)/\delta$, and then apply the moment kernel transform defined in \eqref{eq:moment} (see Appendix \ref{appd:derive_moment_system_Bloch} for the detailed derivation). 

\begin{figure}[H]
     \centering
     \begin{subfigure}[b]{0.495\textwidth}
     \centering
     \includegraphics[width=\textwidth]{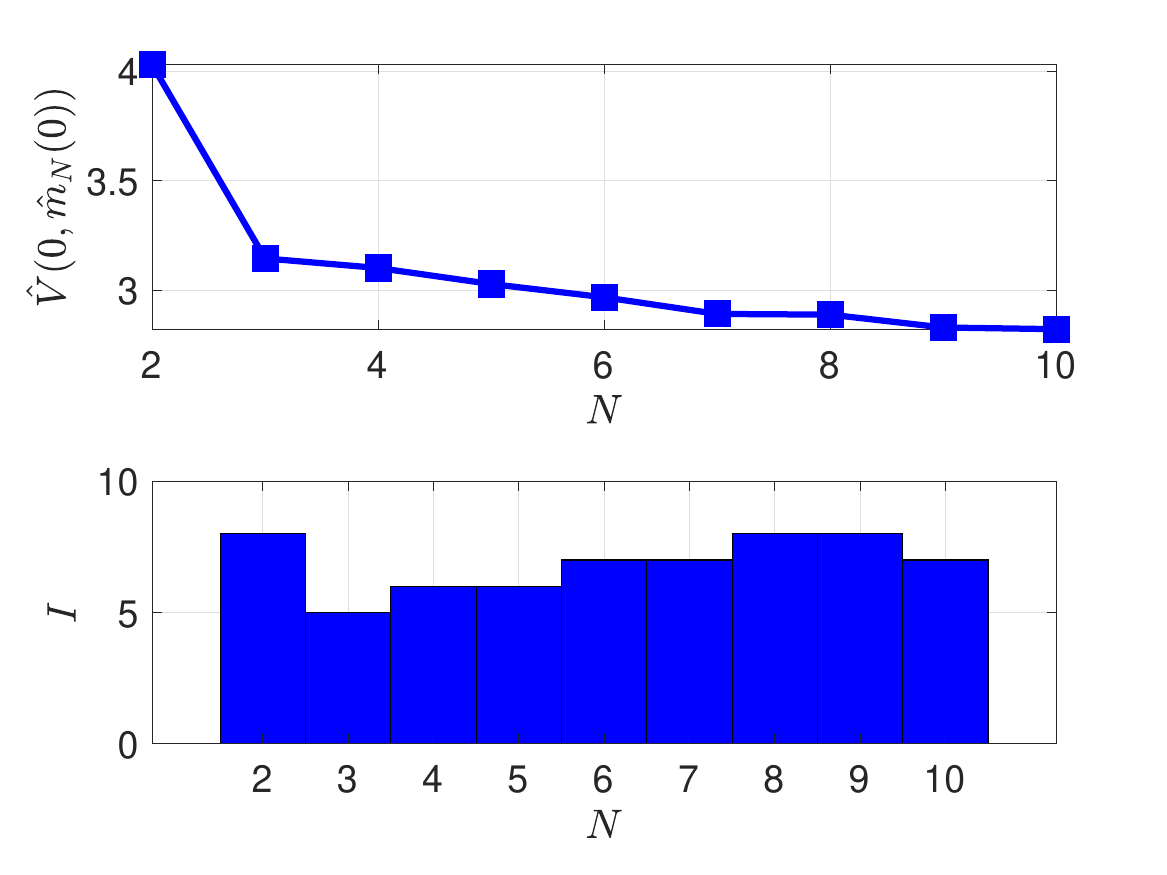}
     \caption{}
     \label{fig:cost_Bloch}
     \end{subfigure}
     \begin{subfigure}[b]{0.495\textwidth}
     \centering
     \includegraphics[width=\textwidth]{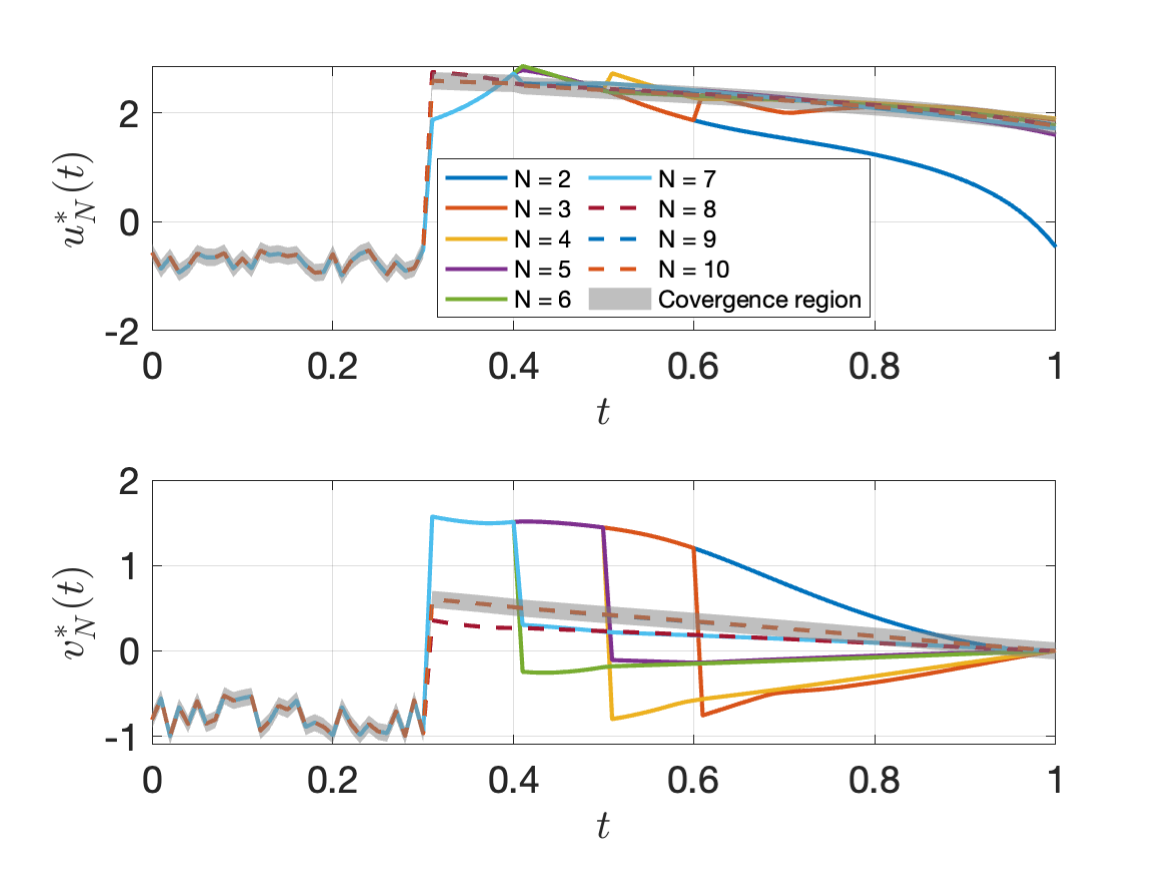}
     \caption{}
      \label{fig:control_Bloch}
     \end{subfigure}
     \caption{\footnotesize Policy learning for robust excitation of the nuclear spin system in \eqref{eq:Bloch} with the presence of rf inhomogeneity using FRL shown in Algorithm \ref{alg:FRL_PIA}. In particular, (a) shows the total reward (top panel) and number of the second-order PS iterations with respect to the moment truncation order, that is, the hierarchy level, $N$ ranging from 2 to 10, and (b) plots the policies learned from each hierarchy of the algorithm.}
     \label{fig:Bloch}
 \end{figure}

We then apply Algorithm \ref{alg:FRL_PIA} to learn the optimal policies for the spin system in \eqref{eq:Bloch}. In the simulation, we consider the maximal rf inhomogeneity $\delta=40\%$ encountered in practice and pick the final time to be $T=1$. Similar to the previous example, we vary the moment truncation order $N$ from 2 to 10, and for each $N$ we approximate the evolution of the truncated moment kernelized system using the sample moments computed from the measurement data for 500 systems in the ensemble with the system parameters uniformly sampled from $[0.6,1.4]$. The results are shown in Figure \ref{fig:Bloch}. In particular, Figure \ref{fig:cost_Bloch} shows the total reward (top panel) and number of second-order PS iterations (bottom panel) with respect to the moment truncation order $N$, that is, the hierarchy level, from which we observe their convergence at $N=9$. Figure \ref{fig:control_Bloch} plots the policies learned from each hierarchy, which stabilize to the corresponding shadowed regions as the truncation order increases. The computational time is 42.30 seconds, which is much longer than that (3.97 seconds) for the LQR problem presented in Section \ref{sec:LQR}. In part, this is because the order-$N$ truncated moment system in this case is of dimension $3N$, 3 times higher than the moment system in the LQR example. Additionally, the nonlinearity of the Bloch system also increase the complexity of this policy learning task. 


\begin{figure}[H]
     \centering
     \begin{subfigure}[b]{0.495\textwidth}
     \centering
     \includegraphics[width=\textwidth]{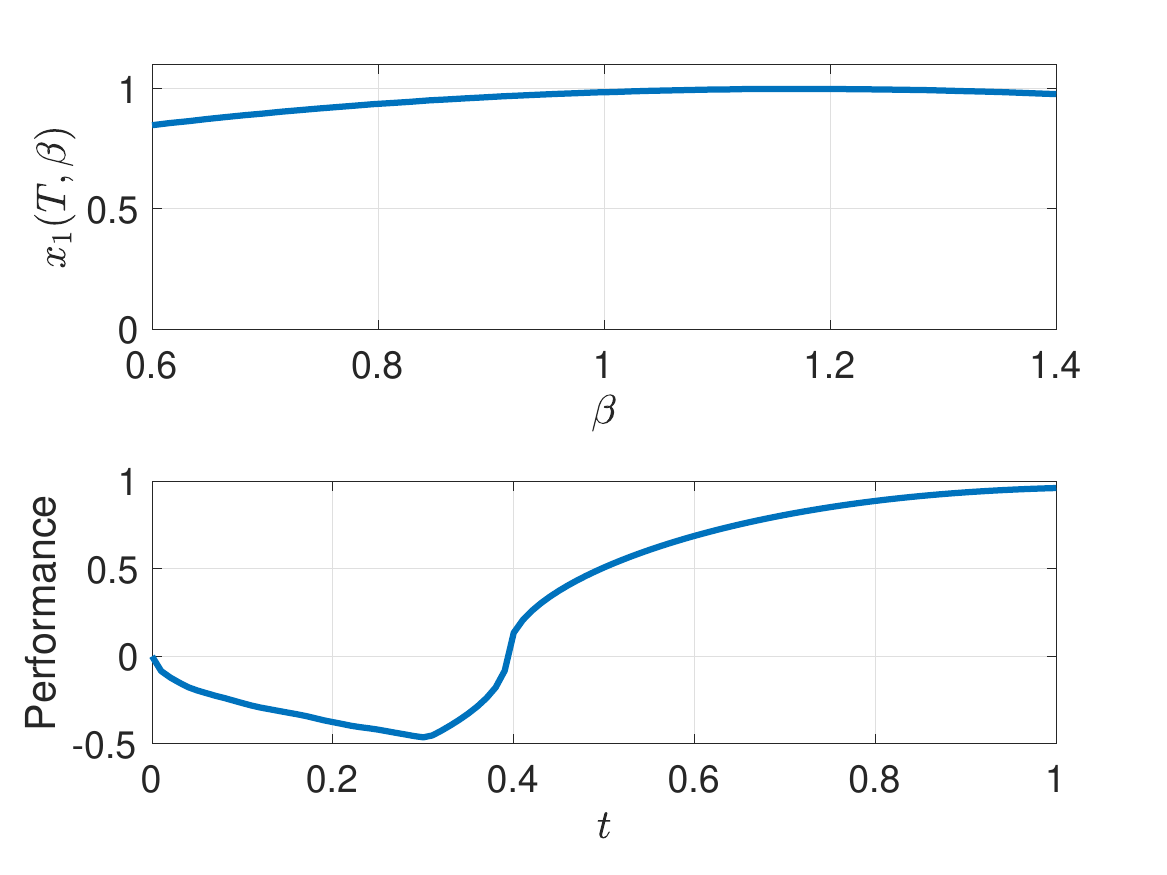}
     \caption{}
     \label{fig:Block_x_1_final}
     \end{subfigure}
     \begin{subfigure}[b]{0.495\textwidth}
     \centering
     \includegraphics[width=\textwidth]{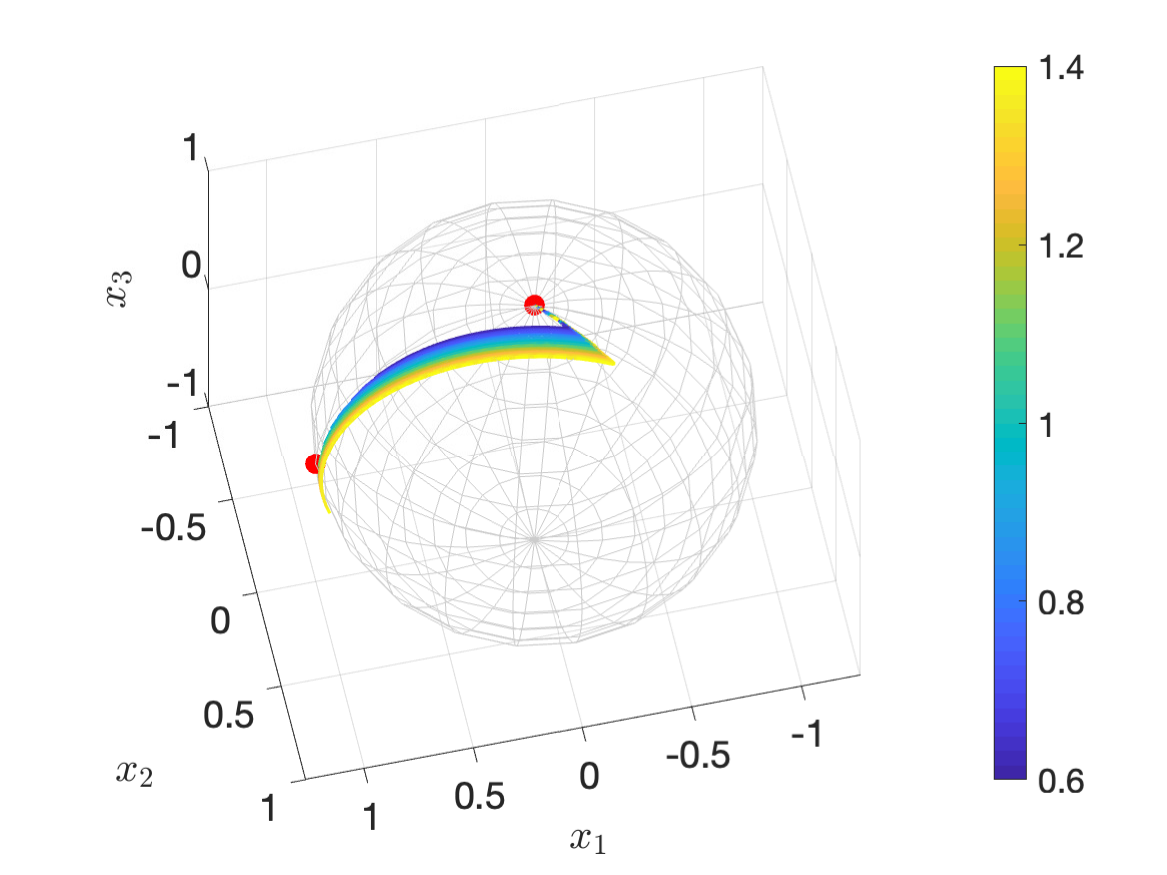}
     \caption{}
      \label{fig:Block_traj}
     \end{subfigure}
      \caption{\footnotesize Excitation performance of the policies learned by using Algorithm \ref{alg:FRL_PIA}. In particular, (a) plots the first component of the final state of the parameterized Bloch system in \eqref{eq:Bloch} with respect to the rf inhomogeneity $\beta$, and (b) plots the entire trajectory of the system driven by the learned policies on the unit sphere.}
     \label{fig:Bloch_NMR}
 \end{figure}

As mentioned previously, from the perspective of quantum physics, a goal of controlling the Bloch system is to steer the spins from the equilibrium state $(0,0,1)'$ to the excited state $(1,0,0)'$ uniformly regardless of the rf inhomogeneity. Note that because $\Omega_x=\left[\begin{array}{ccc} 0 & 0 & 0 \\ 0 & 0 & 1 \\ 0 & -1 & 0 \end{array}\right]$ and $\Omega_y=\left[\begin{array}{ccc} 0 & 0 & -1 \\ 0 & 0 & 0 \\ 1 & 0 & 0 \end{array}\right]$ in the Bloch system $\frac{d}{dt}x(t,\beta)=\beta[u(t)\Omega_y+v(t)\Omega_x]x(t,\beta)$ are skew-symmetric matrices, we have $\frac{d}{dt}|x(t,\beta)|^2=\frac{d}{dt}\big(x'(t,\beta)x(t,\beta)\big)=\frac{d}{dt}x'(t,\beta)\cdot x(t,\beta)+x'(t,\beta)\cdot\frac{d}{dt}x(t,\beta)=x'(t,\beta)(\beta u\Omega_y'+v\Omega_x')x(t,\beta)+x'(t,\beta)(\beta u\Omega_y+v\Omega_x)x(t,\beta)=0$. As a result, starting from $x(0,\beta)=(0,0,1)$, $|x(t,\beta)|=1$ holds for all $t$ and $\beta$, meaning the system trajectory stays on the unit sphere $\mathbb{S}^2=\{x\in\mathbb{R}^3:|x|=1\}$. To evaluate the excitation performance, it then suffices to examine $x_1(T,\beta)$, the first component of the final state $x(T,\beta)$, which is plotted in the top panel of Figure \ref{fig:Block_x_1_final} as a function of $\beta$. Its average value, generally used as the measure of the control performance \citep{Zhang15}, is $\frac{1}{2\delta}\int_{1-\delta}^{1+\delta}x_1(T,\beta)d\beta=0.9613$, which is close to 1, showing the good excitation performance of the learned policies. The bottom panel of Figure \ref{fig:Block_x_1_final} shows the performance measure versus time. Figure \ref{fig:Block_traj} shows the entire trajectory of the Bloch system on the unit sphere, from which we observe that the learned policies indeed steer the system towards the excited state $(1,0,0)'$, regardless of the rf inhomogeneity, as desired.

\subsubsection{Comparison with baseline deep reinforcement learning models} 

We conduct a comparison analysis between 
the proposed FRL architecture with baseline deep RL models. Following our problem formulation, where policies are deterministic and take values in continuous action spaces, we choose the Deep Deterministic Policy Gradient (DDPG) and Twin-Delayed Deep Deterministic Policy Gradient (TD3) as the baseline models. We test their performance on 500 systems in the parametrized linear ensemble in \eqref{eq:challenge_system} as well as 500 systems in the parametrized Bloch ensemble in \eqref{eq:Bloch}, with their system parameters randomly sampled from $[-1,1]$ and $[0.6,1.4]$, respectively. In table \ref{tab:comparison}, we compare the minimal costs and training time of DDPG and TD3 with those of the proposed FRL. We observe that FRL outperforms the baseline deep RL models in both the linear and Bloch system examples. Notably, FRL achieves significant computational efficiency, reducing the training time compared to deep RL models by two orders of magnitude.

\begin{table}[h!]
    \centering
\begin{tabular}{ |p{3cm}||p{3cm}|p{3cm}|p{3cm}|  }
 \hline
 \multicolumn{4}{|c|}{Linear system} \\
 \hline
 & FRL & DDPG & TD3 \\
 \hline
 Minimal cost   &  3.40    &   59.86  &  50.35  \\
 Training time (s) &  3.97    &  1607   &  1575  \\
 \hline
\end{tabular}

\begin{tabular}{ |p{3cm}||p{3cm}|p{3cm}|p{3cm}|  }
 \hline
 \multicolumn{4}{|c|}{Bloch (nonlinear) system} \\
 \hline
 & FRL & DDPG & TD3 \\
 \hline
 Minimal cost   &  2.85    &  28.95   &  32.46  \\
 Training time (s) &   42.30   &  2807   &  2240  \\
 \hline
\end{tabular}
\caption{Comparison of the proposed FRL with baseline deep RL models, DDPG and TD3. The top and bottom panels show the comparison results for the parameterized linear system in \eqref{eq:challenge_system} and (nonlinear) Bloch system in \eqref{eq:Bloch}, respectively.}
\label{tab:comparison}
\end{table}

\begin{figure}[H]
     \centering
     \begin{subfigure}[b]{0.495\textwidth}
     \centering
     \includegraphics[width=\textwidth]{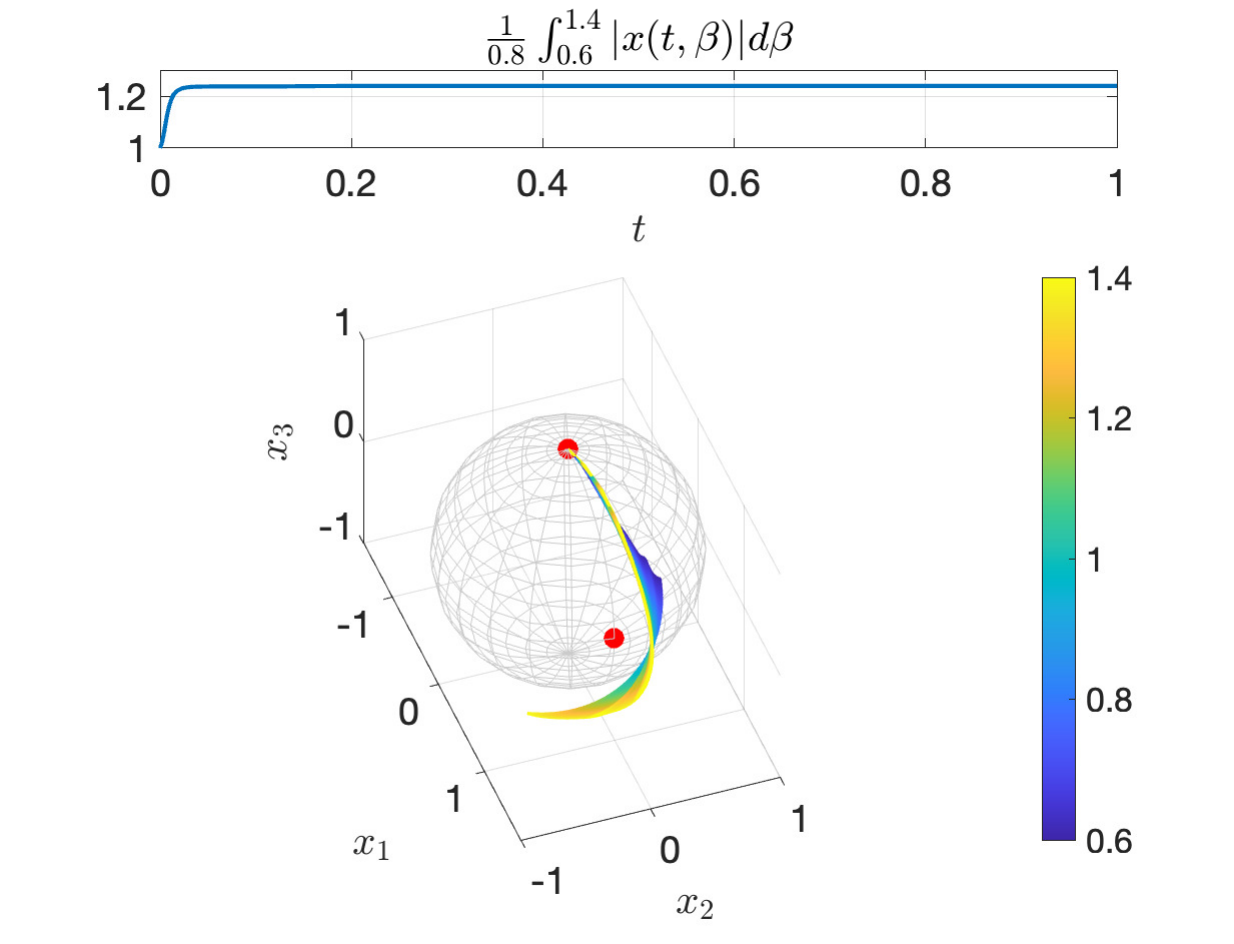}
     \caption{}
     \label{fig:DDPG}
     \end{subfigure}
     \begin{subfigure}[b]{0.495\textwidth}
     \centering
     \includegraphics[width=\textwidth]{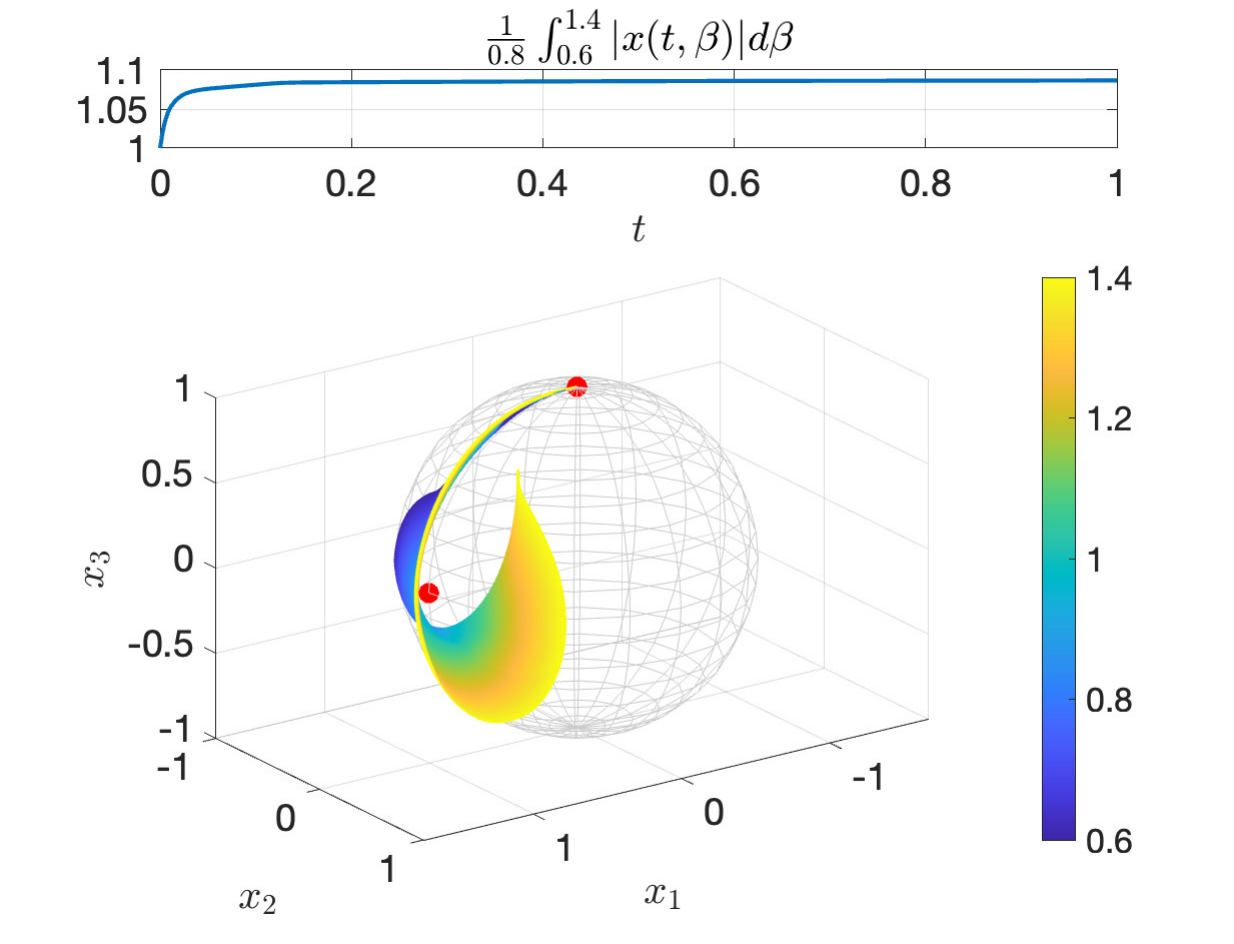}
     \caption{}
      \label{fig:TD3}
     \end{subfigure}
      \caption{\footnotesize Temporal evolution of the parameterized Bloch system in \eqref{eq:Bloch} learned by using DDPG (a) and TD3 (b). In both (a) and (b), the bottom panels show the the learned trajectories and the top panels plot the ``averaged'' norms of the trajectories over the entire ensemble.}
     \label{fig:DRL}
 \end{figure}

In addition, a major drawback of the deep RL models lies in their failure to retain the geometric structure of the Bloch system. Recall that the state of the parameterized Bloch system should stay on the unit sphere $\mathbb{S}^2$. However, as shown in Figure \ref{fig:DRL}, neither the trajectories learned from DDPG and TD3 satisfy this requirement. This indicates that both the DDPG and TD3 agents fail to learn the evolution of the Bloch system.

\section{Conclusion}

In this paper, we propose a novel RL architecture for learning optimal policies of arbitrarily large populations of intelligent agents. In our formulation, such a population is modeled as a parameterized control system defined on an infinite-dimensional function space. To mitigate the challenges arising from the infinite-dimensionality, we develop the moment kernel transform carrying over the parameterized system and its value function to an RKHS consisting of moment sequences, giving rise to a kernel parameterization of the RL problem. We then organize the finite-dimensional truncated moment representations of the RL problem into a filtrated structure and develop a hierarchical policy learning algorithm through this RL filtration, in which each hierarchy consists of an RL problem for a finite-dimensional truncated moment kernelized system. We further investigate early stopping criteria for each hierarchy to improve the computational efficiency of the FRL algorithm, and prove the convergence of the early stopped algorithm by constructing a spectral sequence. Meanwhile, computational and sample efficiency of FRL are also quantitatively analyzed. Examples are provided for demonstration of the excellent performance and high efficiency of the proposed algorithm. 



\acks{This work was supported by the Air Force Office of Scientific Research under the award FA9550-21-1-0335.}


\newpage

\appendix

\section{Proof of Lemma \ref{lem:sequential_continuity_cost}}
\label{appd:sequential_continuity_cost}

Let $x_k(t,\cdot)$ and $x(t,\cdot)$ in $\mathcal{F}(\Omega,M)$ be the trajectories (solutions) of the ensemble system in \eqref{eq:ensemble}, with the initial condition $x(0,\cdot)$, driven by the control inputs $u_k$ and $u$, respectively, for all $k\in\mathbb{N}$, i.e.,
\begin{align*}
\frac{d}{dt}x_k(t,\b)=F\big(t,\b,x_k(t,\b),u_k(t)\big)\quad\text{and}\quad\frac{d}{dt}x(t,\b)=F\big(t,\b,x(t,\b),u(t)\big).
\end{align*}
We claim that $x_k(t,\cdot)$ converges to $x(t,\cdot)$ on $\mathcal{F}(\Omega,M)$ pointwisely as $k\rightarrow\infty$ for any $t\in[0,T]$, which is equivalent to $x_k(t,\b)\rightarrow x(t,\b)$ for any $\b\in\Omega$ on each coordinate chart of $M$ by using a partition of unity on $M$ \citep{Lee12}. Therefore, it suffices to assume that the entire trajectories $x_k([0,T],\b)\subseteq M$ and $x([0,T],\b)\subseteq M$ for all $k\in\mathbb{N}$ and each $\b\in\Omega$ are located in a single coordinate chart of $M$ and hence equivalently in $\mathbb{R}^n$.

To prove the claim, we fix an arbitrary $\b\in\Omega$ and note that
\begin{align*}
\frac{d}{dt}\Big(x_k(t,\b)-x(t,\b)\Big)&=F\big(t,\b,x_k(t,\b),u_k(t))-F(t,\b,x(t,\b),u(t)\big)\\
&\leq\Big|F\big(t,\b,x_k(t,\b),u_k(t))-F(t,\b,x(t,\b),u(t)\big)\Big|\\
&\leq\Big|F\big(t,\b,x_k(t,\b),u_k(t)\big)-F\big(t,\b,x(t,\b),u_k(t)\big)\Big|\\
&\qquad\qquad+\Big|F\big(t,\b,x(t,\b),u_k(t)\big)-F\big(t,\b,x(t,\b),u(t)\big)\Big|,
\end{align*}
where the second inequality follows from the triangle inequality. Because $\frac{d}{dt}\big(x(t,\b)-x_k(t,\b)\big)$ satisfies the same inequality as above, we obtain
\begin{align*}
\frac{d}{dt}\Big|x_k(t,\b)-x(t,\b)\Big|&\leq\Big|F\big(t,\b,x_k(t,\b),u_k(t)\big)-F\big(t,\b,x(t,\b),u_k(t)\big)\Big|\nonumber\\
&\qquad\qquad+\Big|F\big(t,\b,x(t,\b),u_k(t)\big)-F\big(t,\b,x(t,\b),u(t)\big)\Big|\\
&\leq C\Big|x_k(t,\b)-x(t,\b)\Big|+\Big|F\big(t,\b,x(t,\b),u_k(t)\big)-F\big(t,\b,x(t,\b),u(t)\big)\Big|
\end{align*}
where the second inequality follows from the Lipschitz continuity of $F$ in the system state variable according to Assumption S2. Moreover, as solutions of ordinary differential equations, all of $x_k(t,\b)$ and $x(t,\b)$ are Lipschitz continuous, and hence absolutely continuous, then so is $|x_k(t,\b)-x(t,\b)|$. Together with its nonnegativity, Gronwall's inequality can be applied \citep{Evans10}, yielding
\begin{align*}
\Big|x_k(t,\b)-x(t,\b)\Big|\leq e^{Ct}\int_0^t\Big|F\big(s,\b,x(s,\b),u_k(s)\big)-F\big(s,\b,x(s,\b),u(s)\big)\Big|ds
\end{align*}
Because $F$ is continuous in the control policy variable, the pointwise convergence of $u_k(s)$ to $u(s)$ implies that of $F\big(s,\b,x(s,\b),u_k(s)\big)$ to $F\big(s,\b,x(s,\b),u(s)\big)$. Then, by Egoroff's Theorem, there exist sequences of real numbers $\varepsilon_k>0$ and subsets $I_k$ of $[0,T]$ with Lebesgue measure $\varepsilon_k$ such that $\varepsilon_k\rightarrow0$ and $F\big(s,\b,x(s,\b),u_k(s)\big)\rightarrow F\big(s,\b,x(s,\b),u(s)\big)$ uniformly on $I_k^c=[0,T]\backslash I_k$. Consequently, we have
\begin{align*}
\lim_{k\rightarrow\infty}&\Big|x_k(t,\b)-x(t,\b)\Big|\leq\lim_{k\rightarrow\infty}e^{Ct}\int_0^t\Big|F\big(s,\b,x(s,\b),u_k(s)\big)-F\big(s,\b,x(s,\b),u(s)\big)\Big|ds\\
&\leq e^{Ct}\Big\{\lim_{k\rightarrow\infty}\int_{I_k^c}\Big|F\big(s,\b,x(s,\b),u_k(s)\big)-F\big(s,\b,x(s,\b),u(s)\big)\Big|ds\\
&\qquad\qquad\qquad+\lim_{k\rightarrow\infty}\int_{I_k}\Big|F\big(s,\b,x(s,\b),u_k(s)\big)-F\big(s,\b,x(s,\b),u(s)\big)\Big|ds\Big\}\\
&= e^{Ct}\Big\{\int_0^t\lim_{k\rightarrow\infty}\Big|F\big(s,\b,x(s,\b),u_k(s)\big)-F\big(s,\b,x(s,\b),u(s)\big)\Big|\chi_{I_k^c}(s)ds\\
&\qquad\qquad\qquad+\lim_{k\rightarrow\infty}\int_{I_k}\Big|F\big(s,\b,x(s,\b),u_k(s)\big)-F\big(s,\b,x(s,\b),u(s)\big)\Big|ds\Big\}=0,
\end{align*}
in which $\chi_{I_k^c}$ denotes the characteristic function of $I_k^c$, i.e., $\chi_{I_k^c}(s)=1$ for $s\in I_k^c$ and $\chi_{I_k^c}(s)=0$ for $s\not\in I_k^c$, the change of the limit and integral in the first term in the summand follows from the uniform convergence of $F\big(s,\b,x(s,\b),u_k(s)\big)$ to $F\big(s,\b,x(s,\b),u(s)\big)$ on $I_k^c$, and the second integral converges 0 because the Lebesgue of $I_k$ goes to 0. This then proves the claim. 

Now, without loss of generality, we assume that $|J(u)|<\infty$, and the existence of such a $u$ is guaranteed by Assumption C1. Then, we obtain the desired convergence
\begin{align*}
\lim_{k\rightarrow\infty}J(u_k)&=\lim_{k\rightarrow\infty}\Big[\int_\Omega\int_0^Tr(t,x_k(t,\beta),u_k(t))dtd\b+\int_\Omega K(T,x_k(T,\beta))d\beta\Big]\\
&=\int_\Omega\int_0^T\lim_{k\rightarrow\infty}r(x_k(t,\beta),u_k(t))dtd\b+\int_\Omega \lim_{k\rightarrow\infty}K(x_k(T,\beta))d\beta\\
&=\int_\Omega\int_0^Tr(t,\lim_{k\rightarrow\infty}x_k(t,\beta),\lim_{k\rightarrow\infty}u_k(t))dtd\b+\int_\Omega K(T,\lim_{k\rightarrow\infty}x_k(T,\beta))d\beta\\
&=\int_\Omega\int_0^Tr(t,x(t,\beta),u(t))dtd\b+\int_\Omega K(T,x(T,\beta))d\beta=J(u),
\end{align*}
where the second and third equalities follow from the dominated convergence theorem and the continuity of $r$ and $K$, respectively \citep{Folland13}.

\section{Proof of Proposition \ref{prop:V_continuity}}
\label{appd:V_continuity}

Fix $z,\tilde z\in\mathcal{M}$ and $t,\tilde t\in[0,T]$, then by the definition of the infimum, for any $\varepsilon>0$, there exists an ensemble control policy $u\in\mathcal{U}$ such that
\begin{align*}
V^*(\tilde t,\tilde z)+\varepsilon&\geq\int_{\tilde t}^T\bar r(\tau,\tilde m(\tau),u(\tau))d\tau+\bar K(T,\tilde m(T))\\
&=\int_\Omega\int_{\tilde t}^Tr(\tau,\tilde x(\tau,\b),u(\tau))d\tau d\b+\int_\Omega K(T,\tilde x(T,\b))d\b,
\end{align*}
where $\tilde m(\tau)$ satisfies the moment system $\frac{d}{d\tau}\tilde m(\tau)=\bar F(\tau, \tilde m(\tau),u(\tau))$ with the initial condition $\tilde m(s)=\tilde z$ and $\tilde x(t,\cdot)$ is the trajectory of the associated ensmeble system. Next, let $m(\tau)$ be the trajectory of the moment system steered by the same control policy with the initial condition $m(t)=z$ and $x(t,\cdot)$ be the associated ensemble trajectory. Without loss of generality, we assume that $t\leq\tilde t$, and then we have
\begin{align}
V^*(t,z)-V^*(\tilde t,\tilde z)&\leq \int_\Omega\int_{t}^T r(\tau,x(\tau,\b),u(\tau))d\tau d\b+\int_\Omega K(T, x(T,\b))d\b\nonumber\\
&\qquad\quad-\int_\Omega\int_{\tilde t}^Tr(\tau,\tilde x(\tau,\b),u(\tau))d\tau d\b-\int_\Omega K(T,\tilde x(T,\b))d\b+\varepsilon\nonumber\\
&=\int_\Omega\int_{t}^{\tilde t} r(\tau,x(\tau,\b),u(\tau))d\tau d\b+\int_\Omega\int_{\tilde t}^T\Big[r(\tau,x(\tau,\b),u(\tau))\nonumber\\
&\quad\ -r(\tau,\tilde x(\tau,\b),u(\tau))\Big]d\tau d\b+\int_\Omega\Big[K(T, x(T,\b))-K(T,\tilde x(T,\b))\Big]d\b+\varepsilon. \label{eq:V_Lip}
\end{align}
Because finite-time solutions of ordinary differential equations with initial conditions in a bounded set remain bounded, $|r(\tau,x(\tau,\b),u(\tau))|\leq M$ holds for some $M$ over $\tau\in[t,\tilde t]$ so that the first term in \eqref{eq:V_Lip} can be bounded as 
$$\int_\Omega\int_{t}^{\tilde t} r(\tau,x(\tau,\b),u(\tau))d\tau d\b\leq\int_\Omega\int_{t}^{\tilde t} |r(\tau,x(\tau,\b),u(\tau))|d\tau d\b M\leq{\rm Vol}(\Omega)|t-\tilde t|.$$
Then, by using the Lipschtiz continuity of $r$ and $K$ as in Assumption C2, the second and third terms in \eqref{eq:V_Lip} satisfies
\begin{align}
\int_\Omega\int_{\tilde t}^T&\Big[r(\tau,x(\tau,\b),u(\tau))-r(\tau,\tilde x(\tau,\b),u(\tau))\Big]d\tau d\b+\int_\Omega\Big[K(T, x(T,\b))-K(T,\tilde x(T,\b))\Big]d\b\nonumber\\
&\leq\int_\Omega\int_{\tilde t}^TC_r\big|x(\tau,\b)-\tilde x(\tau,\b)\big|d\tau d\b+ \int_\Omega C_K\big|x(T,\b)-\tilde x(T,\b)\big|d\b\nonumber\\
&\leq\int_\Omega(TC_r+C_K)C'\big|x(\tilde t,\b)-\tilde x(\tilde t,\b)\big|d\b\nonumber\\
&\leq(TC_r+C_K)C'\int_\Omega\Big[\big|x(t,\b)-\tilde x(\tilde t,\b)\big|+\big|x(t,\b)-x(\tilde t,\b)\big|\Big]d\b\nonumber\\
&\leq(TC_r+C_K)C'\Big[\int_\Omega\big|x(t,\b)-\tilde x(\tilde t,\b)\big|d\b+\int_\Omega\big|x(t,\b)-x(\tilde t,\b)\big|d\b\Big], \label{eq:V_Lip23}
\end{align}
where the second inequality follows from Gronwall's inequality. Note that the first integral in \eqref{eq:V_Lip23} is exactly the $L^1$-norm of $x(t,\cdot)-\tilde x(t,\cdot)$, which is equal to $\|z-\tilde z\|$, the norm of the associated moment sequences, since the moment transformation is an isometry. For the second term, we use the Lipschitz continuity of solutions to ordinary differential equations to conclude $\int_\Omega\big|x(t,\b)-x(\tilde t,\b)\big|\leq C''{\rm Vol}(\Omega)|t-\tilde t|$ for some $C''>0$. Now, let $C=\max\{(TC_r+C_K)C',M{\rm Vol}(\Omega),{\rm Vol}(\Omega)\}$, then we obtain $V^*(t,z)-V^*(\tilde t,\tilde z)\leq C(|t-\tilde t|+\|z-\tilde z\|)$ since $\varepsilon>0$ is arbitrary. The same argument with the roles of $(t,z)$ and $(\tilde t,\tilde z)$ reversed implies
$$|V^*(t,z)-V^*(\tilde t,\tilde z)|\leq C(|t-\tilde t|+\|z-\tilde z\|),$$
giving the Lipschitz continuity of $V^*$ on $[0,T]\times\mathcal{M}$ as desired.

\section{Derivation of Second-Order Policy Search Update Equations}
\label{appd:PIA_derivation}

\paragraph{Quadratic approximation of Hamilton-Jacobi-Bellman equations.} We pick a nominal policy $\bar u(t)$, which generates a nominal trajectory $\bar m(t)$ by applying $\bar u(t)$ to the moment system in \eqref{eq:moment_system}. Then, the optimal policy and trajectory can be represented as $u^*(t)=\bar u(t)+\delta u(t)$ and $m^*(t)=\bar m(t)+\delta m(t)$, respectively, plugging which into the Hamiltonian-Jacobi-Bellman equation in \eqref{eq:HJB}, i.e., 
\begin{align*}
\frac{\partial}{\partial t}V(t,z)+\min_{a\in U}\big\{\bar r(t,z,a)+\<DV(t,z),\bar F(t,z,a)\>\big\}=0
\end{align*}
yields
\begin{align*}
\frac{\partial}{\partial t}V(t,\bar m+\delta m)+\min_{\delta u}\big\{&\bar r(t,\bar m+\delta m,\bar u+\delta u)\\
&+\<DV(t,\bar m+\delta m),\bar F(t,\bar m+\delta m,\bar u+\delta u)\>\big\}=0,
\end{align*}
where, and in the following as well, we drop the time argument $t$ from the state and control variables for the conciseness of the representation. Now, we assume that the value function $V:[0,T]\times\mathcal{M}\rightarrow\mathbb{R}$ is smooth enough, at least in the region containing the nominal and optimal trajectories, to admit a second-order power series expansion as
\begin{align}
V(t,&\bar m+\delta m)=V(t,\bar m)+\<DV(t,\bar m),\delta m\>+\frac{1}{2}\<D^2 V(t,\bar m)\cdot\delta m,\delta m\>+o(\delta m^2)\nonumber\\
&=\bar V(t,\bar m)+\delta V(t,\bar m)+\<DV(t,\bar m),\delta m\>+\frac{1}{2}\<D^2 V(t,\bar m)\cdot\delta m,\delta m\>+o(\delta m^2),\label{eq:V_second_order}
\end{align}
where we further expand $V(t,\bar m)$ by using the nominal cost $\bar V(t,\bar m)=\int_t^T\bar r(s,\bar m(s),\bar u(s))ds+K(T,\bar m(T))$, that is, the cost obtained by applying the nominal control input $\bar u$ to the system starting from $\bar m(t)$ at time $t$. To explain the second-order term in the above expansion, by the definition, the second derivative $D^2V(t,\bar m)$ of the real-valued function $V(t,\cdot):\mathcal{M}\rightarrow\mathbb{R}$ evaluated at $\bar m$ is a bounded linear map from $\mathcal{M}$ to $\mathcal{M}^*$ \citep{Lang99}. In our notation, $D^2V(t,\bar m)\cdot\delta m$ denotes the evaluation of $D^2V(t,\bar m)$ at $\delta m$, giving an element in $\mathcal{M}^*$ that can be paired with $\delta m\in\mathcal{M}$. Conceptually, $D^2V(t,\bar m)$ is nothing but the infinite-dimensional Hessian matrix with the $(i,j)$-entry given by $\frac{\partial^2V}{\partial z_i\partial z_j}|_{(t,\bar m)}$. If $\delta m$ is small enough to ensure a sufficiently accurate approximation of the value function $V(t,\bar m)$ up to the second-order terms, then we integrate \eqref{eq:V_second_order} with the $o(\delta m^2)$ term neglected into the Hamilton-Jacobi-Bellman equation, yielding
\begin{align}
\frac{\partial}{\partial t}&\bar V(t,\bar m)+\frac{\partial}{\partial t}\delta V(t,\bar m)+\Big\<\frac{\partial}{\partial t}DV(t,\bar m),\delta m\Big\>+\frac{1}{2}\Big\<\frac{\partial}{\partial t}D^2V(t,\bar m)\cdot\delta m,\delta m\Big\>\nonumber\\
&+\min_{\delta u}\Big\{H(t,\bar m+\delta m,\bar u+\delta u,DV(t,\bar m))+\Big\<D^2V(t,\bar m)\cdot\delta m,\bar F(t,\bar m+\delta m,\bar u+\delta u)\Big\>\Big\}=0\label{eq:HJB_update}
\end{align}
with the system Hamiltonian $H(t,m,u,DV)=\bar r(t,m,u)+\<DV(t,m),\bar F(t,m,u)\>$, which is the key equation to the development of the algorithm for successively improving the nominal control policy $\bar u$.

\paragraph{Second-order policy search.} To initialize the algorithm, we start with a nominal control policy $u_0$, applying which to the moment system generates a nominal trajectory $m^{(0)}$. Because the pair $(m^{(0)},u^{(0)})$ satisfies the system, the variation of the trajectory is $\delta m=0$. In this case, the second-order expanded Hamilton-Jacobi-Bellman equation in \eqref{eq:HJB_update} takes the form
\begin{align}
\label{eq:control_update_analytical}
\frac{\partial}{\partial t}\bar V(t,m^{(0)})+\frac{\partial}{\partial t}\delta V(t,m^{(0)})+\min_{\delta u} H(t,m^{(0)}, u^{(0)}+\delta u,DV(t,m^{(0)}))=0,
\end{align}
and in many cases, the minimization of the Hamiltonian can be solved analytically, giving a new control policy ${u^{(0)}}^*$, represented in terms of $m^{(0)}$ and $DV(t,m^{(0)})$ as a feedback control. However, steered by this policy, the system trajectory may not be the nominal one anymore, and hence a variation $\delta m$ is introduced to the nominal trajectory as $m^{(0)}+\delta m$. Correspondingly, the second-order expanded Hamilton-Jacobi-Bellman equation becomes the one in \eqref{eq:HJB_update} with $\bar u$ replaced by ${u^{(0)}}^*$, in which the minimization is taken over $\delta u$ for the function
\begin{align*}
H(t,m^{(0)}+\delta m,{u^{(0)}}^*+\delta u,DV(t,m^{(0)}))+\Big\<D^2V(t,m^{(0)})\cdot\delta m,\bar F(t,m^{(0)}+\delta m,{u^{(0)}}^*+\delta u)\Big\>.
\end{align*}
We further expand this function around $(m^{(0)},{u^{(0)}}^*)$ up to the second-order terms, yielding
\begin{align}
H+\Big\<\frac{\partial H}{\partial u},&\delta u\Big\>+\Big\<DH+D^2V\cdot\bar F,\delta m\Big\>+\Big\<\Big(\frac{\partial DH}{\partial u}+\frac{\partial \bar F'}{\partial u}\cdot D^2V\Big)\cdot\delta m,\delta u\Big\>\nonumber\\
&+\frac{1}{2}\Big\<\frac{\partial^2H}{\partial u^2}\cdot\delta u,\delta u\Big\>+\frac{1}{2}\Big\<\Big(D^2H+D\bar F'\cdot D^2V+D^2V'\cdot DF\Big)\cdot\delta m,\delta m\Big\>,\label{eq:H_expansion}
\end{align}
where the terms involving $H$, $\bar F$, and $V$ are evaluated at $(t,m^{(0)},{u^{(0)}}^*, DV(t,m^{(0)}))$, $(t,m^{(0)}$, ${u^{(0)}}^*)$, and $(t,m^{(0)})$, respectively, and `$\prime$' denotes the dual of a linear operator. For example, because $\bar F(t,m^0,u_0^*)\in\mathcal{M}$ by identifying the tangent space of $\mathcal{M}$ at $m^0$ with $\mathcal{M}$ itself, $D\bar F(t,m^{(0)},{u^{(0)}}^*):\mathcal{M}\rightarrow\mathcal{M}$ is a linear map, then its dual operator is defined as a linear map $D\bar F'(t,m^0,u_0^*):\mathcal{M}^*\rightarrow\mathcal{M}^*$ satisfying $\<L,D\bar F(t,m^{(0)},{u^{(0)}}^*)\cdot z\>=\<D\bar F'(t,m^{(0)},{u^{(0)}}^*)\cdot L,z\>$ for any $z\in\mathcal{M}$ and $L\in\mathcal{M}^*$ \citep{Yosida80}. Conceptually, dual operators are nothing but transpose matrices. Next, to minimize the function in \eqref{eq:H_expansion}, the necessary condition is the vanishing of its derivative with respect to $\delta u$, giving
\begin{align}
\frac{\partial H}{\partial u}+\frac{\partial^2 H}{\partial u^2}\cdot\delta u+\Big(\frac{\partial DH}{\partial u}+\frac{\partial\bar F'}{\partial u}\cdot D^2V\Big)\cdot\delta m=0, \label{eq:opt_condition}
\end{align}
in which it is necessary that $\frac{\partial H}{\partial u}|_{(t,m^{(0)},{u^{(0)}}^*,DV(t,m^{(0)}))}=0$ since $u$ minimizes the Hamiltonian $H(t,m^{(0)},{u^{(0)}}^*,DV(t,m^{(0)}))$ by our choice. 

Recall that our intention is to approximate the Hamiton-Jacobi-Bellman equation up to the second-order term in $\delta m$. Therefore, it is required that $\delta m$ and $\delta u$ are in the same order, meaning, they satisfy a linear relationship; otherwise, say $\delta u$ is quadratic in $\delta m$, then the terms $\Big\<\Big(\frac{\partial DH}{\partial u}+\frac{\partial \bar F'}{\partial u}\cdot D^2V\Big)\cdot\delta m,\delta u\Big\>$ and $\frac{1}{2}\Big\<\frac{\partial^2H}{\partial u^2}\cdot\delta u,\delta u\Big\>$ in \eqref{eq:H_expansion} are of orders higher than $\delta m^2$. Formally, there is a linear map $A:\mathcal{M}\rightarrow\mathbb{R}^m$ such that $\delta u=A\cdot\delta m$. To find $A$, we replace $\delta u$ by $A\cdot\delta m$ in the necessary optimality condition in \eqref{eq:opt_condition}, leading to 
\begin{align*}
A=-\left(\frac{\partial^2 H}{\partial u^2}\right)^{-1}\cdot\Big(\frac{\partial DH}{\partial u}+\frac{\partial\bar F'}{\partial u}\cdot D^2V\Big).
\end{align*}
With this choice of $A$, the function in \eqref{eq:H_expansion} becomes
\begin{align*}
H+\Big\<DH+&D^2V\cdot\bar F,\delta m\Big\>\nonumber\\
&+\frac{1}{2}\Big\<\Big(D^2H+D\bar F'\cdot D^2V+D^2V'\cdot DF-A'\cdot D^2H\cdot A\Big)\cdot\delta m,\delta m\Big\>
\end{align*}
so that the second-order expansion of the Hamilton-Jacobi-Bellman equation in \eqref{eq:HJB_update} takes the form 
\begin{align}
\frac{\partial}{\partial t}\bar V+\frac{\partial}{\partial t}&\delta V+\Big\<\frac{\partial}{\partial t}DV,\delta m\Big\>+\frac{1}{2}\Big\<\frac{\partial}{\partial t}D^2V(t,\bar m)\cdot\delta m,\delta m\Big\>+H+\Big\<DH+D^2V\cdot\bar F,\delta m\Big\>\nonumber\\
&+\frac{1}{2}\Big\<\Big(D^2H+D\bar F'\cdot D^2V+D^2V'\cdot DF-A'\cdot\frac{\partial^2H}{\partial u^2}\cdot A\Big)\cdot\delta m,\delta m\Big\>=0,\label{eq:HJB_update_A}
\end{align}
Because $\delta m$ is arbitrary, the coefficient of each order of $\delta m$ must be 0, which transforms \eqref{eq:HJB_update_A} into a system of three partial differential equations
\begin{align*}
&\frac{\partial}{\partial t}\bar V+\frac{\partial}{\partial t}\delta V-H=0\\
&\frac{\partial}{\partial t}DV-DH-D^2V\cdot\bar F=0\\
&\frac{\partial}{\partial t} D^2V-D^2H-D\bar F'\cdot D^2V-D^2V'\cdot DF+A'\cdot\frac{\partial^2H}{\partial u^2}H\cdot A=0
\end{align*}
with terms involving $H$, $\bar F$, and $V$ are evaluated at $(t,m^{(0)},{u^{(0)}}^*, DV(t,m^{(0)}))$, $(t,m^{(0)},{u^{(0)}}^*)$, and $(t,m^{(0)})$, respectively, as before. Integrating with the chain rule as
\begin{align*}
&\frac{d}{dt}(\bar V+\delta V)=\frac{\partial}{\partial t}\bar V+\frac{\partial}{\partial t}\delta V+\big\<DV,\bar F(t,m^{(0)},{u^{(0)}})\big\>\\
&\frac{d}{dt}DV = \frac{\partial}{\partial t}DV + D^2V\cdot\bar F(t,m^{(0)},{u^{(0)}})\\
&\frac{d}{dt}D^2V = \frac{\partial}{\partial t} D^2V + \frac{1}{2}D^3V\cdot\bar F(t,m^{(0)},{u^{(0)}})+\frac{1}{2}\bar F(t,m^{(0)},{u^{(0)}})'\cdot D^3V'
\end{align*}
then gives three ordinary differential equations
\begin{align}
\frac{d}{dt}\delta V(t,m^{(0)}(t))&=H(t,m^{(0)},{u^{(0)}},DV(t,m^{(0)}))-H(t,m^{(0)},{u^{(0)}}^*,DV(t,m^{(0)})), \label{eq:dynamics_deltaV}\\
\frac{d}{dt}DV(t,m^{(0)}(t))&=-DH-D^2V\cdot(\bar F(t,m^{(0)},{u^{(0)}}^*)-\bar F(t,m^{(0)},{u^{(0)}})),\label{eq:dynamics_DV}\\
\frac{d}{dt}D^2V(t,m^{(0)}(t))&=-D^2H-D\bar F'\cdot D^2V-D^2V'\cdot D\bar F\nonumber\\
&\qquad\quad+\left[\frac{\partial DH}{\partial u}+\frac{\partial \bar F'}{\partial u}\cdot D^2V\right]'\cdot\Big(\frac{\partial^2 H}{\partial u^2}\Big)^{-1}\cdot\left[\frac{\partial DH}{\partial u}+\frac{\partial \bar F'}{\partial u}\cdot D^2V\right],\label{eq:dynamics_D^2V}
\end{align}
where we use $\frac{d}{dt}\bar V(t,m^{(0)}(t))=\frac{d}{dt}\big\{\int_t^T\bar r(s,m^{(0)}(s),u^{(0)}(s))ds+\bar K(T,m^{(0)}(T))\big\}=-\bar r(t$, $m^{(0)}(s),u^{(0)}(s))$ and $A=-\left(\frac{\partial^2 H}{\partial u^2}\right)^{-1}\cdot\Big(\frac{\partial DH}{\partial u}+\frac{\partial\bar F'}{\partial u}\cdot D^2V\Big)$, and omit the third-order terms involving $D^3V$. Moreover, because the value function satisfies $V(T,m^{(0)}(T))=\bar K(T,m^{(0)}(T))$, we have the terminal conditions for the ordinary differential equations in \eqref{eq:dynamics_deltaV} to \eqref{eq:dynamics_DV} as $\delta V(T,m^{(0)}(T))=0$, $DV(T,m^{(0)}(T))=D\bar K(T,m^{(0)}(T))$, and $D^2V(T,m^{(0)}(T))=D^2\bar K(T,m^{(0)}(T))$. In particular, the data $DV(t,m^{(0)}(t))$ obtained from solving the above systems of differential equations is then used to compute the control policy ${u^{(0)}}^*$, which has been represented in terms of $\bar m(t)$ and $DV(t,m^{(0)}(t))$ when minimizing the Hamitonian in \eqref{eq:control_update_analytical} and hence gives rise to an improvement of the nominal control policy $u^{(0)}$. This in turn completes one iteration of the proposed policy search algorithm.

\section{Derivation of Moment Systems}
\label{appd:derive_moment_system}

\subsection{Infinite-dimensional LQR}
\label{appd:derive_moment_system_linear}
In the following, we consider the finite-time horizion LQR problem 
\begin{align}
\begin{cases}
    \frac{d}{dt}x(t,\beta)=\beta x(t,\beta)+u(t),\\
    V(t,x_t) = \int_\Omega\Big[\int_0^T\big(x^2(t,\b)+u^2(t)\big)dtd\b+x^2(T,\beta)\Big]d\b,
\end{cases}
\label{eq:LQR}
\end{align}
where $\Omega=[-1,1]$ and $x_t\in L^2(\Omega)$, the space of real-valued square-integrable functions defined on $\Omega$. 

\paragraph{Moment kernelization.}  
We pick $\{\Phi_k\}_{k\in\mathbb{N}}$ to be the set of Chebyshev polynomials, and by using the recursive relation of Chebyshev polynomials $2\Phi_k=\Phi_{k-1}+\Phi_{k-1}$, we have 
\begin{align*}
\frac{d}{dt}m_k(t)&=\frac{d}{dt}\int_{-1}^1\Phi_k(\b)x(t,\b)d\b=\int_{-1}^1\Phi_k(\b)\frac{d}{dt}x(t,\b)d\b=\int_{-1}^1\Phi_k(\b)\big[\b x(t,\b)+u(t)\big]d\b\\
&=\frac{1}{2}\int_{-1}^1\big[\Phi_{k-1}(\b)+\Phi_{k+1}(\b)\big]x(t,\b)d\b+\int_{-1}^1\Phi_k(\b)d\b\cdot u(t)\\
&=\frac{1}{2}\big[m_{k-1}(t)+m_{k+1}(t)\big]+b_ku(t),
\end{align*}
where the change of the integral and time derivative follows from the dominant convergence theorem \citep{Folland13}, $\Phi_{-1}$, and hence $m_{-1}$, are defined to be identically 0, and $b_k$ is given by $\frac{(-1)^k+1}{1-k^2}$ for $k\neq1$ and 0 otherwise. We further let $L:\ell^2\rightarrow\ell^2$ and $R:\ell^2\rightarrow\ell^2$ denote the left and right shift operators, given by, $(m_0(t),m_1(t),\dots)\mapsto (m_1(t),m_2(t),\dots)$ and $(m_0(t),m_1(t),\dots)\mapsto (0,m_0(t),\dots)$, respectively, then the moment system associated with the linear ensemble system in \eqref{eq:LQR} is a linear system evolving on $\ell^2$ in the form $\frac{d}{dt}m(t)=Am(t)+Bu(t)$
with $A=\frac{1}{2}(L+R)$ and $B\in\ell^2$ whose $k^{\rm th}$ component is $b_k$. On the other hand, to parameterize the cost functional, we note that the moment transformation, that is, the Fourier transform, is a unitary operator from $L^2(\Omega)$ to $\ell^2$, as a result of which $J(u)=\int_0^1\big[\|m(t)\|^2+2u^2(t)\big]dt+\|m(T)\|^2$ in the moment parameterization, where $\|\cdot\|$ denotes the $\ell^2$ norm. In summary, the LQR problem in \eqref{eq:LQR} in the moment kernel parameterization has the form
\begin{align}
\begin{cases}
    \frac{d}{dt}m(t)=Am(t)+Bu(t),\\
    V(t,m(T))=\int_0^1\Big[\|m(t)\|^2+2u^2(t)\Big]dt+\|m(T)\|^2.
\end{cases}
\label{eq:LQR_moment}
\end{align}

\paragraph{Second-order policy search.} The Hamitonian $H:\ell^2\times\mathbb{R}\times\ell^2\rightarrow\mathbb{R}$ of the moment system in \eqref{eq:LQR_moment} is given by $(z,a,p)\mapsto \|z\|^2+2a^2+\<p,Az+Ba\>$, in which $\<\cdot,\cdot\>$ is the $\ell^2$ inner product. Let $V:[0,T]\times\ell^2\rightarrow\mathbb{R}$ be the value function, then along a trajectory $m(t)$ of the system, by setting $\frac{\partial H}{\partial a}|_{(m(t),a,DV(t,m(t)))}=4a+\<DV(m(t)),B\>=0$, we obtain $u^*(t)=-\frac{1}{4}\<DV(t,m(t)),B\>$. Consequently, the differential equations in \eqref{eq:dynamics_deltaV} to \eqref{eq:dynamics_D^2V} for the policy improvement algorithm read
\begin{align*}
\frac{d}{dt}\delta V(t,m(t))&=\Big[u+\frac{1}{4}\<DV,B\>\Big]^2,\\
\frac{d}{dt}DV(t,m(t))&=-2m-A'\cdot DV-D^2V\cdot\Big(u-\frac{1}{4}\<DV,B\>\Big),\\
\frac{d}{dt}D^2V(t,m(t))&=-2I-A'\cdot D^2V-D^2V\cdot A+\frac{1}{4}D^2V\cdot B\cdot B'\cdot D^2V,
\end{align*}
where $I$ denotes the identity operator on $\ell^2$ and we use the fact that $D^2V$ is a self-adjoint operator on $\ell^2$. More concretely, when applying Algorithm \ref{alg:FRL_PIA} to a truncated moment system, say of truncation order $N$, then $A$ and $B$ in the above system of differential equations are replaced by
\begin{align*}
    \widehat A_N=\frac{1}{2}\left[\begin{array}{ccccc}
    0&1&& & \\
    1&0&1&  &  \\
    &1&0&  \ddots & \\
    &&\ddots&\ddots&1 \\
    &&&1&0
    \end{array}\right]\in\mathbb{R}^{(N+1)\times(N+1)}\quad\text{and}\quad\widehat B_N=\left[\begin{array}{c}b_0 \\ b_1 \\ b_2 \\ \vdots \\ b_N\end{array}\right]\in\mathbb{R}^{N+1},
\end{align*}
respectively, and the operator dual is essentially the matrix transpose.  

\subsection{Moment kernelization of nuclear spin systems}
\label{appd:derive_moment_system_Bloch}
The policy learning problem for the nuclear spin systems in \eqref{eq:Bloch} is given by
\begin{align*}
\begin{cases}
    \frac{d}{dt}x(t,\beta)=\beta\big[u(t)\Omega_y+v(t)\Omega_x\big]x(t,\beta),\\
    V(t,x_t)=\int_t^T\big(u^2(t)+v^2(t)\big)dt+\int_{1-\delta}^{1+\delta}|x(T,\beta)-x_F(\beta)|^2d\beta,
\end{cases}    
\end{align*}
where the Bloch system is defined on $L^2(\Omega,\mathbb{R}^3)$ with $\Omega=[1-\delta,1+\delta]$ for some $0<\delta<1$.

\paragraph{Moment kernelization.} Similar to the LQR case presented above, we still define the moments by using the set of Chebyshev polynomials $\{\Phi_k\}_{k\in\mathbb{N}}$. However, in order to fully utilize the orthonormal property of Chebyshev polynomials, which only holds on $\Omega'=[-1,1]$, we consider the transformation $\psi:\Omega\rightarrow\Omega'$, given by $\beta\mapsto(\beta-1)/\delta$, and defined the moments by
\begin{align*}
m_k(t)=\int_{\Omega}\Phi_k\circ\psi\cdot x_td\lambda=\int_{\Omega'}\Phi\cdot x_t\circ\psi^{-1}d\psi_{\#}\lambda,
\end{align*}
where $\lambda$ denotes the Lebesgue measure on $\mathbb{R}$, and $\psi_\#\lambda$ is the pushforward measure of $\lambda$ by $\psi$, that is, $\psi_\#\lambda(I)=\lambda(\psi^{-1}(I))=\delta\lambda(I)$ is satisfied for $I\subseteq\mathbb{R}$, equivalently $d\psi_\#\lambda=\delta d\lambda$. This directly implies that the cost functional in the moment parameterization takes the form $J(u,v)=\int_0^T(u^2(t)+v^2(t))dt+\|m(T)\|^2$, where $\|\cdot\|$ denotes the $\ell^2$-norm on the $\mathbb{R}^3$-valued sequences, given by, $\|m(T)\|^2=\sum_{k=0}^\infty|m_k(T)|^2$. Next, we compute the moment parameterization of the Bloch system in \eqref{eq:Bloch} as follows
\begin{align*}
\frac{d}{dt}m_k(t)&=\frac{d}{dt}\int_{-1}^1\Phi_k(\eta)\cdot x(t,\psi^{-1}(\eta))d\psi_\#\lambda(\eta)=\int_{-1}^1\Phi_k(\eta)\frac{d}{dt}x(t,\psi^{-1}(\eta))d\psi_\#\lambda(\eta)\\
&=\int_{-1}^1\Phi_k(\eta)(\delta\eta+1)\big[u(t)\Omega_y+v(t)\Omega_x\big]x(t,\psi^{-1}(\eta))d\psi_\#\lambda(\eta)\\
&=\big[u(t)\Omega_y+v(t)\Omega_x\big]\cdot\Big\{\frac{\delta}{2}\big[m_{k-1}(t)+m_{k+1}(t)\big]+m_k(t)\Big\},
\end{align*}
where we use the recursive relation satisfied by Chebyshev polynomials, i.e., $2\Phi_k=\Phi_{k-1}+\Phi_{k+1}$, together with $\Phi_{-1}=0$. Let $L$ and $R$ be the left and right shift operators for real-valued sequences as introduced in Section \ref{sec:LQR}, the moment parameterization of the Bloch ensemble system is given by $\frac{d}{dt}m(t)=\Big[\frac{\delta}{2}(R+L)+I\Big]\otimes\big[u(t)\Omega_y+v(t)\Omega_x\big]m(t)$, where $I$ denotes the identity operator for real-valued sequences and $\otimes$ be the tensor product of linear operators. As a result, we have obtained the moment kernel parameterization of the RL problem for the Bloch system as 
\begin{align}
\begin{cases}
    \frac{d}{dt}m(t)=\big[u(t)B_y+v(t)B_x\big]m(t),\\
    V(t,m(t))=\int_0^T\big(u^2(t)+v^2(t)\big)dt+\|m(T)-m_F\|^2
\end{cases}
\label{eq:pulse_design_moment}
\end{align}
where we define $B_x=\big[\delta(R+L)/2+I\big]\otimes\Omega_x$ and $B_y=\big[\delta(R+L)/2+I\big]\otimes\Omega_y$ to simplify the notations, and $m_0$ and $m_F$ are the moment sequences of the constant functions $x_0(\beta)=(0,0,1)'$ and $x_F(\beta)=(1,0,0)'$, respectively. 

\paragraph{Second-order policy search.} The Hamiltonian of the moment system in \eqref{eq:pulse_design_moment} is given by $H:(\ell^2)^3\times\mathbb{R}\times\mathbb{R}\times(\ell^2)^3\rightarrow\mathbb{R}$, $(z,a,b,p)\mapsto a^2+b^2+\<p,\big[aB_y+bB_x\big]z\>$. Let $V:[0,T]\times\ell^2\rightarrow\mathbb{R}$ be the value function, then along a moment trajectory, by setting $\frac{\partial H}{\partial a}|_{(m(t),a,b,DV(t,m(t)))}=2a+\<DV,B_ym(t)\>=0$ and $\frac{\partial H}{\partial b}|_{(m(t),a,b,DV(t,m(t)))}=2b+\<DV,B_xm(t)\>=0$, we obtain the optimal policies $u^*(t)=-\<DV,B_ym(t)\>/2$ and $v^*(t)=-\<DV,B_ym(t)\>/2$. Integrating them into the system of differential equations in \eqref{eq:dynamics_deltaV} to \eqref{eq:dynamics_D^2V} for the policy improvement algorithm yields
\begin{align*}
\frac{d}{dt}\delta V(t,m(t))&=(u-u^*)^2+(v-v^*)^2,\\
\frac{d}{dt}DV(t,m(t))&=-DV\cdot(u^*B_y+v^*B_x)-D^2V\cdot\big[(u^*-u)B_y+(v^*-v)B_x\big]m,\\
\frac{d}{dt}D^2V(t,m(t))&=-(u^*B_y+v^*B_x)'\cdot D^2V-D^2V\cdot(u^*B_y+v^*B_x),\\
&\quad\quad+\frac{1}{2}(DV\cdot B_y+m'\cdot B_y'\cdot D^2V)'\cdot(DV\cdot B_y+m'\cdot B_y'\cdot D^2V)\\
&\quad\quad\quad\quad+\frac{1}{2}(DV\cdot B_x+m'\cdot B_x'\cdot D^2V)'\cdot(DV\cdot B_x+m'\cdot B_x'\cdot D^2V).
\end{align*}
Specifically, when applying Algorithm \ref{alg:FRL_PIA} to the order $N$ truncated problem, $B_x$ and $B_y$ are replaced by the $\mathbb{R}^{3(N+1)\times3(N+1)}$ block matrices
\begin{align*}
    (\widehat{B_x})_N=\frac{1}{2}\left[\begin{array}{ccccc}
    0&\Omega_x&& & \\
   \Omega_x&0&\Omega_x&  &  \\
    &\Omega_x&0&  \ddots & \\
    &&\ddots&\ddots&\Omega_x \\
    &&&\Omega_x&0
    \end{array}\right]\quad\text{ and }\quad
    (\widehat{B_y})_N=\frac{1}{2}\left[\begin{array}{ccccc}
    0&\Omega_y&& & \\
   \Omega_y&0&\Omega_y&  &  \\
    &\Omega_y&0&  \ddots & \\
    &&\ddots&\ddots&\Omega_y \\
    &&&\Omega_y&0
    \end{array}\right],
\end{align*}
respectively.


\vskip 0.2in
\bibliography{NRL}

\begin{thebibliography}{111}
\providecommand{\natexlab}[1]{#1}
\providecommand{\url}[1]{\texttt{#1}}
\expandafter\ifx\csname urlstyle\endcsname\relax
  \providecommand{\doi}[1]{doi: #1}\else
  \providecommand{\doi}{doi: \begingroup \urlstyle{rm}\Url}\fi

\bibitem[Albrecht et~al.(2024)Albrecht, Christianos, and
  Sch{\"a}fer]{Albrecht2024}
S.V. Albrecht, F.~Christianos, and L.~Sch{\"a}fer.
\newblock \emph{Multi-Agent Reinforcement Learning: Foundations and Modern
  Approaches}.
\newblock MIT Press, 2024.

\bibitem[Arnold(1978)]{Arnold78}
Vladimir~I. Arnold.
\newblock \emph{Ordinary Differential Equations}.
\newblock MIT Press, 1978.

\bibitem[Baker et~al.(2020)Baker, Kanitscheider, Markov, Wu, Powell, McGrew,
  and Mordatch]{Baker2020}
Bowen Baker, Ingmar Kanitscheider, Todor Markov, Yi~Wu, Glenn Powell, Bob
  McGrew, and Igor Mordatch.
\newblock Emergent tool use from multi-agent autocurricula.
\newblock In \emph{International Conference on Learning Representations}, 2020.

\bibitem[Becker and Bretl(2012)]{Becker12}
Aaron Becker and Timothy Bretl.
\newblock Approximate steering of a unicycle under bounded model perturbation
  using ensemble control.
\newblock \emph{IEEE Transactions on Robotics}, 28\penalty0 (3):\penalty0
  580--591, 2012.

\bibitem[Bellemare et~al.(2020)Bellemare, Candido, Castro, Gong, Machado,
  Moitra, Ponda, and Wang]{Bellemare2020}
Marc~G. Bellemare, Salvatore Candido, Pablo~Samuel Castro, Jun Gong, Marlos~C.
  Machado, Subhodeep Moitra, Sameera~S. Ponda, and Ziyu Wang.
\newblock Autonomous navigation of stratospheric balloons using reinforcement
  learning.
\newblock \emph{Nature}, 588\penalty0 (7836):\penalty0 77--82, 2020.
\newblock \doi{10.1038/s41586-020-2939-8}.
\newblock URL \url{https://doi.org/10.1038/s41586-020-2939-8}.

\bibitem[Bellman(1961)]{Bellman1961}
R.~Bellman.
\newblock \emph{Adaptive Control Processes: A Guided Tour}.
\newblock Princeton Legacy Library. Princeton University Press, 1961.

\bibitem[Bellman et~al.(1957)Bellman, Corporation, and Collection]{Bellman1957}
R.~Bellman, Rand Corporation, and Karreman Mathematics~Research Collection.
\newblock \emph{Dynamic Programming}.
\newblock Rand Corporation research study. Princeton University Press, 1957.

\bibitem[Bensoussan et~al.(2013)Bensoussan, Frehse, and Yam]{Bensoussan2013}
A.~Bensoussan, J.~Frehse, and P.~Yam.
\newblock \emph{Mean Field Games and Mean Field Type Control Theory}.
\newblock SpringerBriefs in Mathematics. Springer New York, 2013.
\newblock ISBN 9781461485087.

\bibitem[Bertsekas and Tsitsiklis(1996)]{Bertsekas1996}
D.~Bertsekas and J.N. Tsitsiklis.
\newblock \emph{Neuro-Dynamic Programming}.
\newblock Athena Scientific, 1996.

\bibitem[Bertsekas(2019)]{Bertsekas19}
Dimitri Bertsekas.
\newblock \emph{Reinforcement Learning and Optimal Control}.
\newblock Athena Scientific, 2019.

\bibitem[Billingsley(1995)]{Billingsley95}
P.~Billingsley.
\newblock \emph{Probability and Measure}, volume 245 of \emph{Wiley Series in
  Probability and Statistics}.
\newblock Wiley, 3 edition, 1995.

\bibitem[Brockett(2015)]{Brockett15}
Roger~W. Brockett.
\newblock \emph{Finite Dimensional Linear Systems}, volume~74 of \emph{Classics
  in Applied Mathematics}.
\newblock Society for Industrial and Applied Mathematics, 2015.

\bibitem[Bukov et~al.(2018)Bukov, Day, Sels, Weinberg, Polkovnikov, and
  Mehta]{Bukov2018}
Marin Bukov, Alexandre G.~R. Day, Dries Sels, Phillip Weinberg, Anatoli
  Polkovnikov, and Pankaj Mehta.
\newblock Reinforcement learning in different phases of quantum control.
\newblock \emph{Phys. Rev. X}, 8:\penalty0 031086, Sep 2018.

\bibitem[Bu{\c{s}}oniu et~al.(2010)Bu{\c{s}}oniu, Babu{\v{s}}ka, and
  De~Schutter]{Lucian2010}
Lucian Bu{\c{s}}oniu, Robert Babu{\v{s}}ka, and Bart De~Schutter.
\newblock \emph{Multi-agent Reinforcement Learning: An Overview}, pages
  183--221.
\newblock Springer Berlin Heidelberg, Berlin, Heidelberg, 2010.

\bibitem[Carmona et~al.(2019)Carmona, Lauri{\`e}re, and Tan]{Carmona2019}
Ren{\'e} Carmona, Mathieu Lauri{\`e}re, and Zongjun Tan.
\newblock Linear-quadratic mean-field reinforcement learning: convergence of
  policy gradient methods.
\newblock \emph{arXiv preprint arXiv:1910.04295}, 2019.

\bibitem[Carmona et~al.(2020)Carmona, Hamidouche, Laurière, and
  Tan]{Carmona2020}
Ren\'{e} Carmona, Kenza Hamidouche, Mathieu Laurière, and Zongjun Tan.
\newblock Policy optimization for linear-quadratic zero-sum mean-field type
  games.
\newblock In \emph{2020 59th IEEE Conference on Decision and Control (CDC)},
  pages 1038--1043, 2020.

\bibitem[Cavanagh et~al.(2010)Cavanagh, Skelton, Fairbrother, Rance, and Arthur
  G.~Palmer]{Cavanagh10}
John Cavanagh, Nicholas~J. Skelton, Wayne~J. Fairbrother, Mark Rance, and III
  Arthur G.~Palmer.
\newblock \emph{Protein NMR Spectroscopy: Principles and Practice}.
\newblock Elsevier, 2 edition, 2010.

\bibitem[Chen et~al.(2014)Chen, Dong, Long, Petersen, and Rabitz]{Chen14}
Chunlin Chen, Daoyi Dong, Ruixing Long, Ian~R. Petersen, and Herschel~A.
  Rabitz.
\newblock Sampling-based learning control of inhomogeneous quantum ensembles.
\newblock \emph{Phys. Rev. A}, 89:\penalty0 023402, Feb 2014.
\newblock \doi{10.1103/PhysRevA.89.023402}.
\newblock URL \url{https://link.aps.org/doi/10.1103/PhysRevA.89.023402}.

\bibitem[Chen et~al.(2011)Chen, Torrontegui, Stefanatos, Li, and
  Muga]{Li_PRA_Transport11}
X.~Chen, E.~Torrontegui, D.~Stefanatos, J.-S. Li, and J.~G. Muga.
\newblock Optimal trajectories for efficient atomic transport without final
  excitation.
\newblock \emph{Physical Review A}, 84:\penalty0 043415, 2011.

\bibitem[Ching and Ritt(2013)]{Ching13}
ShiNung Ching and Jason~T. Ritt.
\newblock Control strategies for underactuated neural ensembles driven by
  optogenetic stimulation.
\newblock \emph{Front Neural Circuits}, 7:\penalty0 54, 2013.

\bibitem[Dong et~al.(2008)Dong, Chen, Li, and Tarn]{Dong08}
Daoyi Dong, Chunlin Chen, Hanxiong Li, and Tzyh-Jong Tarn.
\newblock Quantum reinforcement learning.
\newblock \emph{IEEE Transactions on Systems, Man, and Cybernetics, Part B
  (Cybernetics)}, 38\penalty0 (5):\penalty0 1207--1220, 2008.
\newblock \doi{10.1109/TSMCB.2008.925743}.

\bibitem[Evans et~al.(2020)Evans, Periera, Boutselis, and Theodorou]{Evans2020}
Ethan~N. Evans, Marcus~A. Periera, George~I. Boutselis, and Evangelos~A.
  Theodorou.
\newblock Variational optimization based reinforcement learning for infinite
  dimensional stochastic systems.
\newblock In Leslie~Pack Kaelbling, Danica Kragic, and Komei Sugiura, editors,
  \emph{Proceedings of the Conference on Robot Learning}, volume 100 of
  \emph{Proceedings of Machine Learning Research}, pages 1231--1246. PMLR, 30
  Oct--01 Nov 2020.

\bibitem[Evans(2010)]{Evans10}
Lawrence~C. Evans.
\newblock \emph{Partial Differential Equations}, volume~19 of \emph{Graduate
  Studies in Mathematics}.
\newblock American Mathematical Society, 2nd edition, 2010.

\bibitem[Foerster et~al.(2016)Foerster, Assael, de~Freitas, and
  Whiteson]{Lee2016}
Jakob Foerster, Ioannis~Alexandros Assael, Nando de~Freitas, and Shimon
  Whiteson.
\newblock Learning to communicate with deep multi-agent reinforcement learning.
\newblock In D.~Lee, M.~Sugiyama, U.~Luxburg, I.~Guyon, and R.~Garnett,
  editors, \emph{Advances in Neural Information Processing Systems}, volume~29.
  Curran Associates, Inc., 2016.

\bibitem[Folland(2013)]{Folland13}
Gerald~B. Folland.
\newblock \emph{Real Analysis: Modern Techniques and Their Applications},
  volume~40 of \emph{Pure and Applied Mathematics: A Wiley Series of Texts,
  Monographs and Tracts}.
\newblock John Wiley \& Sons, 2 edition, 2013.

\bibitem[Fran{\c c}ois-Lavet et~al.(2018)Fran{\c c}ois-Lavet, Henderson, Islam,
  Bellemare, and Pineau]{Vincent2018}
Vincent Fran{\c c}ois-Lavet, Peter Henderson, Riashat Islam, Marc~G. Bellemare,
  and Joelle Pineau.
\newblock An introduction to deep reinforcement learning.
\newblock \emph{Foundations and Trends{\textregistered} in Machine Learning},
  11\penalty0 (3-4):\penalty0 219--354, 2018.
\newblock ISSN 1935-8237.
\newblock \doi{10.1561/2200000071}.

\bibitem[Fu et~al.(2020)Fu, Yang, Chen, and Wang]{Fu2020}
Zuyue Fu, Zhuoran Yang, Yongxin Chen, and Zhaoran Wang.
\newblock Actor-critic provably finds nash equilibria of linear-quadratic
  mean-field games.
\newblock In \emph{International Conference on Learning Representations}, 2020.

\bibitem[Gelada et~al.(2019)Gelada, Kumar, Buckman, Nachum, and
  Bellemare]{Chaudhuri2019}
Carles Gelada, Saurabh Kumar, Jacob Buckman, Ofir Nachum, and Marc~G.
  Bellemare.
\newblock {D}eep{MDP}: Learning continuous latent space models for
  representation learning.
\newblock In Kamalika Chaudhuri and Ruslan Salakhutdinov, editors,
  \emph{Proceedings of the 36th International Conference on Machine Learning},
  volume~97 of \emph{Proceedings of Machine Learning Research}, pages
  2170--2179. PMLR, 09--15 Jun 2019.

\bibitem[Glaser et~al.(1998)Glaser, Schulte-{Herbr\"uggen}, Sieveking,
  O.~Schedletzky, {S{\o}rensen}, and Griesinger]{Glaser98}
S.~J. Glaser, T.~Schulte-{Herbr\"uggen}, M.~Sieveking, N.~C.~Nielsen
  O.~Schedletzky, O.~W. {S{\o}rensen}, and C.~Griesinger.
\newblock Unitary control in quantum ensembles, maximizing signal intensity in
  coherent spectroscopy.
\newblock \emph{Science}, 280:\penalty0 421--424, 1998.

\bibitem[Gupta et~al.(2017)Gupta, Egorov, and Kochenderfer]{Gupta2017}
Jayesh~K. Gupta, Maxim Egorov, and Mykel Kochenderfer.
\newblock Cooperative multi-agent control using deep reinforcement learning.
\newblock In Gita Sukthankar and Juan~A. Rodriguez-Aguilar, editors,
  \emph{Autonomous Agents and Multiagent Systems}, pages 66--83, Cham, 2017.
  Springer International Publishing.
\newblock ISBN 978-3-319-71682-4.

\bibitem[Hamburger(1920)]{Hamburger20}
Hans~L. Hamburger.
\newblock \"{U}ber eine erweiterung des stieltjesschen momentenproblems.
\newblock \emph{Mathematische Annalen}, 82:\penalty0 120--164, 1920.

\bibitem[Hamburger(1921{\natexlab{a}})]{Hamburger20_2}
Hans~L. Hamburger.
\newblock \"{U}ber eine erweiterung des stieltjesschen momentenproblems.
\newblock \emph{Mathematische Annalen}, 82:\penalty0 168--187,
  1921{\natexlab{a}}.

\bibitem[Hamburger(1921{\natexlab{b}})]{Hamburger21}
Hans~L. Hamburger.
\newblock \"{U}ber eine erweiterung des stieltjesschen momentenproblems.
\newblock \emph{Mathematische Annalen}, 81:\penalty0 235--319,
  1921{\natexlab{b}}.

\bibitem[Hastie et~al.(2009)Hastie, Tibshirani, and Friedman]{Hastie2009}
T.~Hastie, R.~Tibshirani, and J.H. Friedman.
\newblock \emph{The Elements of Statistical Learning: Data Mining, Inference,
  and Prediction}.
\newblock Springer series in statistics. Springer, 2009.

\bibitem[Haug et~al.(2021)Haug, Dumke, Kwek, Miniatura, and Amico]{Haug2021}
Tobias Haug, Rainer Dumke, Leong-Chuan Kwek, Christian Miniatura, and Luigi
  Amico.
\newblock Machine-learning engineering of quantum currents.
\newblock \emph{Phys. Rev. Res.}, 3:\penalty0 013034, Jan 2021.

\bibitem[Hausdorff(1923)]{Hausdorff23}
Felix Hausdorff.
\newblock Momentprobleme f{\"u}r ein endliches intervall.
\newblock \emph{Mathematische Zeitschrift}, 16\penalty0 (1):\penalty0 220--248,
  1923.

\bibitem[Herculano-Houzel(2012)]{Suzana2012}
Suzana Herculano-Houzel.
\newblock The remarkable, yet not extraordinary, human brain as a scaled-up
  primate brain and its associated cost.
\newblock \emph{Proceedings of the National Academy of Sciences}, 109\penalty0
  (supplement\_1):\penalty0 10661--10668, 2012.

\bibitem[Heredia et~al.(2020)Heredia, Ghadialy, and Mou]{Heredia20}
Paulo Heredia, Hasan Ghadialy, and Shaoshuai Mou.
\newblock Finite-sample analysis of distributed q-learning for multi-agent
  networks.
\newblock In \emph{2020 American Control Conference (ACC)}, pages 3511--3516,
  2020.

\bibitem[Heredia et~al.(2022)Heredia, George, and Mou]{Heredia2022}
Paulo Heredia, Jemin George, and Shaoshuai Mou.
\newblock Distributed offline reinforcement learning.
\newblock In \emph{2022 IEEE 61st Conference on Decision and Control (CDC)},
  pages 4621--4626, 2022.

\bibitem[Heredia and Mou(2019)]{Mou2019}
Paulo~C. Heredia and Shaoshuai Mou.
\newblock Distributed multi-agent reinforcement learning by actor-critic
  method.
\newblock \emph{IFAC-PapersOnLine}, 52\penalty0 (20):\penalty0 363--368, 2019.
\newblock ISSN 2405-8963.
\newblock \doi{https://doi.org/10.1016/j.ifacol.2019.12.182}.
\newblock URL
  \url{https://www.sciencedirect.com/science/article/pii/S240589631932035X}.
\newblock 8th IFAC Workshop on Distributed Estimation and Control in Networked
  Systems NECSYS 2019.

\bibitem[Jacobson and Mayne(1970)]{Jacobson70}
David~H. Jacobson and David~Q. Mayne.
\newblock \emph{Differential Dynamic Programming}.
\newblock Modern Analytic and Computational Mehtods in Sciencen and
  Mathematics. American Elsevier Publishing Company, 1970.

\bibitem[Jiang et~al.(2021)Jiang, Narayanan, and Li]{Jing21}
Wei-Cheng Jiang, Vignesh Narayanan, and Jr-Shin Li.
\newblock Model learning and knowledge sharing for cooperative multiagent
  systems in stochastic environment.
\newblock \emph{IEEE Transactions on Cybernetics}, 51\penalty0 (12):\penalty0
  5717--5727, 2021.

\bibitem[Kaiser et~al.(2020)Kaiser, Babaeizadeh, Milos, Osinski, Campbell,
  Czechowski, Erhan, Finn, Kozakowski, Levine, Mohiuddin, Sepassi, Tucker, and
  Michalewski]{Kaiser2020}
Lukasz Kaiser, Mohammad Babaeizadeh, Piotr Milos, Blazej Osinski, Roy~H
  Campbell, Konrad Czechowski, Dumitru Erhan, Chelsea Finn, Piotr Kozakowski,
  Sergey Levine, Afroz Mohiuddin, Ryan Sepassi, George Tucker, and Henryk
  Michalewski.
\newblock Model based reinforcement learning for atari.
\newblock In \emph{International Conference on Learning Representations}, 2020.

\bibitem[Koppel et~al.(2021)Koppel, Warnell, Stump, Stone, and
  Ribeiro]{Alec2021}
Alec Koppel, Garrett Warnell, Ethan Stump, Peter Stone, and Alejandro Ribeiro.
\newblock Policy evaluation in continuous mdps with efficient kernelized
  gradient temporal difference.
\newblock \emph{IEEE Transactions on Automatic Control}, 66\penalty0
  (4):\penalty0 1856--1863, 2021.

\bibitem[Kwakernaak and Sivan(1972)]{Kwakernaak72}
Huibert Kwakernaak and Raphel Sivan.
\newblock \emph{Linear Optimal Control Systems}.
\newblock Wiley-Interscience, 1972.

\bibitem[Lamata(2017)]{Lamata2017}
Lucas Lamata.
\newblock Basic protocols in quantum reinforcement learning with
  superconducting circuits.
\newblock \emph{Scientific Reports}, 7\penalty0 (1):\penalty0 1609, 2017.
\newblock \doi{10.1038/s41598-017-01711-6}.
\newblock URL \url{https://doi.org/10.1038/s41598-017-01711-6}.

\bibitem[Lang(1999)]{Lang99}
Serge Lang.
\newblock \emph{Fundamentals of Differential Geometry}, volume 191 of
  \emph{Graduate Texts in Mathematics}.
\newblock Springer New York, NY, 1999.

\bibitem[Laskin et~al.(2020)Laskin, Srinivas, and Abbeel]{Daume2020}
Michael Laskin, Aravind Srinivas, and Pieter Abbeel.
\newblock {CURL}: Contrastive unsupervised representations for reinforcement
  learning.
\newblock In Hal~Daum{\'e} III and Aarti Singh, editors, \emph{Proceedings of
  the 37th International Conference on Machine Learning}, volume 119 of
  \emph{Proceedings of Machine Learning Research}, pages 5639--5650. PMLR,
  13--18 Jul 2020.

\bibitem[Lauri{\`e}re et~al.(2022)Lauri{\`e}re, Perrin, Girgin, Muller, Jain,
  Cabannes, Piliouras, P'erolat, {\'E}lie, Pietquin, and Geist]{Laurire2022}
Mathieu Lauri{\`e}re, Sarah Perrin, Sertan Girgin, Paul Muller, Ayush Jain,
  Th{\'e}ophile Cabannes, Georgios Piliouras, Julien P'erolat, Romuald
  {\'E}lie, Olivier Pietquin, and Matthieu Geist.
\newblock Scalable deep reinforcement learning algorithms for mean field games.
\newblock In \emph{International Conference on Machine Learning}, 2022.

\bibitem[Le et~al.(2022)Le, Rathour, Yamazaki, Luu, and Savvides]{Le2022}
Ngan Le, Vidhiwar~Singh Rathour, Kashu Yamazaki, Khoa Luu, and Marios Savvides.
\newblock Deep reinforcement learning in computer vision: a comprehensive
  survey.
\newblock \emph{Artificial Intelligence Review}, 55\penalty0 (4):\penalty0
  2733--2819, 2022.
\newblock \doi{10.1007/s10462-021-10061-9}.
\newblock URL \url{https://doi.org/10.1007/s10462-021-10061-9}.

\bibitem[Lee(2012)]{Lee12}
John~M. Lee.
\newblock \emph{Introduction to Smooth Manifolds}, volume 218 of \emph{Graduate
  Texts in Mathematics}.
\newblock Springer New York, NY, 2nd edition, 2012.

\bibitem[Lever and Stafford(2015)]{Guy2015}
Guy Lever and Ronnie Stafford.
\newblock Modelling policies in mdps in reproducing kernel hilbert space.
\newblock In Guy Lebanon and S.~V.~N. Vishwanathan, editors, \emph{Proceedings
  of the Eighteenth International Conference on Artificial Intelligence and
  Statistics}, volume~38 of \emph{Proceedings of Machine Learning Research},
  pages 590--598, San Diego, California, USA, 09--12 May 2015. PMLR.

\bibitem[Li(2006)]{Li_thesis}
Jr-Shin Li.
\newblock Control of inhomogeneous ensembles, May 2006.

\bibitem[Li(2011)]{Li_TAC11}
Jr-Shin Li.
\newblock Ensemble control of finite-dimensional time-varying linear systems.
\newblock \emph{IEEE Transactions on Automatic Control}, 56\penalty0
  (2):\penalty0 345--357, 2011.

\bibitem[Li and Khaneja(2009)]{Li09}
Jr-Shin Li and Navin Khaneja.
\newblock Ensemble control of bloch equations.
\newblock \emph{IEEE Transactions on Automatic Control}, 54\penalty0
  (3):\penalty0 528--536, 2009.

\bibitem[Li et~al.(2011)Li, Ruths, Yu, Arthanari, and Wagner]{Li_PNAS11}
Jr-Shin Li, Justin Ruths, Tsyr-Yan Yu, Haribabu Arthanari, and Gerhard Wagner.
\newblock Optimal pulse design in quantum control: A unified computational
  method.
\newblock \emph{Proceedings of the National Academy of Sciences}, 108\penalty0
  (5):\penalty0 1879--1884, 2011.

\bibitem[Li et~al.(2013)Li, Dasanayake, and Ruths]{Li13}
Jr-Shin Li, Isuru Dasanayake, and Justin Ruths.
\newblock Control and synchronization of neuron ensembles.
\newblock \emph{IEEE Transactions on Automatic Control}, 58\penalty0
  (8):\penalty0 1919--1930, 2013.
\newblock \doi{10.1109/TAC.2013.2250112}.

\bibitem[Li et~al.(2022)Li, Zhang, and Kuan]{Li22}
Jr-Shin Li, Wei Zhang, and Yuan-Hung Kuan.
\newblock Moment quantization of inhomogeneous spin ensembles.
\newblock \emph{Annual Reviews in Control}, 54:\penalty0 305--313, 2022.
\newblock ISSN 1367-5788.

\bibitem[Li et~al.(2025)Li, Kuan, and Zhang]{Li_SICON25}
Jr-Shin Li, Yuan-Hung Kuan, and Wei Zhang.
\newblock Optimal quantum control using ensemble quantization.
\newblock \emph{SIAM Journal on Control and Optimization}, 63\penalty0
  (1):\penalty0 S107--S127, 2025.

\bibitem[Littman(1994)]{Cohen1994}
Michael~L. Littman.
\newblock Markov games as a framework for multi-agent reinforcement learning.
\newblock In William~W. Cohen and Haym Hirsh, editors, \emph{Machine Learning
  Proceedings 1994}, pages 157--163. Morgan Kaufmann, San Francisco (CA), 1994.

\bibitem[Liu et~al.(2022)Liu, Lever, Wang, Merel, Eslami, Hennes, Czarnecki,
  Tassa, Omidshafiei, Abdolmaleki, Siegel, Hasenclever, Marris,
  Tunyasuvunakool, Song, Wulfmeier, Muller, Haarnoja, Tracey, Tuyls, Graepel,
  and Heess]{Liu2022}
Siqi Liu, Guy Lever, Zhe Wang, Josh Merel, S.~M.~Ali Eslami, Daniel Hennes,
  Wojciech~M. Czarnecki, Yuval Tassa, Shayegan Omidshafiei, Abbas Abdolmaleki,
  Noah~Y. Siegel, Leonard Hasenclever, Luke Marris, Saran Tunyasuvunakool,
  H.~Francis Song, Markus Wulfmeier, Paul Muller, Tuomas Haarnoja, Brendan
  Tracey, Karl Tuyls, Thore Graepel, and Nicolas Heess.
\newblock From motor control to team play in simulated humanoid football.
\newblock \emph{Science Robotics}, 7\penalty0 (69):\penalty0 eabo0235, 2022.

\bibitem[Long et~al.(2018)Long, Fan, Liao, Liu, Zhang, and Pan]{Long2018}
Pinxin Long, Tingxiang Fan, Xinyi Liao, Wenxi Liu, Hao Zhang, and Jia Pan.
\newblock Towards optimally decentralized multi-robot collision avoidance via
  deep reinforcement learning.
\newblock In \emph{2018 IEEE International Conference on Robotics and
  Automation (ICRA)}, pages 6252--6259, 2018.

\bibitem[Lu et~al.(2024)Lu, Zhou, and Mou]{Lu2024}
Zehui Lu, Tianyu Zhou, and Shaoshuai Mou.
\newblock Real-time multi-robot mission planning in cluttered environment.
\newblock \emph{Robotics}, 13\penalty0 (3), 2024.

\bibitem[Mackey(1980)]{Mackey80}
George~W. Mackey.
\newblock Harmonic analysis as the exploitation of symmetry - a historical
  survey.
\newblock \emph{Bulletin (New Series) of the American Mathematical Society.},
  3\penalty0 (1):\penalty0 543--698, 1980.

\bibitem[Marks(2005)]{Marks05}
W.J. Marks.
\newblock William j. marks.
\newblock \emph{Current Treatment Options in Neurology}, 7:\penalty0 237--243,
  2005.

\bibitem[Matari{\'c}(1997)]{Mataric1997}
Maja~J. Matari{\'c}.
\newblock Reinforcement learning in the multi-robot domain.
\newblock \emph{Autonomous Robots}, 4\penalty0 (1):\penalty0 73--83, 1997.
\newblock \doi{10.1023/A:1008819414322}.
\newblock URL \url{https://doi.org/10.1023/A:1008819414322}.

\bibitem[Mayne(1966)]{Mayne66}
David~Q. Mayne.
\newblock A second-order gradient method for determining optimal trajectories
  of non-linear discrete-time systems.
\newblock \emph{International Journal of Control}, 3\penalty0 (1):\penalty0
  85--95, 1966.

\bibitem[Mnih et~al.(2015)Mnih, Kavukcuoglu, Silver, Rusu, Veness, Bellemare,
  Graves, Riedmiller, Fidjeland, Ostrovski, Petersen, Beattie, Sadik,
  Antonoglou, King, Kumaran, Wierstra, Legg, and Hassabis]{Mnih2015}
Volodymyr Mnih, Koray Kavukcuoglu, David Silver, Andrei~A. Rusu, Joel Veness,
  Marc~G. Bellemare, Alex Graves, Martin Riedmiller, Andreas~K. Fidjeland,
  Georg Ostrovski, Stig Petersen, Charles Beattie, Amir Sadik, Ioannis
  Antonoglou, Helen King, Dharshan Kumaran, Daan Wierstra, Shane Legg, and
  Demis Hassabis.
\newblock Human-level control through deep reinforcement learning.
\newblock \emph{Nature}, 518\penalty0 (7540):\penalty0 529--533, 2015.

\bibitem[Momennejad et~al.(2017)Momennejad, Russek, Cheong, Botvinick, Daw, and
  Gershman]{Momennejad2017}
I.~Momennejad, E.~M. Russek, J.~H. Cheong, M.~M. Botvinick, N.~D. Daw, and
  S.~J. Gershman.
\newblock The successor representation in human reinforcement learning.
\newblock \emph{Nature Human Behaviour}, 1\penalty0 (9):\penalty0 680--692,
  2017.
\newblock \doi{10.1038/s41562-017-0180-8}.
\newblock URL \url{https://doi.org/10.1038/s41562-017-0180-8}.

\bibitem[Munkres(2000)]{Munkres00}
James~R. Munkres.
\newblock \emph{Topology}.
\newblock Prentice Hall, Incorporated, 2000.

\bibitem[Nakamura-Zimmerer et~al.(2021)Nakamura-Zimmerer, Gong, and
  Kang]{Tenavi2021}
Tenavi Nakamura-Zimmerer, Qi~Gong, and Wei Kang.
\newblock Adaptive deep learning for high-dimensional hamilton--jacobi--bellman
  equations.
\newblock \emph{SIAM Journal on Scientific Computing}, 43\penalty0
  (2):\penalty0 A1221--A1247, 2021.

\bibitem[Narayanan et~al.(2024)Narayanan, Zhang, and Li]{Narayanan2024}
Vignesh Narayanan, Wei Zhang, and Jr-Shin Li.
\newblock Duality of ensemble systems through moment representations.
\newblock \emph{IEEE Transactions on Automatic Control}, pages 1--8, 2024.
\newblock \doi{10.1109/TAC.2024.3397159}.

\bibitem[Nishimura et~al.(2001)Nishimura, Fu, and Cross]{Nishimura2001}
Katsuyuki Nishimura, Riqiang Fu, and Timothy~A. Cross.
\newblock The effect of rf inhomogeneity on heteronuclear dipolar recoupling in
  solid state nmr: Practical performance of sfam and redor.
\newblock \emph{Journal of Magnetic Resonance}, 152\penalty0 (2):\penalty0
  227--233, 2001.
\newblock ISSN 1090-7807.

\bibitem[{\O}ksendal(2003)]{Oksendal2003}
B.~{\O}ksendal.
\newblock \emph{Stochastic Differential Equations: An Introduction with
  Applications}.
\newblock Universitext (1979). Springer Berlin, Heidelberg, 6 edition, 2003.

\bibitem[OpenAI et~al.(2019)OpenAI, :, Berner, Brockman, Chan, Cheung, Debiak,
  Dennison, Farhi, Fischer, Hashme, Hesse, J{\'o}zefowicz, Gray, Olsson,
  Pachocki, Petrov, d.~O.~Pinto, Raiman, Salimans, Schlatter, Schneider, Sidor,
  Sutskever, Tang, Wolski, and Zhang]{OpenAI2019}
OpenAI, :, Christopher Berner, Greg Brockman, Brooke Chan, Vicki Cheung,
  Przemyslaw Debiak, Christy Dennison, David Farhi, Quirin Fischer, Shariq
  Hashme, Chris Hesse, Rafal J{\'o}zefowicz, Scott Gray, Catherine Olsson,
  Jakub Pachocki, Michael Petrov, Henrique~P. d.~O.~Pinto, Jonathan Raiman, Tim
  Salimans, Jeremy Schlatter, Jonas Schneider, Szymon Sidor, Ilya Sutskever,
  Jie Tang, Filip Wolski, and Susan Zhang.
\newblock Dota 2 with large scale deep reinforcement learning, 2019.
\newblock URL \url{https://arxiv.org/abs/1912.06680}.

\bibitem[P{\'a}sztor et~al.(2023)P{\'a}sztor, Krause, and
  Bogunovic]{Pasztor2023}
Barna P{\'a}sztor, Andreas Krause, and Ilija Bogunovic.
\newblock Efficient model-based multi-agent mean-field reinforcement learning.
\newblock \emph{Transactions on Machine Learning Research}, 2023.
\newblock ISSN 2835-8856.
\newblock URL \url{https://openreview.net/forum?id=gvcDSDYUZx}.

\bibitem[Pathria and Beale(2021)]{Pathria2021}
R.K. Pathria and P.D. Beale.
\newblock \emph{Statistical Mechanics}.
\newblock Elsevier Science, 2021.

\bibitem[Paulsen and Raghupathi(2016)]{Paulsen2016}
V.I. Paulsen and M.~Raghupathi.
\newblock \emph{An Introduction to the Theory of Reproducing Kernel Hilbert
  Spaces}.
\newblock Cambridge Studies in Advanced Mathematics. Cambridge University
  Press, 2016.

\bibitem[Roos and Moelmer(2004)]{Molmer04}
I.~Roos and K.~Moelmer.
\newblock Quantum computing with an inhomogeneously broadened ensemble of ions:
  Suppression of errors from detuning variations by specially adapted pulses
  and coherent population trapping.
\newblock \emph{Physical Review A}, 69:\penalty0 022321, 2004.

\bibitem[Rudin(1976)]{Rudin76}
Walter Rudin.
\newblock \emph{Principles of Mathematical Analysis}.
\newblock McGraw-Hill, 3 edition, 1976.

\bibitem[Sarang and Poullis(2023)]{Sarang23}
Nima Sarang and Charalambos Poullis.
\newblock Tractable large-scale deep reinforcement learning.
\newblock \emph{Computer Vision and Image Understanding}, 232:\penalty0 103689,
  2023.
\newblock ISSN 1077-3142.
\newblock \doi{https://doi.org/10.1016/j.cviu.2023.103689}.
\newblock URL
  \url{https://www.sciencedirect.com/science/article/pii/S1077314223000693}.

\bibitem[Schrittwieser et~al.(2020)Schrittwieser, Antonoglou, Hubert, Simonyan,
  Sifre, Schmitt, Guez, Lockhart, Hassabis, Graepel, Lillicrap, and
  Silver]{Schrittwieser2020}
Julian Schrittwieser, Ioannis Antonoglou, Thomas Hubert, Karen Simonyan,
  Laurent Sifre, Simon Schmitt, Arthur Guez, Edward Lockhart, Demis Hassabis,
  Thore Graepel, Timothy Lillicrap, and David Silver.
\newblock Mastering atari, go, chess and shogi by planning with a learned
  model.
\newblock \emph{Nature}, 588\penalty0 (7839):\penalty0 604--609, 2020.
\newblock \doi{10.1038/s41586-020-03051-4}.
\newblock URL \url{https://doi.org/10.1038/s41586-020-03051-4}.

\bibitem[Schulman et~al.(2015)Schulman, Levine, Abbeel, Jordan, and
  Moritz]{Schulman15}
John Schulman, Sergey Levine, Pieter Abbeel, Michael Jordan, and Philipp
  Moritz.
\newblock Trust region policy optimization.
\newblock In Francis Bach and David Blei, editors, \emph{Proceedings of the
  32nd International Conference on Machine Learning}, volume~37 of
  \emph{Proceedings of Machine Learning Research}, pages 1889--1897, Lille,
  France, 07--09 Jul 2015. PMLR.

\bibitem[Schulman et~al.(2017)Schulman, Wolski, Dhariwal, Radford, and
  Klimov]{Schulman2017}
John Schulman, Filip Wolski, Prafulla Dhariwal, Alec Radford, and Oleg Klimov.
\newblock Proximal policy optimization algorithms, 2017.

\bibitem[Shahrokhi et~al.(2018)Shahrokhi, Lin, Ertel, Wan, and
  Becker]{Shahrokhi2018}
Shiva Shahrokhi, Lillian Lin, Chris Ertel, Mable Wan, and Aaron~T. Becker.
\newblock Steering a swarm of particles using global inputs and swarm
  statistics.
\newblock \emph{IEEE Transactions on Robotics}, 34\penalty0 (1):\penalty0
  207--219, 2018.

\bibitem[Shoeb(2009)]{Shoeb09}
Ali~H. Shoeb.
\newblock Application of machine learning to epileptic seizure onset detection
  and treatment, September 2009.

\bibitem[Silver et~al.(2017)Silver, Schrittwieser, Simonyan, Antonoglou, Huang,
  Guez, Hubert, Baker, Lai, Bolton, Chen, Lillicrap, Hui, Sifre, van~den
  Driessche, Graepel, and Hassabis]{Silver2017}
David Silver, Julian Schrittwieser, Karen Simonyan, Ioannis Antonoglou, Aja
  Huang, Arthur Guez, Thomas Hubert, Lucas Baker, Matthew Lai, Adrian Bolton,
  Yutian Chen, Timothy Lillicrap, Fan Hui, Laurent Sifre, George van~den
  Driessche, Thore Graepel, and Demis Hassabis.
\newblock Mastering the game of go without human knowledge.
\newblock \emph{Nature}, 550\penalty0 (7676):\penalty0 354--359, 2017.
\newblock \doi{10.1038/nature24270}.
\newblock URL \url{https://doi.org/10.1038/nature24270}.

\bibitem[Silver et~al.(2018)Silver, Hubert, Schrittwieser, Antonoglou, Lai,
  Guez, Lanctot, Sifre, Kumaran, Graepel, Lillicrap, Simonyan, and
  Hassabis]{Silver2018}
David Silver, Thomas Hubert, Julian Schrittwieser, Ioannis Antonoglou, Matthew
  Lai, Arthur Guez, Marc Lanctot, Laurent Sifre, Dharshan Kumaran, Thore
  Graepel, Timothy Lillicrap, Karen Simonyan, and Demis Hassabis.
\newblock A general reinforcement learning algorithm that masters chess, shogi,
  and go through self-play.
\newblock \emph{Science}, 362\penalty0 (6419):\penalty0 1140--1144, 2018.

\bibitem[Silver et~al.(1985)Silver, Joseph, and Hoult]{Silver85}
M.~S. Silver, R.~I. Joseph, and D.~I. Hoult.
\newblock Selective spin inversion in nuclear magnetic resonance and coherent
  optics through an exact solution of the bloch-riccati equation.
\newblock \emph{Physical Review A}, 31\penalty0 (4):\penalty0 2753--2755, 1985.

\bibitem[Sontag(1998)]{Sontag98}
Eduardo~D. Sontag.
\newblock \emph{Mathematical Control Theory: Deterministic Finite Dimensional
  Systems}.
\newblock Springer New York, NY, 2nd edition, 1998.

\bibitem[Stefanatos and Li(2011)]{Li_SICON11}
D.~Stefanatos and J.-S. Li.
\newblock Minimum-time frictionless atom cooling in harmonic traps.
\newblock \emph{SIAM Journal on Control and Optimization}, 49:\penalty0
  2440--2462, 2011.

\bibitem[Stefanatos and Li(2014)]{Li_TAC14}
D.~Stefanatos and J.-S. Li.
\newblock Minimum-time quantum transport with bounded trap velocity.
\newblock \emph{IEEE Transactions on Automatic Control}, 59\penalty0
  (3):\penalty0 733--738, 2014.

\bibitem[Stieltjes(1993)]{Stieltjes93}
Thomas~J. Stieltjes.
\newblock \emph{{\OE}uvres Compl{\`e}tes II - Collected Papers II}.
\newblock Springer Collected Works in Mathematics. Springer Berlin, Heidelberg,
  1993.

\bibitem[Subramanian and Mahajan(2019)]{Subramanian2019}
Jayakumar Subramanian and Aditya Mahajan.
\newblock Reinforcement learning in stationary mean-field games.
\newblock Richland, SC, 2019. International Foundation for Autonomous Agents
  and Multiagent Systems.

\bibitem[Sutton and Barto(2018)]{Sutton18}
Richard~S. Sutton and Andrew~G. Barto.
\newblock \emph{Reinforcement Learning An Introduction}.
\newblock MIT Press, 2018.

\bibitem[Vinyals et~al.(2019)Vinyals, Babuschkin, Czarnecki, Mathieu, Dudzik,
  Chung, Choi, Powell, Ewalds, Georgiev, Oh, Horgan, Kroiss, Danihelka, Huang,
  Sifre, Cai, Agapiou, Jaderberg, Vezhnevets, Leblond, Pohlen, Dalibard,
  Budden, Sulsky, Molloy, Paine, Gulcehre, Wang, Pfaff, Wu, Ring, Yogatama,
  W{\"u}nsch, McKinney, Smith, Schaul, Lillicrap, Kavukcuoglu, Hassabis, Apps,
  and Silver]{Vinyals2019}
Oriol Vinyals, Igor Babuschkin, Wojciech~M. Czarnecki, Micha{\"e}l Mathieu,
  Andrew Dudzik, Junyoung Chung, David~H. Choi, Richard Powell, Timo Ewalds,
  Petko Georgiev, Junhyuk Oh, Dan Horgan, Manuel Kroiss, Ivo Danihelka, Aja
  Huang, Laurent Sifre, Trevor Cai, John~P. Agapiou, Max Jaderberg,
  Alexander~S. Vezhnevets, R{\'e}mi Leblond, Tobias Pohlen, Valentin Dalibard,
  David Budden, Yury Sulsky, James Molloy, Tom~L. Paine, Caglar Gulcehre, Ziyu
  Wang, Tobias Pfaff, Yuhuai Wu, Roman Ring, Dani Yogatama, Dario W{\"u}nsch,
  Katrina McKinney, Oliver Smith, Tom Schaul, Timothy Lillicrap, Koray
  Kavukcuoglu, Demis Hassabis, Chris Apps, and David Silver.
\newblock Grandmaster level in starcraft ii using multi-agent reinforcement
  learning.
\newblock \emph{Nature}, 575\penalty0 (7782):\penalty0 350--354, 2019.
\newblock \doi{10.1038/s41586-019-1724-z}.
\newblock URL \url{https://doi.org/10.1038/s41586-019-1724-z}.

\bibitem[Vu et~al.(2024)Vu, Singhal, Zeng, and Li]{Vu2024}
Minh Vu, Bharat Singhal, Shen Zeng, and Jr-Shin Li.
\newblock {Data-driven control of oscillator networks with population-level
  measurement}.
\newblock \emph{Chaos: An Interdisciplinary Journal of Nonlinear Science},
  34\penalty0 (3):\penalty0 033138, 03 2024.

\bibitem[Wilson(2005)]{Wilson05}
Scott~B. Wilson.
\newblock A neural network method for automatic and incremental learning
  applied to patient-dependent seizure detection.
\newblock \emph{Clinical Neurophysiology}, 116\penalty0 (8):\penalty0
  1785--1795, August 2005.

\bibitem[Xie et~al.(2024)Xie, Mou, and Sundaram]{Xie2024}
Yijing Xie, Shaoshuai Mou, and Shreyas Sundaram.
\newblock Communication-efficient and resilient distributed q-learning.
\newblock \emph{IEEE Transactions on Neural Networks and Learning Systems},
  35\penalty0 (3):\penalty0 3351--3364, 2024.

\bibitem[Yang and Gu(2004)]{Yang2024}
Erfu Yang and Dongbing Gu.
\newblock Multiagent reinforcement learning for multi-robot systems: A survey.
\newblock Technical report, tech. rep, 2004.

\bibitem[Yang and Wang(2019)]{Yang2019}
Lin~F. Yang and Mengdi Wang.
\newblock Sample-optimal parametric q-learning using linearly additive
  features.
\newblock In \emph{International Conference on Machine Learning}, 2019.

\bibitem[Yang et~al.(2020{\natexlab{a}})Yang, Juntao, and Lingling]{Yang2020}
Yang Yang, Li~Juntao, and Peng Lingling.
\newblock Multi-robot path planning based on a deep reinforcement learning dqn
  algorithm.
\newblock \emph{CAAI Transactions on Intelligence Technology}, 5\penalty0
  (3):\penalty0 177--183, 2020{\natexlab{a}}.

\bibitem[Yang et~al.(2018)Yang, Luo, Li, Zhou, Zhang, and Wang]{Yang2018}
Yaodong Yang, Rui Luo, Minne Li, Ming Zhou, Weinan Zhang, and Jun Wang.
\newblock Mean field multi-agent reinforcement learning.
\newblock In Jennifer Dy and Andreas Krause, editors, \emph{Proceedings of the
  35th International Conference on Machine Learning}, volume~80 of
  \emph{Proceedings of Machine Learning Research}, pages 5567--5576,
  Stockholmsmässan, Stockholm Sweden, 10--15 Jul 2018. PMLR.

\bibitem[Yang et~al.(2020{\natexlab{b}})Yang, Jin, Wang, Wang, and
  Jordan]{Yang2020_NIPS}
Zhuoran Yang, Chi Jin, Zhaoran Wang, Mengdi Wang, and Michael~I. Jordan.
\newblock On function approximation in reinforcement learning: optimism in the
  face of large state spaces.
\newblock In \emph{Proceedings of the 34th International Conference on Neural
  Information Processing Systems}, NIPS '20, Red Hook, NY, USA,
  2020{\natexlab{b}}. Curran Associates Inc.

\bibitem[Yazdanbakhsh et~al.(2020)Yazdanbakhsh, Chen, and
  Zheng]{Yazdanbakhsh20}
Amir Yazdanbakhsh, Junchao Chen, and Yu~Zheng.
\newblock Menger: Massively large-scale distributed reinforcement learning.
\newblock \emph{NeurIPS, Beyond Backpropagation Workshop, 2020}, 2020.
\newblock URL \url{https://beyondbackprop.github.io/}.

\bibitem[Yosida(1980)]{Yosida80}
K{\=o}saku Yosida.
\newblock \emph{Functional Analysis}, volume 123 of \emph{Grundlehren der
  mathematischen Wissenschaften}.
\newblock Springer Berlin, Heidelberg, 6 edition, 1980.

\bibitem[Yu et~al.(2023)Yu, Zhang, O'Gara, Li, and Chang]{Yu2023}
Yao-Chi Yu, Wei Zhang, David O'Gara, Jr-Shin Li, and Su-Hsin Chang.
\newblock A moment kernel machine for clinical data mining to inform medical
  decision making.
\newblock \emph{Scientific Reports}, 13\penalty0 (1):\penalty0 10459, 2023.
\newblock \doi{10.1038/s41598-023-36752-7}.
\newblock URL \url{https://doi.org/10.1038/s41598-023-36752-7}.

\bibitem[Zhang et~al.(2021{\natexlab{a}})Zhang, McAllister, Calandra, Gal, and
  Levine]{Zhang2021_RL}
Amy Zhang, Rowan~Thomas McAllister, Roberto Calandra, Yarin Gal, and Sergey
  Levine.
\newblock Learning invariant representations for reinforcement learning without
  reconstruction.
\newblock In \emph{International Conference on Learning Representations},
  2021{\natexlab{a}}.

\bibitem[Zhang et~al.(2021{\natexlab{b}})Zhang, Yang, and
  Ba{\c{s}}ar]{Zhang2021}
Kaiqing Zhang, Zhuoran Yang, and Tamer Ba{\c{s}}ar.
\newblock \emph{Multi-Agent Reinforcement Learning: A Selective Overview of
  Theories and Algorithms}.
\newblock Springer International Publishing, Cham, 2021{\natexlab{b}}.

\bibitem[Zhang and Li(2015)]{Zhang15}
Wei Zhang and Jr-Shin Li.
\newblock Uniform and selective excitations of spin ensembles with rf
  inhomogeneity.
\newblock In \emph{2015 54th IEEE Conference on Decision and Control (CDC)},
  pages 5766--5771, 2015.
\newblock \doi{10.1109/CDC.2015.7403125}.

\bibitem[Zlotnik and Li(2012)]{Zlotnik12}
A.~Zlotnik and J.-S. Li.
\newblock Optimal entrainment of neural oscillator ensembles.
\newblock \emph{Journal of Neural Engineering}, 9\penalty0 (4):\penalty0
  046015, July 2012.

\end{thebibliography}

\end{document}